\documentclass[%
 aip,
 amsmath,amssymb,
 reprint,%
]{revtex4-1}

\usepackage{graphicx}
\usepackage{dcolumn}
\usepackage{bm}

\usepackage[utf8]{inputenc}
\usepackage[T1]{fontenc}
\usepackage{mathptmx}
\usepackage{booktabs} 

\usepackage{algorithm,algpseudocode}
\algnewcommand{\To}{\textbf{To }}
\algnewcommand\Input{\item[\textbf{Input:}]}%
\algnewcommand\Output{\item[\textbf{Output:}]}%
\graphicspath{{Images/}} 

\begin{document}

\preprint{AIP/123-QED}

\title[Simulated tempering with irreversible Gibbs sampling techniques]{Simulated tempering with irreversible Gibbs sampling techniques}

\author{Fahim Faizi}
\email{fahim.faizi@kcl.ac.uk}
\affiliation{Department of Mathematics, King's College London, Strand, WC2R 2LS, London, U.K.}

\author{Pedro J. Buigues}%
\affiliation{Department of Chemistry, King's College London, 7 Trinity street,  SE1 1DB, London, U.K.}
\affiliation{Department of Physics and Astronomy, University College London, Gower St, Kings Cross, WC1E 6BT, London, UK.}
 
\author{George Deligiannidis}
\email{deligian@stats.ox.ac.uk}
\affiliation{Department of Statistics, University of Oxford, 24-29 St Giles', OX1 3LB, Oxford, U.K.}

\author{Edina Rosta}
\email{e.rosta@ucl.ac.uk}
\affiliation{Department of Chemistry, King's College London, 7 Trinity street,  SE1 1DB, London, U.K.}
\affiliation{Department of Physics and Astronomy, University College London, Gower St, Kings Cross, WC1E 6BT, London, UK.}

\date{\today}

\begin{abstract}
We present here two novel algorithms for simulated tempering simulations, which break detailed balance condition (DBC) but satisfy the skewed detailed balance to ensure invariance of the target distribution. The irreversible methods we present here are based on Gibbs sampling and concern breaking DBC at the update scheme of the temperature swaps. We utilise three systems as a test bed for our methods: an MCMC simulation on a simple system described by a 1D double well potential, the Ising model and MD simulations on Alanine pentapeptide (ALA5). The relaxation times of inverse temperature, magnetic susceptibility and energy density for the Ising model indicate clear gains in sampling efficiency over conventional Gibbs sampling techniques with DBC and also over the conventionally used simulated tempering with Metropolis-Hastings (MH) scheme. Simulations on ALA5 with large number of temperatures indicate distinct gains in mixing times for inverse temperature and consequently the energy of the system compared to conventional MH. With no additional computational overhead, our methods were found to be more efficient alternatives to conventionally used simulated tempering methods with DBC. Our algorithms should be particularly advantageous in simulations of large systems with many temperature ladders, as our algorithms showed a more favorable constant scaling in Ising spin systems as compared with both reversible and irreversible MH algorithms. In future applications, our irreversible methods can also be easily tailored to utilize a given dynamical variable other than temperature to flatten rugged free energy landscapes.  
\end{abstract}

\maketitle

\section*{Introduction}
Algorithms based on Markov Chain Monte Carlo (MCMC) techniques are the most commonly used in Monte Carlo (MC) simulations. The broadly applicable Metropolis-Hastings (MH) algorithm \cite{Metropolis, Hastings} has been implemented in various fields including physics,\cite{MCMC_physics1, MCMC_physics2} chemical and biological sciences,\cite{MCMC_biochemistry, MCMC_biochemistry2} and economics.\cite{MCMC_finance} In most cases one is interested in sampling from intractable multi-dimensional probability distributions with the intention to  estimate the expectation value of an observable with respect to the given distribution. However when we consider the simulation of complex physical systems, we often find that it remains difficult to efficiently sample them from a target distribution with conventional MCMC algorithms such as the Metropolis-Hastings \cite{Metropolis, Hastings} (MH) and the Gibbs sampler.\cite{Gibbs_sampler} Particularly, systems with multiple minimum energy states, such as bio-polymers and spin glasses, can often get trapped in local minima.

Extended ensemble MCMC techniques \cite{Generalized_ensemble} such as multi-canonical methods,\cite{MUCA_1, MUCA_2} simulated \cite{STM} and parallel tempering \cite{REM_1, REM_2, REM_3} provide a solution to explore the state space more efficiently than it is possible with conventional MCMC methods. In this paper we focus on the simulated tempering method.\cite{STM} In simulated tempering, unlike in conventional methods, the temperature in the Gibbs-Boltzmann distribution is also treated as a dynamical variable along with the configuration. A single replica of the system is therefore simulated with conventional MCMC or MD, while a temperature change is attempted periodically from among a predetermined discrete set of values. Indeed, at higher temperatures free energy barriers are lower, it is therefore more probable that at a higher temperature the system may cross a free energy barrier and then, upon cooling off again, visit a different energy minimum. The exploration of the temperature space therefore allows the system to escape local minimum energy states by simply transitioning to higher temperatures, this allows broad sampling of the state space at high temperatures and thorough sampling around local energy minima at low temperatures.  

Convergence to the correct \emph{enlarged} target distribution, and therefore invariance  at all chosen temperatures, can be ensured by a careful construction of the transition rate for temperature change. The most commonly used criteria for temperature change is the MH algorithm,\cite{Hastings} which ensures invariance through the detailed balance condition (DBC). However DBC is not a strict requirement for invariance\cite{BC_sufficiency1, BC_sufficiency2, BC_sufficiency3}. Several studies have shown that breaking it enhances sampling efficiency \cite{Diaconis,Chen, Barkema, Suwa-Todo, Turitsyn, Weigel, Sakai_Hukushima_1D, Sakai_Hukushima_eigenvalue, Sakai_Hukushima_simulated_tempering, ECMC_continuous_spins, ECMC_heisenberg} and may speed up convergence to the target distribution.\cite{hot_topic_1, hot_topic_2} The lifting framework,\cite{Diaconis} which violates DBC, has been implemented for several systems.\cite{Barkema, Turitsyn, Weigel, Sakai_Hukushima_1D, Sakai_Hukushima_eigenvalue, Sakai_Hukushima_simulated_tempering, ECMC_continuous_spins, ECMC_heisenberg} One of the earlier applications of the lifting technique to one-dimensional random walk showed a square root reduction in the mixing time,\cite{Diaconis} which may be an optimal improvement through the lifting framework.\cite{Chen} In simulated tempering \cite{STM}, the dynamics of the system in temperature space with $K$ predetermined temperatures can be comparable to a random walk on one-dimensional lattice with $K$ sites. In this light, Sakai and Hukushima have implemented the lifting framework with skewed detailed balance condition (SDBC) to the update scheme of the inverse temperature \cite{Sakai_Hukushima_simulated_tempering} and have demonstrated (with the Ising model as a test system) a considerable improvement in the relaxation dynamics of the inverse temperature compared to the standard updating scheme of MH with DBC.

In simulated tempering, the temperature update scheme with the Gibbs sampler (GS)\cite{Gibbs_sampler} and its variant, the Metropolized Gibbs sampler (MGS)\cite{MGS}, has been suggested in some literature. \cite{Rosta_1, Gibbs_ST, ST_Suwa-Todo} The transition rates for both GS and MGS satisfy the strict DBC, however we recently proposed their irreversible counter-parts with SDBC, namely the irreversible Gibbs sampler (IGS) and the irreversible Metropolized-Gibbs sampler (IMGS) respectively.\cite{EIMCS} In this paper we implement IGS and IMGS to the update scheme of inverse temperature in simulated tempering. We apply our simulated tempering methods to three test systems: MCMC simulations on a simple system described by a 1D double well potential and the Ising model and MD simulations on Alanine pentapeptide (ALA5). Applications to the Ising model show that the update scheme of inverse temperature $\beta$ with IGS and IMGS can improve the relaxation dynamics of $\beta$ when compared to their respective reversible counterparts with DBC. Furthermore the gain in relaxation dynamics of $\beta$ gets exceedingly better with increasing domain size $K$ (i.e. the number of temperatures within a fixed range) when compared to both the conventionally used MH algorithm and irreversible Metropolis-Hastings (IMH) with SDBC, as implemented by Sakai and Hukushima.\cite{Sakai_Hukushima_simulated_tempering}. We further demonstrate that both IGS and IMGS reduce the integrated autocorrelation times on magnetic susceptibility and energy density by a considerable factor compared to their reversible counterparts, and significantly so compared to both MH and IMH in large temperature domains. The MD simulations on ALA5 indicate distinct gains in the mixing time of inverse temperature and of total energy for large temperature domain size, but modest gains in the mixing time of the slowest dihedral angles when compared to the conventional simulated tempering with MH.

Assuming  a constant specific heat capacity across the predetermined set of discrete temperatures, the mean energy $\bar{\mathcal{E}}$ of a system can be assumed to scale as $\bar{\mathcal{E}} \sim \mathcal{N}k_{B}T$ with the the degrees of freedom $\mathcal{N}$. \cite{HRPM_01} The difference in mean energy $\Delta \bar{\mathcal{E}} = \bar{\mathcal{E}}(T_2) - \bar{\mathcal{E}}(T_1)$ at two temperatures $T_1$ and $T_2$ therefore scales as $\Delta \bar{\mathcal{E}} \sim \mathcal{N}k_B\Delta T$. In conventional simulated tempering methods with the MH scheme, for systems with large degrees of freedom the temperature spacing is therefore required to be small to ensure overlap of energy distributions at $T_1$ and $T_2$ for reasonable acceptance probability. For simulations of large systems at a fixed temperature range sampling of the temperature space becomes inefficient with the conventional simulated tempering, as one would expect of random walks in domains of increasing size. With our methods the mixing time of inverse temperature and system observables is particularly improved in large temperature domain sizes when compared to conventional methods. We argue that our methods can therefore be more efficient alternatives for the simulation of large systems.

\section*{The simulated tempering method}
One is often interested in using MCMC methods to estimate expectation values under probability distributions with very large dimensions. We may consider a physical system with state space $\Omega $. In classical statistical mechanics the conditional probability of finding the system in a given configuration $\bm{\sigma}\in\Omega$ is given by the Gibbs-Boltzmann distribution,
\begin{equation}\label{Gibbs-Boltz distr}
\pi(\bm{\sigma}\vert \beta) = \frac{1}{Z(\beta)}e^{-\beta H(\bm{\sigma})},
\end{equation}
where $Z(\beta) = \sum_{\Omega}e^{-\beta H(\bm{\sigma})}$ is the partition function for a given inverse temperature $\beta$ and $H(\bm{\sigma})$ is the Hamiltonian of the system. In conventional MCMC methods, such as the Metropolis-Hastings algorithm, configurations are sampled from the Gibbs-Boltzmann distribution at fixed $\beta$. However in simulated tempering $\beta$ is allowed to vary from among a predetermined set of $K$ discrete values $\beta \in \lbrace \beta_1,...,\beta_K  \rbrace$. In simulated tempering  both $\beta$ and $\bm{\sigma} \in \Omega$ are therefore stochastic variables. The original state space is enlarged to $\bar{\Omega}:= \Omega \times \lbrace 1,...,K \rbrace$ and the probability of finding the system in a given state $(\bm{\sigma}, \beta_k) \in \bar{\Omega}$ is given by the joint probability
\begin{equation}\label{enlarged target}
\pi(\bm{\sigma},\beta_k) = \frac{1}{\mathcal{Z}}e^{-\beta_k H(\bm{\sigma}) + w_k},
\end{equation}
where the functions $w_k = w(\beta_k)$ for $k = 1,...,K$ are the weighting factors determined so that the marginal probability distribution, denoted by the probability vector $\bm{\pi}(\beta) = \left(\pi(\beta_1), \pi(\beta_2), ..., \pi(\beta_K)\right)$, is uniform in $\beta$. We will demonstrate this in equation \eqref{marginal distribution} shortly. In simulated tempering, in the update scheme of inverse temperature at fixed $\bm{\sigma}$, we therefore wish to sample from the target probability distribution
\begin{equation}\label{original target}
\bm{\bar{\pi}} = \left(\pi(\beta_1 \vert \bm{\sigma}),\pi(\beta_2 \vert \bm{\sigma}), ..., \pi(\beta_K \vert \bm{\sigma})\right) \,\,\,\,\,\,\,\,\, \forall \,\, \bm{\sigma} \in \Omega.
\end{equation}
where $\bm{\bar{\pi}}$ is a probability vector so that $\pi\left(\beta_k \vert \bm{\sigma}\right) > 0$ and $\sum_{k = 1}^{K}\pi\left(\beta_k \vert \bm{\sigma}\right) = 1$. In essence, simulated tempering therefore involves alternately sampling from the two conditional distributions $\pi(\bm{\sigma}\vert \beta)$ and $\pi(\beta \vert \bm{\sigma})$.
The generalized partition function $\mathcal{Z}$ is given by
\begin{align}\label{generalised partition func}
\mathcal{Z} &= \sum_{\bar{\Omega}}e^{-\beta_k H(\bm{\sigma}) + w_k}\\ \nonumber
            &= \sum\limits_{k}\sum_{\Omega}e^{-\beta_k H(\bm{\sigma}) + w_k}\\ \nonumber
            &= \sum\limits_{k}Z(\beta_k)e^{w_k}.
\end{align}
From equation \eqref{generalised partition func} we notice that the partition functions $Z(\beta_k)$ are weighted differently for the given temperatures, where $e^{w_k}$ indicates the weight corresponding to the $k^{\text{th}}$ temperature and $w_k$ the corresponding logarithmic weight. In this paper we will refer to $w_k$ as simply the weights. In simulated tempering one wishes to avoid confinement of the system in a sub-space of the temperature space, therefore typically a uniform sampling of the temperature space is desired. The determination of the weighs $w_k$ are therefore dictated by the requirement that the probability distribution of temperature is flat. This is ideally achieved by setting $w_k = -\text{ln}\, Z(\beta_k)$, in which case the marginal probability $\pi(\beta_k)$ for a given $\beta_k$ becomes constant:
\begin{align}\label{marginal distribution}
\pi(\beta_k) &= \sum\limits_{\Omega}\pi(\bm{\sigma},\beta_k)\\ \nonumber
             &= \frac{Z(\beta_k)}{\mathcal{Z}}e^{w_k}\\ \nonumber
             & = \frac{1}{K}
\end{align}
Notice that $w_k = -\text{ln}\, Z(\beta_k)$  is proportional to the Helmholtz free energy $F$ of the system at $\beta_k$, which is given by $\beta_k F = -\text{ln}\, Z(\beta_k)$. The determination of the free energies and therefore of the weights $w_k$ are generally difficult to achieve for large complex systems. However even if the weights are estimated approximately using one of the several iterative methods,\cite{weights_1, weights_2, weights_3, weights_4, weights_5, weights_6} a uniform sampling of the temperature space can be realised to a good approximation.

 In Algorithm \ref{ST} we give a general execution of the simulated tempering method where we have used the notation $X^{(t,\tau)}$ as a state of enlarged state space $\bar{\Omega}$ after $t$ iterations of $\beta$ update and $\tau$ iterations of $\bm{\sigma}$ update. $T(\bm{\sigma'}, \beta_l \vert \bm{\sigma}, \beta_k)$ denotes the transition probability from state $(\bm{\sigma}, \beta_k) \in \bar{\Omega}$ to $(\bm{\sigma'}, \beta_l) \in \bar{\Omega}$. Once the weights are estimated by either short trial simulations (see e.g. Park and Pande \cite{weights_4}) or continually adjusted throughout the main simulation (see e.g. Nguyen et al.\cite{weights_5}), a simulated tempering simulation is then executed by alternately performing MC or MD simulations at a fixed $\beta$ (i.e. sampling from the conditional distribution $\pi(\bm{\sigma}\vert \beta)$ at step 3) and a Monte Carlo step to update $\beta$ at fixed $\bm{\sigma}$ (that is, sampling from the conditional distribution $\pi(\beta \vert \bm{\sigma})$ at step 5). The focus of this paper is on transition probabilities for updating $\beta$ at a fixed configuration $\bm{\sigma}$: $T(\bm{\sigma}, \beta_l \vert \bm{\sigma}, \beta_k)$. In order to ensure  convergence to the correct target distribution in equation \eqref{enlarged target} the transition matrix $T(\bm{\sigma}, \beta_l \vert \bm{\sigma}, \beta_k)$ must satisfy the balance condition 
\begin{equation}
\pi(\bm{\sigma}, \beta_k) = \sum\limits_{l = 1}^{K}\pi(\bm{\sigma}, \beta_l)T(\bm{\sigma}, \beta_k \vert \bm{\sigma}, \beta_l), \,\,\,\,\,\,\,\,\,\,\, \forall \,\, (\bm{\sigma}, \beta_k) \in \bar{\Omega}.
\end{equation}
In some conventional simulated tempering methods the BC condition is satisfied through the DBC:
\begin{equation}\label{DBC condition}
 \pi(\bm{\sigma}, \beta_k)T(\bm{\sigma}, \beta_l \vert \bm{\sigma}, \beta_k) = \pi(\bm{\sigma}, \beta_l)T(\bm{\sigma}, \beta_k \vert \bm{\sigma}, \beta_l)
\end{equation}
Markov chains that satisfy DBC are reversible chains while those that violate DBC are irreversible chains. Perhaps the most widely used transition probability for updating $\beta$ is that of the Metropolis-Hastings criterion \cite{Hastings}, which we  discuss in the next section. 
\begin{algorithm}[H]
  \begin{algorithmic}[1]
    \Input{Initialize $X^{(0,0)} = \left(\bm{\sigma}, \beta^{(0)}\right)$}
       \State \textbf{For} $t = 0,...,\mathcal{T}-1$
       \State \textbf{For} $\tau = 0,...,\Gamma - 1$
          \State Sample from $\pi\left(\bm{\sigma}\vert \beta\right):$ Perform an MC or MD simulation to update $X^{(t,\tau)} = \left(\bm{\sigma}, \beta^{(t)}\right)$ to $X^{(t, \tau + 1)} = \left(\bm{\sigma'}, \beta^{(t)}\right), \bm{\sigma},\bm{\sigma'} \in \Omega$.
          \State \textbf{end for}
          \State Sample from $\pi\left(\beta \vert \bm{\sigma}\right):$ Assuming $X^{(t,\Gamma)} = \left(\bm{\sigma}, \beta_k \right)$, assign $X^{(t+1, \Gamma)} = \left(\bm{\sigma}, \beta_{l}\right), \beta_l \in \lbrace \beta_1, ..., \beta_K \rbrace$ with the transition probability
          \begin{equation}\label{generic transition}
          T\left( \bm{\sigma}, \beta_l\vert \bm{\sigma}, \beta_k \right).           
           \end{equation}  
           \State  $X^{(t+1, \Gamma)}  = \left( \bm{\sigma}, \beta^{(t+1)}\right)\rightarrow X^{(t+1,0)} $.         
       \State \textbf{end for}
  \end{algorithmic}
  \caption{Simulated tempering}
  \label{ST}
\end{algorithm}

\section*{Updating inverse temperature with DBC}
\subsection*{The Metropolis-Hastings scheme for updating $\beta$}
In conventional simulated tempering the Metropolis-Hastings\cite{Hastings} type of transition probability is often used for updating $\beta$. The Metropolis-Hastings algorithm enforces the detailed balance condition by requiring that the stochastic flow $v(\bm{\sigma}, \beta_l \vert \bm{\sigma}, \beta_k) = \pi(\bm{\sigma}, \beta_k)P(\bm{\sigma}, \beta_l \vert \bm{\sigma}, \beta_k)$ is balanced out by its inverse flow $v(\bm{\sigma}, \beta_k \vert \bm{\sigma}, \beta_l) = \pi(\bm{\sigma}, \beta_l)P(\bm{\sigma}, \beta_k \vert \bm{\sigma}, \beta_l)$. The transition probability $T(\bm{\sigma}, \beta_l \vert \bm{\sigma}, \beta_k)$ from state $(\bm{\sigma}, \beta_k)$ to $(\bm{\sigma}, \beta_l)$ (i.e. for updating $\beta$ at a fixed configuration $\bm{\sigma}$) can be written as
\begin{widetext}
\begin{align}\label{general transition}
T(\bm{\sigma}, \beta_l \vert \bm{\sigma}, \beta_k) &= Q(\bm{\sigma}, \beta_l \vert \bm{\sigma}, \beta_k)A(\bm{\sigma}, \beta_l \vert \bm{\sigma}, \beta_k) \,\,\,\,\, \forall \,\, \beta_l \neq \beta_k \in \lbrace \beta_1,...,\beta_K \rbrace, \nonumber \\
T(\bm{\sigma}, \beta_k \vert \bm{\sigma}, \beta_k) &= 1 - \sum\limits_{\beta_l \neq \beta_k }T(\bm{\sigma}, \beta_l \vert \bm{\sigma}, \beta_k),
\end{align} 
\end{widetext}
where $Q(\bm{\sigma}, \beta_l \vert \bm{\sigma}, \beta_k)$ and $A(\bm{\sigma}, \beta_l \vert \bm{\sigma}, \beta_k)$ denote the  proposal and acceptance probabilities respectively. Hereafter, we assume that the set of inverse temperatures $\lbrace \beta_1,...,\beta_K \rbrace$ are equally spaced and ordered such that $\beta_1 < \beta_2 < ... < \beta_K$. Assuming a symmetric proposal, the MH acceptance probability $A(\bm{\sigma}, \beta_l \vert \bm{\sigma}, \beta_k)_{\text{MH}}$ is then given by 
\begin{align} \label{MH acceptance}
A(\bm{\sigma}, \beta_l \vert \bm{\sigma}, \beta_k)_{\text{MH}} &= \text{min}\left[1, \frac{Q(\bm{\sigma}, \beta_k \vert \bm{\sigma}, \beta_l) \pi(\bm{\sigma}, \beta_l)}{Q(\bm{\sigma}, \beta_l \vert \bm{\sigma}, \beta_k) \pi(\bm{\sigma}, \beta_k)} \right]  \,\,\,\,\, \forall \,\, \beta_l \neq \beta_k \\ \nonumber
&= \text{min}\left[1, e^{-\Delta}\right]
\end{align}
where $\Delta = (\beta_l - \beta_k)H(\bm{\sigma}) - (w_l -w_k)$. We immediately notice that the MH acceptance probability to transition from $\beta_k$ to $\beta_l$ reduces for large values of $\beta_l - \beta_k$. Therefore, in practice the proposal $\beta_l$ is often chosen from among $\lbrace\beta_{k-1}, \beta_{k+1} \rbrace$ such that $Q\left(\bm{\sigma}, \beta_{k+1} \vert \bm{\sigma}, \beta_k\right)$ = $Q\left(\bm{\sigma}, \beta_{k-1} \vert \bm{\sigma}, \beta_k\right) = 1/2$, $Q\left(\bm{\sigma}, \beta_{2} \vert \bm{\sigma}, \beta_1\right)$ = $Q\left(\bm{\sigma}, \beta_{K-1} \vert \bm{\sigma}, \beta_K\right) = 1$ and zero otherwise. Now, to update the inverse temperature with the Metropolis-Hastings transition, in step (5) of Algorithm \ref{ST}, we simply makes use of equation \eqref{general transition} with the MH acceptance given in \eqref{MH acceptance}. We show this explicitly in algorithm \ref{MH_ST}.

\begin{algorithm}[H]
  \begin{algorithmic}[1]
    \Input{Initialize $X^{(0,0)} = \left(\bm{\sigma}, \beta^{(0)}\right)$}
       \State \textbf{For} $t = 0,...,\mathcal{T}-1$
       \State \textbf{For} $\tau = 0,...,\Gamma - 1$
          \State Sample from $\pi\left(\bm{\sigma}\vert \beta\right):$ Perform an MC or MD simulation to update $X^{(t,\tau)} = \left(\bm{\sigma}, \beta^{(t)}\right)$ to $X^{(t,\tau + 1)}= \left(\bm{\sigma'}, \beta^{(t)}\right), \bm{\sigma}, \bm{\sigma'} \in \Omega$.
          \State \textbf{end for}
          \State Sample from $\pi\left(\beta \vert \bm{\sigma}\right):$ Assuming $X^{(t,\Gamma)} = \left(\bm{\sigma}, \beta_k \right)$, propose $X^{(t+1, \Gamma)} = \left(\bm{\sigma}, \beta_{l}\right), \beta_l \neq \beta_k \in \lbrace \beta_1,...,\beta_K \rbrace$ with the probability $Q\left(\bm{\sigma}, \beta_{l} \vert \bm{\sigma}, \beta_{k} \right)$ and accept it with the probability  $A\left( \bm{\sigma}, \beta_l\vert \bm{\sigma}, \beta_k \right)_{\text{MH}}$. If the proposal is rejected assign $X^{(t+1, \Gamma)} = X^{(t,\Gamma)}$.
          \State $X^{(t+1, \Gamma)}  = \left( \bm{\sigma}, \beta^{(t+1)}\right)\rightarrow X^{(t+1,0)} $.
       \State \textbf{end for}
  \end{algorithmic}
  \caption{Simulated tempering with Metropolis-Hastings}
  \label{MH_ST}
\end{algorithm}
It is a simple exercise to demonstrate that the MH transition matrix, $T\left(\bm{\sigma}, \beta_l \vert \bm{\sigma}, \beta_k\right)_{\text{MH}} = Q\left(\bm{\sigma}, \beta_l \vert \bm{\sigma}, \beta_k \right)A\left(\bm{\sigma}, \beta_l \vert \bm{\sigma}, \beta_k \right)_{\text{MH}}$ satisfies the DBC condition in \eqref{DBC condition} and therefore ensures the invariance of the target distribution.

\subsection*{The Gibbs sampler for updating $\beta$}
In the Gibbs sampler (GS),\cite{Gibbs_sampler} also known as the Heat-bath algorithm in statistical physics, the inverse temperature is updated whereby a new $\beta_l \in \lbrace \beta_1, ..., \beta_K \rbrace$  is drawn from its \emph{conditional} distribution $\pi( \cdot \, \vert \bm{\sigma})$. We let $G\left(\bm{\sigma}, \beta_l \vert \bm{\sigma}, \beta_k\right)$ to denote the Gibbs transition probability from $\beta_k$ to $\beta_l$, which is simply the conditional distribution given $\bm{\sigma}$:
\begin{equation} \label{Gibbs ST}
G\left(\bm{\sigma}, \beta_l \vert \bm{\sigma}, \beta_k \right) = \frac{\pi\left(\bm{\sigma}, \beta_l \right)}{\sum\limits_{r = 1}^{K}\pi\left(\bm{\sigma}, \beta_r\right)} \,\,\,\,\,\, \forall \,\, (\bm{\sigma}, \beta_l) \in \bar \Omega
\end{equation}
Notice that the Gibbs transition to the new value $\beta_l$ is independent of the current value $\beta_k$. The execution of simulated tempering with the Gibbs sampler is then straightforward: In step (5) of Algorithm \ref{ST} the generic transition probability in \eqref{generic transition} is now replaced with the Gibbs transition given in \eqref{Gibbs ST}. Note that for a fixed configuration $\bm{\sigma}$ the computational cost of the summation in \eqref{Gibbs ST} is next to negligible even for excessively large $K$ values. The Gibbs sampler is in fact a special case of the Metropolis-Hastings algorithm whereby every proposal is accepted. This is easily demonstrated by letting the proposal $Q\left(\bm{\sigma}, \beta_l \vert \bm{\sigma}, \beta_k \right) = \pi(\beta_l \vert \bm{\sigma})$, in which case the MH acceptance probability for every proposal is then exactly one:
\begin{equation}
A\left(\bm{\sigma}, \beta_l \vert \bm{\sigma}, \beta_k\right) = \text{min}\left[1, \frac{\pi\left(\beta_k \vert \bm{\sigma}\right) \pi\left(\bm{\sigma}, \beta_l\right)}{\pi\left(\beta_l \vert \bm{\sigma}\right) \pi\left(\bm{\sigma}, \beta_k\right)}\right] = 1.
\end{equation} 
As a special case of the Metropolis-Hastings criteria, the Gibbs sampler therefore ensures the invariance of the target distribution. The Metropolis-Hastings acceptance in \eqref{MH acceptance} is dependent on the spacing $(\beta_l - \beta_k)$, the acceptance probabilities for large jumps in temperature space are therefore small. However, the Gibbs transition probability is independent of the current inverse temperature $\beta_k$, it is therefore capable of providing a more efficient global exploration of temperature space than is possible with the standard Metropolis-Hastings method. Simulated tempering with the Gibbs sampler, whereby the Gibbs transition in \eqref{Gibbs ST} is used to update the inverse temperature, has been implemented in several studies \cite{Rosta_1, Gibbs_ST, ST_Suwa-Todo} that demonstrate better performance compared with the conventional method of updating $\beta$ with the Metropolis-Hastings method. However for reasons unclear it is not as widely in practice as the standard Metropolis-Hastings method.

\subsection*{The Metropolized-Gibbs sampler for updating $\beta$}
We now briefly introduce a variant of the Gibbs sampler, namely the \emph{Metropolized-Gibbs sampler} (MGS), which was originally introduced by Liu \cite{MGS} as a modification of the random scan Gibbs sampler with improved mixing rate. The MGS transition matrix, which also satisfies DBC, provides an improved sampling of the state space compared to the standard Gibbs transition in \eqref{Gibbs ST}. The development of the Metropolized-Gibbs sampler was directly motivated by Peskun's theorem: Given that both transition matrices $(T^{A}_{ij})_{i,j \in \Omega}$ and $(T^{B}_{ij})_{i,j \in \Omega}$ satisfy DBC and $T^{A}_{ii} < T^{B}_{ii}$ $\forall \,\, i \in \Omega$, then a Markov chain with the transition matrix $T^{A}_{ij}$ returns estimates with smaller asymptotic variance that a Markov chain with transition matrix $T^{B}_{ij}$. In other words minimising the probability of remaining in the current state increases mobility in, and therefore provides a more efficient sampling of, the state space. In the context of simulated tempering for updating $\beta$ a new candidate, $\beta_l \neq \beta_k$, is proposed with the probability
\begin{equation}\label{MG proposal}
Q\left(\bm{\sigma}, \beta_l \vert \bm{\sigma}, \beta_k \right) = \frac{G\left(\bm{\sigma}, \beta_l \vert \bm{\sigma}, \beta_k \right)}{1 - G\left(\bm{\sigma}, \beta_k \vert \bm{\sigma}, \beta_l \right)} \,\,\,\,\,\, \forall \,\, \beta_l \neq \beta_k \in \lbrace \beta_1, ..., \beta_K \rbrace,
\end{equation} 
and accepted with the Metropolis-Hastings acceptance  probability (eq.\eqref{MH acceptance}):
\begin{equation}
A(\bm{\sigma}, \beta_l \vert \bm{\sigma}, \beta_k) = \text{min}\left[1, \frac{1 - G\left(\bm{\sigma}, \beta_k \vert \bm{\sigma}, \beta_l \right)}{1 - G\left(\bm{\sigma}, \beta_l \vert \bm{\sigma}, \beta_k \right)}\right] \,\,\,\,\,\, \forall \,\, \beta_l \neq \beta_k,
\end{equation}  
where upon rejected the current state $\left(\bm{\sigma}, \beta_k\right)$ is retained. The reversible transition matrix $M\left(\bm{\sigma}, \beta_l \vert \bm{\sigma}, \beta_k\right)$  for the Metropolized-Gibbs sampler can then be written as
\begin{widetext}
\begin{align}\label{MGS ST}
M\left(\bm{\sigma}, \beta_l \vert \bm{\sigma}, \beta_k \right) &= \text{min}\left[\frac{G\left(\bm{\sigma}, \beta_l \vert \bm{\sigma}, \beta_k \right)}{1 - G\left(\bm{\sigma}, \beta_k \vert \bm{\sigma}, \beta_l \right)}, \frac{G\left(\bm{\sigma}, \beta_l \vert \bm{\sigma}, \beta_k \right)}{1 - G\left(\bm{\sigma}, \beta_l \vert \bm{\sigma}, \beta_k \right)} \right] \,\,\,\,\,\, \forall \,\, \beta_l \neq \beta_k \in \lbrace \beta_1, ..., \beta_K \rbrace,\\ \nonumber
M\left(\bm{\sigma}, \beta_k \vert \bm{\sigma}, \beta_k \right) &= 1 - \sum\limits_{\beta_l \neq \beta_k}M\left(\bm{\sigma}, \beta_l \vert \bm{\sigma}, \beta_k \right),
\end{align}
\end{widetext}
which satisfies DBC. A few points here merit some elaboration; we point out that for a two state solution, $K = 2$, The Gibbs transition probability in \eqref{Gibbs ST} becomes equivalent to Barker's method,\cite{Barker} whereas the MGS transition in \eqref{MGS ST} decomposes to the standard Metropolis-Hastings transition. Peskun had demonstrated that, within DBC, the Metropolis-Hastings criteria is superior to Barker's method as the former returns a smaller probability of remaining in the current state and therefore increases mobility in the state space. This argument applies more generally to the MGS sampler. By minimising the probability of retaining the current state the MGS transition, $M\left(\bm{\sigma}, \beta_l \vert \bm{\sigma}, \beta_k \right)$, is more efficient at sampling of the state space than the Gibbs transition $G\left(\bm{\sigma}, \beta_l \vert \bm{\sigma}, \beta_k \right)$. This has been numerically demonstrated in some studies. \cite{EIMCS,Gibbs_ST,Metropolized-Gibbs} In practice, to update the inverse temperature with the MGS sampler one simply replaces the generic transition probability in \eqref{generic transition} with that of the MGS transition given in \eqref{MGS ST}. 

\section*{Updating inverse temperature with SDBC}
\subsection*{The lifting framework}
In the lifting framework, as introduced by Diaconis et al.,\cite{Diaconis} the state space is enlarged by effectively replicating a duplicate copy of the original space. Each replica, which is characterised by a lifting variable $\varepsilon \in \lbrace -1, +1 \rbrace$, consists of all configurations $\bm{\sigma} \in \Omega$ as in the original space. The system now explores an extended state space, that is, in addition to \emph{intra-replica} transition between configurations $\bm{\sigma} \rightarrow \bm{\sigma}'$ as in the original space, the system can now also perform \emph{inter-replica} transition $\left(\bm{\sigma},\varepsilon\right) \rightarrow \left(\bm{\sigma},-\varepsilon \right)$ between duplicate copies of a given configuration. A Markov chain propagated in this enlarged state space breaks DBC but ensures convergence to the target distribution by satisfying BC.\cite{Diaconis,Turitsyn,Weigel,Sakai_Hukushima_eigenvalue,EIMCS} In this section we utilize the lifting framework with skewed detailed balance condition (SDBC) as proposed by Turitsyn et al.\cite{Turitsyn}. In particular, we implement the lifting framework in the updating scheme of inverse temperature in simulated tempering. Sakai and Hukushima\cite{Sakai_Hukushima_simulated_tempering} have already implemented lifting with SDBC to the Metropolis-Hastings transition $T\left(\bm{\sigma}, \beta_l \vert \bm{\sigma}, \beta_k\right)_{\text{MH}} = Q\left(\bm{\sigma}, \beta_l \vert \bm{\sigma}, \beta_k \right)A\left(\bm{\sigma}, \beta_l \vert \bm{\sigma}, \beta_k \right)_{\text{MH}}$ for updating $\beta$. The authors applied their algorithm to the simulation of the 2D Ising model and demonstrated that, when compared with the conventional Metropolis-Hastings method with DBC, their algorithm provides significant improvement in the relaxation dynamics of $\beta$ and the magnetisation of the model. Hereafter we will refer to Sakai and Hukushima's algorithm \cite{Sakai_Hukushima_simulated_tempering} as irreversible Metropolis-Hastings (IMH). We had recently proposed the irreversible counter-parts of the Gibbs sampler and the Metropolized-Gibbs sampler, namely IGS and IMGS that both satisfy SDBC.\cite{EIMCS} Here we demonstrate that both IGS and IMGS can be adapted for the update scheme of inverse temperature. Numerical simulations in the next section show that, when compared with their respective reversible counterparts, both IGS and IMGS improve the relaxation dynamics of $\beta$ and consequently that of some system observables. Furthermore our results also show considerable improvement over the IMH algorithm in the mixing time of $\beta$ and that of some system observables for large temperature domains $K$.

\subsection*{SDBC in the context of simulated tempering}
The lifting variable $\varepsilon  \in \lbrace +1, -1 \rbrace$ is introduced to double the state space $\bar{\Omega}$ so that the extended state space $\tilde{\Omega}:= \bar{\Omega} \times \lbrace +1, -1 \rbrace$ now consist of two replicas characterized by $\varepsilon = \pm$. The probability of finding the system in state $(\bm{\sigma}, \beta_k, \varepsilon)$ is given by
\begin{equation}\label{extended original targ relat}
\tilde{\pi}\left(\bm{\sigma}, \beta_k, \varepsilon\right) = \frac{1}{2}\pi\left(\bm{\sigma}, \beta_k \right).
\end{equation}
To update $\beta$ we now wish to sample from the \emph{extended} target distribution given by the probability vector
\begin{align}
\bm{\tilde{\pi}} &= \left( \tilde{\pi}\left(\beta_1 \vert \bm{\sigma}, +\right), ... , \tilde{\pi}\left(\beta_K \vert \bm{\sigma}, +\right), \tilde{\pi}\left(\beta_1 \vert \bm{\sigma}, - \right), ... , \tilde{\pi}\left(\beta_K \vert \bm{\sigma}, - \right)\right) \\ \nonumber
                 &=\frac{1}{2}\left(\bm{\bar{\pi}}, \bm{\bar{\pi}}\right)
\end{align}
where the original target $\bm{\bar{\pi}}$ is given in \eqref{original target}. Notably, $\tilde{\pi}\left(\bm{\sigma}, \beta_k, \varepsilon\right) = \tilde{\pi}\left(\bm{\sigma}, \beta_k, -\varepsilon\right)$ and the marginal $\sum_{\varepsilon'}\tilde{\pi}\left(\bm{\sigma}, \beta_k, \varepsilon' \right) = \pi\left(\bm{\sigma}, \beta_k \right)$. The transition matrix $\bm{\tilde{T}}$ of the Markov chain on the extended state space $\tilde{\Omega}$ is now given by 
\begin{equation}\label{extended transition}
 \widetilde{\bm{T}}=
  \left( {\begin{array}{cc}
   T(\bm{\sigma}, \beta_l,+ \vert \bm{\sigma}, \beta_k, +) & \Lambda(\bm{\sigma},\beta_k,- \vert \bm{\sigma}, \beta_k, +) \\
   \Lambda(\bm{\sigma},\beta_k, + \vert \bm{\sigma}, \beta_k, -) & T(\bm{\sigma}, \beta_l,- \vert \bm{\sigma}, \beta_k, -) \\
  \end{array} } \right),
\end{equation}
where $T(\bm{\sigma}, \beta_l,\pm \vert \bm{\sigma}, \beta_k, \pm) \geq 0$ denotes the \emph{intra-replica} transition probability from state $\left(\bm{\sigma}, \beta_k \right) \in \bar{\Omega}$ to $\left(\bm{\sigma}, \beta_l\right) \in \bar{\Omega}$ in respective $\varepsilon = \pm$ replicas. The \emph{inter-replica} transition probability from state $\left(\bm{\sigma}, \beta_k, \pm\right)$ to $ \left( \bm{\sigma}, \beta_k, \mp\right)$ is denoted by  $\Lambda(\bm{\sigma},\beta_k,\mp \vert \bm{\sigma}, \beta_k, \pm) \geq 0$. Note that $\Lambda(\bm{\sigma},\beta_l,\mp \vert \bm{\sigma}, \beta_k, \pm) = 0 \,\,\,\, \forall \,\,\beta_l \neq \beta_k \in \lbrace \beta_1, ..., \beta_K \rbrace$.  The normalization of probability in the extended state space can be written as
\begin{equation}
\sum\limits_{l = 1}^{K} T(\bm{\sigma}, \beta_l,\varepsilon \vert \bm{\sigma}, \beta_k, \varepsilon) + \Lambda\left(\bm{\sigma}, \beta_k, -\varepsilon \vert \bm{\sigma}, \beta_k, \varepsilon \right) = 1.
\end{equation}
Assuming that the transition matrix $\bm{\widetilde{T}}$ is ergodic, it must then satisfy the balance condition $\bm{\tilde{\pi}} = \bm{\tilde{\pi}}\bm{\widetilde{T}}$ to ensure invariance of the target distribution $\bm{\tilde{\pi}}$. The balance condition takes the form
\begin{widetext}
\begin{align}\label{BC ST}
\sum_{l = 1}^{K}  T\left(\bm{\sigma}, \beta_l, \varepsilon \vert \bm{\sigma}, \beta_k, \varepsilon \right)\tilde{\pi}\left(\bm{\sigma}, \beta_k, \varepsilon\right) + \Lambda\left(\bm{\sigma}, \beta_k, -\varepsilon \vert \bm{\sigma}, \beta_k, \varepsilon\right)\tilde{\pi}\left(\bm{\sigma}, \beta_k, \varepsilon \right)
= \sum\limits_{l = 1}^{K}  &T\left(\bm{\sigma}, \beta_k, \varepsilon \vert \bm{\sigma}, \beta_l, \varepsilon \right)\tilde{\pi}\left(\bm{\sigma}, \beta_l, \varepsilon \right)\\ \nonumber
&+ \Lambda\left(\bm{\sigma}, \beta_k, \varepsilon \vert \bm{\sigma}, \beta_k, -\varepsilon\right)\tilde{\pi}\left(\bm{\sigma}, \beta_k, -\varepsilon \right),
\end{align} 
\end{widetext}
$\forall \,\, \beta_k \in \lbrace \beta_1, ..., \beta_K \rbrace$. The balance condition can be satisfied by imposing the skewed detailed balance condition (SDBC) on the transition matrix:

\begin{align}\label{SDBC ST}
\tilde{\pi}\left(\bm{\sigma}, \beta_k, \varepsilon\right)&T\left(\bm{\sigma}, \beta_l, \varepsilon \vert \bm{\sigma}, \beta_k, \varepsilon \right) \\ \nonumber
& = \tilde{\pi}\left(\bm{\sigma}, \beta_l, -\varepsilon \right)T\left(\bm{\sigma}, \beta_k, -\varepsilon \vert \bm{\sigma}, \beta_l, -\varepsilon \right).
\end{align}

The SDBC requires that the stochastic flow from state $(\bm{\sigma}, \beta_k) \rightarrow (\bm{\sigma}, \beta_l)$ in one replica is balanced by reverse flow $(\bm{\sigma}, \beta_l) \rightarrow (\bm{\sigma}, \beta_k)$ in the other replica. The SDBC therefore by definition breaks the detailed balance condition:    $\tilde{\pi}\left(\bm{\sigma}, \beta_k, \varepsilon\right)T\left(\bm{\sigma}, \beta_l, \varepsilon \vert \bm{\sigma}, \beta_k, \varepsilon \right) \neq \tilde{\pi}\left(\bm{\sigma}, \beta_l, \varepsilon \right)T\left(\bm{\sigma}, \beta_k,\varepsilon \vert \bm{\sigma}, \beta_l, \varepsilon \right)$. Note that by imposing SDBC on the transition matrix we can obtain a condition for the construction of the \emph{inter-replica} transition probability  $\Lambda\left(\bm{\sigma}, \beta_k, -\varepsilon \vert \bm{\sigma}, \beta_k, \varepsilon\right)$, we see this clearly once we insert \eqref{SDBC ST} into \eqref{BC ST} to obtain
\begin{align}\label{lambda condition}
\Lambda & \left(\bm{\sigma}, \beta_k, -\varepsilon \vert \bm{\sigma}, \beta_k, \varepsilon\right) - \Lambda\left(\bm{\sigma}, \beta_k, \varepsilon \vert \bm{\sigma}, \beta_k, -\varepsilon\right)  \\ \nonumber
&= \sum\limits_{l \neq k} \left[T\left(\bm{\sigma}, \beta_l, -\varepsilon \vert \bm{\sigma}, \beta_k, -\varepsilon \right) - T\left(\bm{\sigma}, \beta_l, \varepsilon \vert \bm{\sigma}, \beta_k, \varepsilon \right)\right]
\end{align}
A particular solution of \eqref{lambda condition}, which was originally proposed by Turitsyn et al., \cite{Turitsyn} is of the form
\begin{align}\label{TCV solution}
& \Lambda\left(\bm{\sigma}, \beta_k, -\varepsilon \vert \bm{\sigma}, \beta_k, \varepsilon\right)\\ \nonumber
& = \text{max}\left[0, \sum\limits_{l\neq k} \left(T\left(\bm{\sigma}, \beta_l, -\varepsilon \vert \bm{\sigma}, \beta_k, -\varepsilon \right) - T\left(\bm{\sigma}, \beta_l, \varepsilon \vert \bm{\sigma}, \beta_k, \varepsilon \right)\right) \right],
\end{align} 
which is known as the Turitsyn-Chertkov-Vucelja (TCV) solution. However several alternative solutions of \eqref{lambda condition} has been proposed and studied,\cite{Sakai_Hukushima_eigenvalue} as for example, the alternative choice known as the Sakai-Hukushima 1 (SH1) solution, which is given by 
\begin{equation}
\Lambda\left(\bm{\sigma}, \beta_k, -\varepsilon \vert \bm{\sigma}, \beta_k, \varepsilon\right) = \sum\limits_{l \neq k} T\left(\bm{\sigma}, \beta_l, -\varepsilon \vert \bm{\sigma}, \beta_k, -\varepsilon \right),
\end{equation}
has been studied in the context of the 1D Ising model.\cite{Sakai_Hukushima_1D}

Our task  at hand is now to construct an \emph{intra-replica} transition matrix $T\left(\bm{\sigma}, \beta_l, \varepsilon \vert \bm{\sigma}, \beta_k, \varepsilon \right)$ that satisfies the SDBC given in \eqref{SDBC ST}. With this in mind, we follow the same procedure we had outlined recently; \cite{EIMCS} which involves modifying a generic transition matrix $T\left(\bm{\sigma}, \beta_l \vert \bm{\sigma}, \beta_k \right)$, that satisfies DBC in \eqref{DBC condition}, with the \emph{skewness} function $\Theta(\beta_l, \varepsilon \vert \beta_k, \varepsilon)$. We therefore define 
\begin{equation}\label{skewed transition}
T\left(\bm{\sigma}, \beta_l, \varepsilon \vert \bm{\sigma}, \beta_k, \varepsilon \right) = \Theta(\beta_l, \varepsilon \vert \beta_k, \varepsilon)T\left(\bm{\sigma}, \beta_l \vert \bm{\sigma}, \beta_k \right),
\end{equation}
where the skewness function has the properties:
\begin{equation}\label{property 1}
 0 \leq \Theta(\beta_l, \varepsilon \vert \beta_k, \varepsilon) \leq 1
\end{equation} 
and 
\begin{equation}\label{property 2}
\Theta(\beta_l, \varepsilon \vert \beta_k, \varepsilon) = \Theta(\beta_k, -\varepsilon \vert \beta_l, -\varepsilon).
\end{equation}
With this definition we note that the transition matrix $T\left(\bm{\sigma}, \beta_l, \varepsilon \vert \bm{\sigma}, \beta_k, \varepsilon \right)$ in \eqref{skewed transition} now satisfies the SDBC in \eqref{SDBC ST}. This completes our description of the extended transition matrix $\tilde{\bm{T}}$, as defined in \eqref{extended transition}. An irreversible Markov chain can therefore be propagated on the extended state space $\tilde{\Omega}$, whereby the stationary distribution of the chain will, by the arguments above, converge to the invariant target distribution. Next, we introduce an explicit form of a suitable skewness function and demonstrate how to adapt the IGS and IMGS \cite{EIMCS} to the update scheme of inverse temperature. 

\subsection*{Irreversible Gibbs sampler for updating $\beta$} 
In principle any skewness function that meets conditions \eqref{property 1} and \eqref{property 2} will suffice to construct a transition matrix  $T\left(\bm{\sigma}, \beta_l, \varepsilon \vert \bm{\sigma}, \beta_k, \varepsilon \right)$. However certain choices of the skewness function may lead to more efficient sampling of the temperature space than others. Here we make use of the skewness function we had recently proposed, \cite{EIMCS} which we express in general formulation in the context of simulated tempering,
\begin{equation}\label{skewness function}
\Theta\left(f_l, \varepsilon \vert f_k, \varepsilon\right) = \varphi\left(1 + \delta \varepsilon \Phi(f)\right),
\end{equation}
where $f \in \lbrace f_1,...,f_K \rbrace$ is the lifting coordinate and the function $\Phi(f) = \text{sgn}\left( f_l - f_k \right)$ with the sign function given by
\[
    \text{sgn}(x)= 
\begin{cases}
    -1,& \text{if}\,\,\,\, x < 0,\\
     \,\,\,\,0,& \text{if}\,\,\,\, x = 0,\\
     +1,& \text{if}\,\,\,\,x > 0.
\end{cases}
\]
The constant $\varphi = 1/(1+\delta)$. The deviation parameter $\delta \in [0,1]$ determines the extent to which DBC is violated, DBC is recovered by setting $\delta = 0$. Notably, with $\delta = 0$ the transition $T\left(\bm{\sigma}, \beta_l, \varepsilon \vert \bm{\sigma}, \beta_k, \varepsilon \right)$ in \eqref{skewed transition} becomes uniform in $\varepsilon$, and consequently the SDBC in \eqref{SDBC ST} reduces to DBC in \eqref{DBC condition}.

In general applications of the lifting framework with SDBC, the lifting coordinate $f$ is often taken to be an observation of interest, as for example in the context of spin systems this could be the magnetisation \cite{Sakai_Hukushima_1D, Sakai_Hukushima_2D, Turitsyn, Weigel, EIMCS} or energy of the system.\cite{EIMCS, Weigel} Often $f$ is assigned to be an observable with  slow relaxation dynamics, so that by \emph{lifting} it a more efficient sampling of the state space can be realised along the coordinates of this particular observable. In temperature simulated tempering, particularly in the update scheme of $\beta$,  one is often interested in improving the mixing rate in temperature space. We therefore assign the lifting coordinate $f$ to be the inverse temperature $\beta$. However, considering a simulated tempering equivalent of Hamiltonian replica exchange\cite{Hamiltonian_REM} one may assign a dynamical variable other than $\beta$ as the lifting coordinate, such as an interaction parameter (e.g. different strengths of an external magnetic field) in the Hamiltonian of spin systems.

The skewness function in \eqref{skewness function} effectively introduces a bias in the way the variable $f$ is sampled. To better understand this we set the lifting coordinate $f$ as the inverse temperature and consider two distinct cases: $\left(\varepsilon = \pm, \Phi(\beta) = \pm \right)$ and $\left(\varepsilon = \pm, \Phi(\beta) = \mp \right)$. The transition probability $T(\bm{\sigma}, \beta_l, \varepsilon \vert \bm{\sigma}, \beta_k, \varepsilon)$ in \eqref{skewed transition} then breaks down to $ T\left(\bm{\sigma}, \beta_l \vert \bm{\sigma}, \beta_k \right)$ for $ \left( \varepsilon = \pm 1, \Phi(\beta) = \pm 1 \right)$ and $(1-\delta/1+\delta)T\left(\bm{\sigma}, \beta_l \vert \bm{\sigma}, \beta_k \right)$ for $\left( \varepsilon = \pm 1, \Phi(\beta) = \mp 1 \right)$. Considering the $\varepsilon = +1$ replica and $\delta \neq 0$ as an example, we observe that Monte Carlo moves that increase $\beta$ ( i.e., $\Phi(\beta) = +1$) are stochastically favoured over  moves that decrease $\beta$ ( i.e., $\Phi(\beta) = -1$), while the opposite is true in the $\varepsilon = -1$ replica. In the two replicas the system therefore stochastically favours opposing and fixed directions in temperature space.

Notably, inserting the Metropolis-Hastings transition $T(\bm{\sigma}, \beta_l \vert \bm{\sigma}, \beta_k)_{\text{MH}}$ on the right hand side of \eqref{skewed transition} (and setting $\varphi = 1$) leads to the IMH algorithm of Sakai and Hukushima.\cite{Sakai_Hukushima_simulated_tempering} The IMH algorithm in particular introduces a bias in the proposal $Q(\bm{\sigma}, \beta_l \vert \bm{\sigma}, \beta_k)$ in the nearest neighbour exchange scheme for $\beta$. In this scheme, the dynamical behaviour of a fixed configuration $\bm{\sigma}$ is comparable to a simple random walk in temperature space with $K$ states. Sakai and Hukushima have numerically demonstrated with the Ising model that  the IMH algorithm suppresses the diffusive behaviour of $\bm{\sigma}$ on the temperature space, so that the mixing time of $\beta$ now scales on the order of $\mathcal{O}(K)$ with the IMH algorithm as opposed to $\mathcal{O}(K^2)$ with conventional MH with DBC. However as we have seen in previous sections, other than the conventional Metropolis-Hastings algorithm, there are several alternative choices with DBC to sample from the conditional distribution $\pi\left(\beta \vert \bm{\sigma}\right)$. We argue here that the irreversible counter-parts of the Gibbs sampler and Metropolized-Gibbs sampler can be constructed to sample from the conditional $\pi(\beta \vert \bm{\sigma})$ through satisfying SDBC. As we will see in the next section, the resulting IGS and IMGS schemes for updating $\beta$ provide an improvement in the relaxation dynamics of $\beta$ and system observables over their respective reversible counterparts with DBC, and also when compared with the IMH algorithm.

We proceed to  adapt the irreversible Gibbs sampler \cite{EIMCS} for the update scheme of $\beta$. We do this by inserting the Gibbs transition, $G\left(\bm{\sigma}, \beta_l \vert \bm{\sigma}, \beta_k \right)$ given in \eqref{Gibbs ST}, on the right hand side of \eqref{skewed transition}. The IGS \emph{intra-replica} transition matrix $\mathcal{G}\left(\bm{\sigma}, \beta_l, \varepsilon \vert \bm{\sigma}, \beta_k, \varepsilon \right)$ is then defined as
\begin{widetext}
\begin{align}
\mathcal{G}\left(\bm{\sigma}, \beta_l, \varepsilon \vert \bm{\sigma}, \beta_k, \varepsilon \right) &= \Theta\left(\beta_l, \varepsilon \vert \beta_k, \varepsilon \right)G\left(\bm{\sigma}, \beta_l \vert \bm{\sigma}, \beta_k \right) \,\,\,\,\,\, \forall \,\, \beta_l \neq \beta_k \in \lbrace \beta_1, ..., \beta_K \rbrace, \\ \nonumber
\mathcal{G}\left(\bm{\sigma}, \beta_k, \varepsilon \vert \bm{\sigma}, \beta_k, \varepsilon \right)  &= 1 - \sum\limits_{l \neq k} \mathcal{G}\left(\bm{\sigma}, \beta_l, \varepsilon \vert \bm{\sigma}, \beta_k, \varepsilon \right). 
\end{align}
\end{widetext}
We recover the \emph{inter-replica} transition probability of the TCV solution from the general formulation in \eqref{TCV solution}:
\begin{align}
&\Lambda\left(\bm{\sigma}, \beta_k, -\varepsilon \vert \bm{\sigma}, \beta_k, \varepsilon\right)_{\text{IGS}}\\ \nonumber
& = \text{max}\left[0, \sum\limits_{l\neq k} \left(\mathcal{G}\left(\bm{\sigma}, \beta_l, -\varepsilon \vert \bm{\sigma}, \beta_k, -\varepsilon \right) - \mathcal{G}\left(\bm{\sigma}, \beta_l, \varepsilon \vert \bm{\sigma}, \beta_k, \varepsilon \right)\right) \right].
\end{align}
This completes our description of the irreversible Gibbs sampler for the update scheme of inverse temperature in simulated tempering. The general execution of the IGS to update $\beta$ is given in Algorithm \ref{IGS ST algo} where we have used the notation $\tilde{X}^{(t,\tau)}$ as a state of the \emph{extended} state space $\tilde{\Omega}$ after $t$ iterations of $\beta$ updates and $\tau$ iterations of $\bm{\sigma}$ updates.

The implementation of the irreversible Gibbs sampler as given in Algorithm \ref{IGS ST algo} differs from that of the conventional Gibbs sampler of equation \eqref{Gibbs ST} with DBC in steps 5, 6 and 7. Therefore to verify that the steps shown in Algorithm \ref{IGS ST algo} leads to sampling from the correct target distribution, it suffices to demonstrate that the conditional $\tilde{\pi}\left(\beta, \varepsilon \vert \bm{\sigma}\right) = 1/2 \, \pi(\beta \vert \bm{\sigma})$ satisfies the balance condition. We show this in Appendix A.

\subsection*{Irreversible Metropolized-Gibbs sampler for updating $\beta$} 

Likewise, we briefly demonstrate that the IMGS\cite{EIMCS} can be adapted to sample from the conditional distribution $\pi(\beta \vert \bm{\sigma})$. We insert the Metropolized-Gibbs transition $M(\bm{\sigma}, \beta_l \vert \bm{\sigma}, \beta_k)$, as given in \eqref{MGS ST}, on the right hand side of \eqref{skewed transition} to obtain the transition matrix $\mathcal{M}\left(\bm{\sigma}, \beta_l, \varepsilon \vert \bm{\sigma}, \beta_k, \varepsilon \right)$ for IMGS: 
\begin{widetext}
\begin{align}\label{IMGS SDBC transition}
\mathcal{M}\left(\bm{\sigma}, \beta_l, \varepsilon \vert \bm{\sigma}, \beta_k, \varepsilon \right) &= \Theta\left(\beta_l, \varepsilon \vert \beta_k, \varepsilon \right)M\left(\bm{\sigma}, \beta_l \vert \bm{\sigma}, \beta_k \right) \,\,\, \forall \,\, \beta_l \neq \beta_k  \in \lbrace \beta_1, ..., \beta_K \rbrace,\\ \nonumber
\mathcal{M}\left(\bm{\sigma}, \beta_k, \varepsilon \vert \bm{\sigma}, \beta_k, \varepsilon \right)  &= 1 - \sum\limits_{l \neq k} \mathcal{M}\left(\bm{\sigma}, \beta_l, \varepsilon \vert \bm{\sigma}, \beta_k, \varepsilon \right). 
\end{align}
\end{widetext}
The \emph{inter-replica} transition probability of the TCV solution is then given by:
\begin{align}\label{IMGS SDBC inter-replica transition}
& \Lambda\left(\bm{\sigma}, \beta_k, -\varepsilon \vert \bm{\sigma}, \beta_k, \varepsilon\right)_{\text{IMGS}} \\ \nonumber
& = \text{max}\left[0, \sum\limits_{l\neq k} \left(\mathcal{M}\left(\bm{\sigma}, \beta_l, -\varepsilon \vert \bm{\sigma}, \beta_k, -\varepsilon \right) - \mathcal{M}\left(\bm{\sigma}, \beta_l, \varepsilon \vert \bm{\sigma}, \beta_k, \varepsilon \right)\right) \right].
\end{align}
To sample from $\pi\left(\beta \vert \bm{\sigma}\right)$ with the IMGS, one follows the same steps as in Algorithm \eqref{IGS ST algo}, except for making use of \eqref{IMGS SDBC transition} and \eqref{IMGS SDBC inter-replica transition} in steps (5) and (6). We have demonstrated\cite{EIMCS} in the context of 1D Potts model that the optimality of MGS over the standard Gibbs sampler is modestly replicated in their irreversible counterparts with SDBC. It is therefore of interest to inspect if similar improvement is replicated in the context of simulated tempering. 

\begin{algorithm}[H]
  \begin{algorithmic}[1]
    \Input{Initialize $\tilde{X}^{(0,0)} = \left(\bm{\sigma}, \beta^{(0)}, \varepsilon^{(0)} \right)$}
       \State \textbf{For} $t = 0,...,\mathcal{T}-1$
       \State \textbf{For} $\tau = 0,...,\Gamma - 1$
          \State Sample from $\tilde{\pi}\left(\bm{\sigma}\vert \beta, \varepsilon \right)$: Perform an MC or MD simulation to update $\tilde{X}^{(t,\tau)} = \left(\bm{\sigma}, \beta^{(t)}, \varepsilon^{(t)}\right)$ to $\tilde{X}^{(t,\tau + 1)} = \left(\bm{\sigma'}, \beta^{(t)}, \varepsilon^{(t)}\right), \bm{\sigma},\bm{\sigma'} \in \Omega$.
          \State \textbf{end for}
          \State Sample from $\tilde{\pi}\left(\beta \vert \bm{\sigma}, \varepsilon \right)$: Suppose that $\tilde{X}^{(t, \Gamma)} = \left(\bm{\sigma}, \beta_{k}, \varepsilon \right)$,  assign $\tilde{X}^{(t+1, \Gamma)} = \left(\bm{\sigma}, \beta_{l}, \varepsilon\right), \beta_l \in \lbrace \beta_1, ..., \beta_K \rbrace$, with the transition probability
          \begin{align}
          \mathcal{G}\left( \bm{\sigma}, \beta_l, \varepsilon \vert \bm{\sigma}, \beta_k, \varepsilon \right) &= \Theta\left(\beta_l, \varepsilon \vert \beta_k, \varepsilon \right)G\left(\bm{\sigma}, \beta_l \vert \bm{\sigma}, \beta_k \right) \,\,\,\,\,\, \forall \,\, \beta_l \neq \beta_k, \\ \nonumber           
\mathcal{G}\left(\bm{\sigma}, \beta_k, \varepsilon \vert \bm{\sigma}, \beta_k, \varepsilon \right)  &= 1 - \sum\limits_{l \neq k} \mathcal{G}\left(\bm{\sigma}, \beta_l, \varepsilon \vert \bm{\sigma}, \beta_k, \varepsilon \right).
           \end{align}    
           \State If $\beta_l = \beta_k$, assign $\tilde{X}^{(t+1,\Gamma)} = \left(\bm{\sigma}, \beta_{k}, -\varepsilon \right)$ with the transition probability  
           \begin{equation}
           P\left(\bm{\sigma}, \beta_k, -\varepsilon \vert \bm{\sigma}, \beta_k, \varepsilon\right) = \frac{\Lambda\left(\bm{\sigma}, \beta_k, -\varepsilon \vert \bm{\sigma}, \beta_k, \varepsilon\right)_{\text{IGS}}}{1 - \sum\limits_{r\neq k}\mathcal{G}\left( \bm{\sigma}, \beta_r, \varepsilon \vert \bm{\sigma}, \beta_k, \varepsilon \right)}.
           \end{equation}   
           \State If this is also rejected then set $\tilde{X}^{(t+1, \Gamma)} = \tilde{X}^{(t, \Gamma)}$.
           \State $\tilde{X}^{(t+1, \Gamma)} = \left(\bm{\sigma}, \beta^{(t+1)}, \varepsilon^{(t+1)}\right)  \rightarrow \tilde{X}^{(t+1, 0)} $
       \State \textbf{end for}
  \end{algorithmic}
  \caption{Simulated tempering with IGS}
  \label{IGS ST algo}
\end{algorithm}

\section*{Performance analysis with MCMC simulations}
We carry out simulated tempering simulations to test the performance of IGS and IMGS when sampling from $\tilde{\pi}\left( \beta \vert \bm{\sigma}, \varepsilon \right)$. In this section we perform MCMC simulations to sample from $\tilde{\pi}\left(\bm{\sigma}\vert \beta, \varepsilon \right)$. Performance analysis whereby MD simulations are used to sample from $\tilde{\pi}\left(\bm{\sigma}\vert \beta, \varepsilon \right)$ are provided in the next section.

To demonstrate the validity of our algorithms we first consider a system described by a one dimensional double well potential whose exacts weights $w_k$ for $k = 1,..., K$ are known. We show that both IGS and IMGS sample from the correct distribution at a given $\beta$.
In the subsequent subsection we consider a more complicated test system, the Ising model. For the Ising model we demonstrate that both IGS and IMGS improve the relaxation dynamics of inverse temperature and some system observables when compared to their respective reversible counterparts. We also show that the relaxation dynamics  of inverse temperature and system observables can be significantly better than those of MH and IMH algorithms. In this section we use $\beta = 1/k_BT$, whereby the Boltzmann constant $k_B$ is set to 1.

In this paper we define the integrated autocorrelation time $\tau_{int,f}$ for a system observable $f$ as 
\begin{equation}
\tau_{int,f} = 1 + 2\sum_{t = 1}^{\infty} C_f(t)
\end{equation}
where $C_f(t)$ is the autocorrelation function given by
\begin{equation}
C_f(t) = \frac{\text{E}_{\pi}\left[f(t' + t)f(t')\right] - \text{E}_{\pi}\left[f(t')\right]^2}{\text{E}_{\pi}[f^2(t')] - \text{E}_{\pi}\left[f(t')\right]^2},
\end{equation}
and we set $t'$ sufficiently large for equilibration to estimate $C_f(t)$. $\text{E}_{\pi}\left[...\right]$ is understood to be the expectation value with respect to the target distribution $\pi\left(\bm{\sigma},\beta\right)$. Note that the expectation value $\text{E}_{\tilde{\pi}}[f]$ with respect to the extended target distribution $\tilde{\pi}(\bm{\sigma},\beta, \varepsilon)$ is equivalent to the expectation value with respect to the original distribution $\pi(\bm{\sigma},\beta)$: 
\begin{align}
\text{E}_{\tilde{\pi}}\left[f\right] &= \sum\limits_{\tilde{\Omega}}\tilde{\pi}\left(\bm{\sigma}, \beta, \varepsilon\right)f\left(\bm{\sigma}, \beta, \varepsilon \right) \nonumber \\
&= \sum_{\varepsilon'}\sum_{\bar{\Omega}}\tilde{\pi}\left(\bm{\sigma}, \beta, \varepsilon' \right)f\left(\bm{\sigma},\beta,\varepsilon'\right) \nonumber \\
&= \sum_{\varepsilon'}\sum_{\bar{\Omega}}\frac{1}{2}\pi(\bm{\sigma}, \beta)f(\bm{\sigma}, \beta) \nonumber \\
&= \sum\limits_{\bar{\Omega}}\pi(\bm{\sigma}, \beta)f(\bm{\sigma},\beta) \nonumber \\
&= \text{E}_{\pi}\left[f\right],
\end{align}
where we have made use of \eqref{extended original targ relat} and have assumed $f(\bm{\sigma}, \beta, \varepsilon) = f(\bm{\sigma}, \beta, -\varepsilon) = f(\bm{\sigma}, \beta)$. Consider the measurements $f_1,..., f_N$ of the the observable $f$. A large integrated autocorrelation time of the observable $f$ is indicative of a large corresponding asymptotic variance on the expectation value for $f$. As often the relationship $\sigma_f^2 = \sigma_{0,f}^2\tau_{int,f}$ is used to compute the asymptotic variance $\sigma_f^2$. $\sigma_{0,f}^2$ is the \emph{naive} variance on the expectation value of $f$, that is, the variance on the expectation value of $f$ by treating all the realisations $f_1,...,f_N$ as though they were independently sampled.

 In simulated and parallel tempering simulations we are often interested in computing expectation values of system observables under the distribution $\pi(\bm{\sigma}\vert \beta)$ given in \eqref{Gibbs-Boltz distr} at a single temperature of interest, often the coolest temperature. The simplest method of achieving this is to discard all measurements made at temperatures other than the temperature of interest and thereby make use of only a fraction of the data generated. However perhaps a more cost effective method is to use one of the re-weighting techniques \cite{Chodera, WHAM} to properly weight the data generated at all temperatures in order to compute expectations under $\pi(\bm{\sigma} \vert \beta)$ for any given temperature of interest. In other words, all configurations sampled from the joint distribution $\pi(\bm{\sigma}, \beta)$ in \eqref{enlarged target} can be reweighed to compute expectations at a given temperature of interest. Within a fixed computational time, it is therefore of interest to collect as large amount as possible of \emph{uncorrelated} samples from the joint $\pi(\bm{\sigma}, \beta)$ in order to compute (using reweighing techniques) expectation values with small variance at a given temperature of interest. The quantity $N/\tau_{int}$ is often used to establish the effective sample size i.e. number of uncorrelated samples in the time series data consisting of $N$ measurements. The integrated autocorrelation time therefore not only quantifies the relaxation dynamics of a given system observable, but it can also be used to test the \emph{sampling efficiency} of a given algorithm. Among other comparison tools we will therefore make use of the integrated autocorrelation time to provide comparison of sampling efficiency of our methods to other methods currently in use.
 
  \begin{figure*}[t!]
\centering

\includegraphics[width=.32\textwidth, height=.25\textwidth, ]{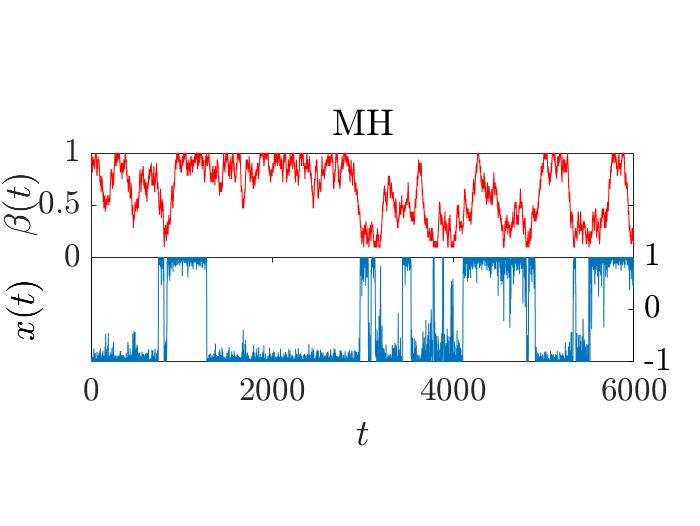}\hfill
\includegraphics[width=.32\textwidth, height=.25\textwidth, ]{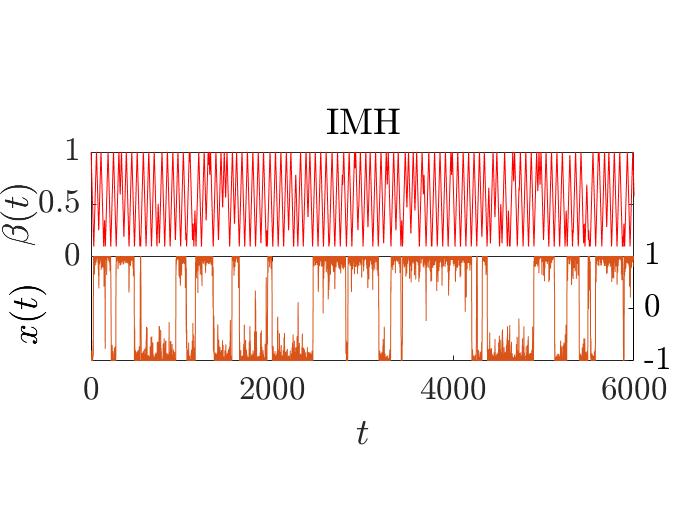}\hfill
\includegraphics[width=.32\textwidth, height=.25\textwidth, ]{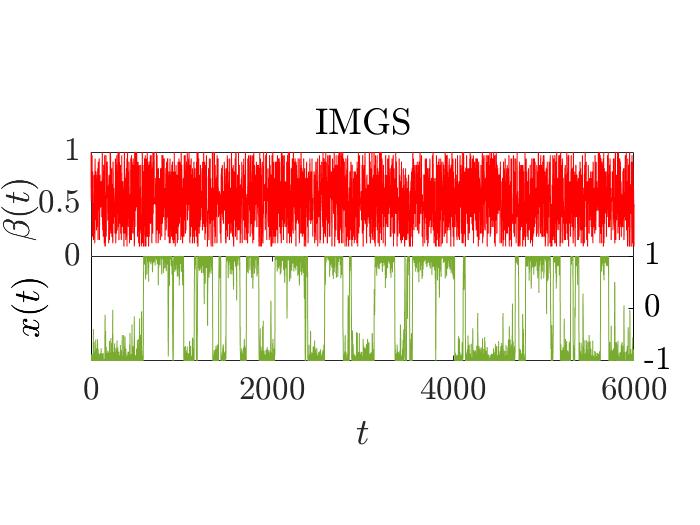}

\includegraphics[width=.32\textwidth, height=.25\textwidth, ]{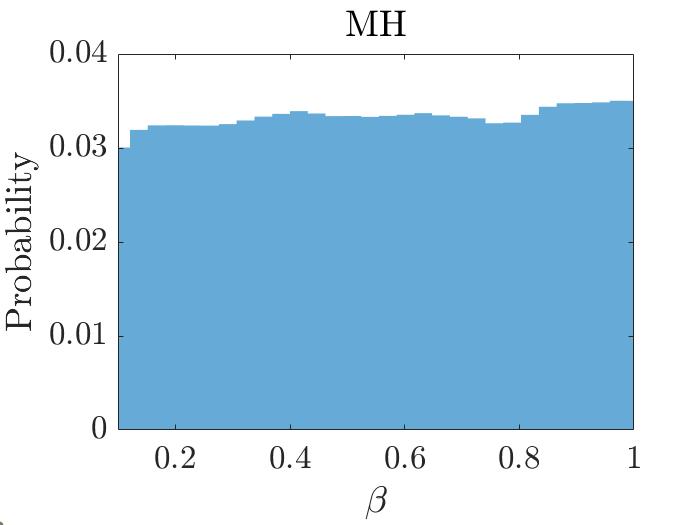}\hfill
\includegraphics[width=.32\textwidth, height=.25\textwidth, ]{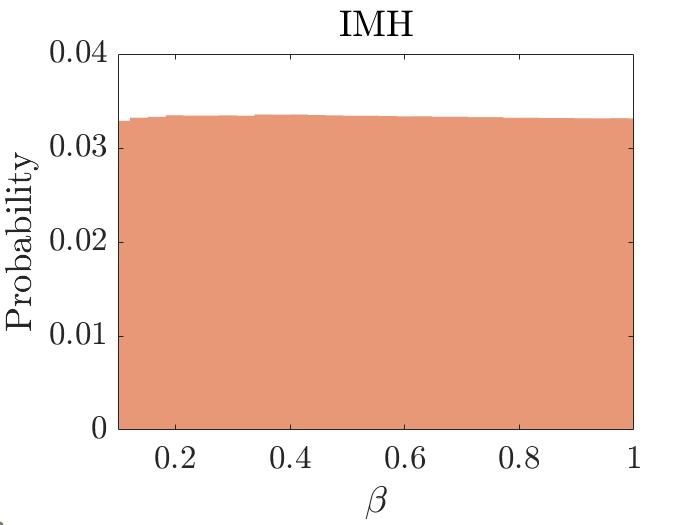}\hfill
\includegraphics[width=.32\textwidth, height=.25\textwidth, ]{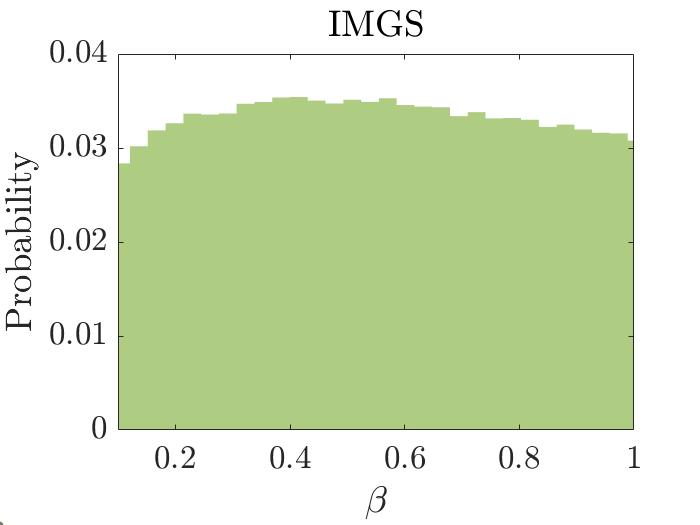}

\includegraphics[width=.32\textwidth, height=.25\textwidth, ]{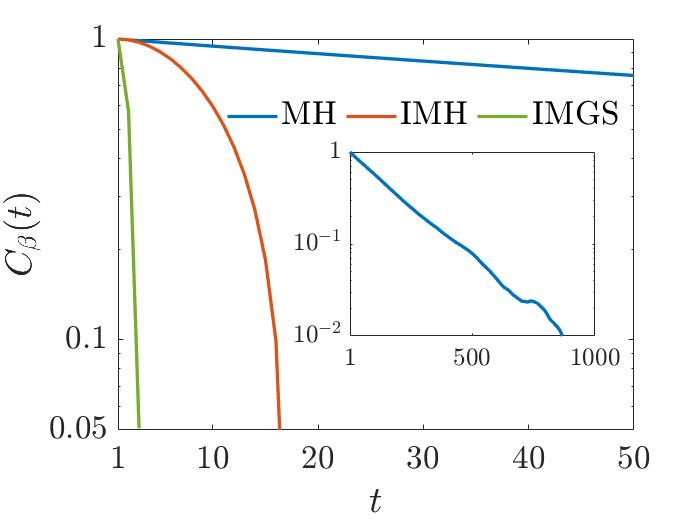}\hfill
\includegraphics[width=.32\textwidth, height=.25\textwidth, ]{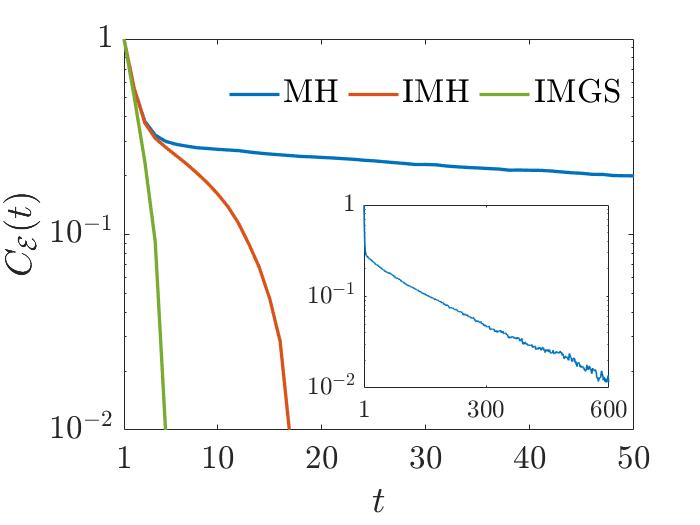}\hfill
\includegraphics[width=.32\textwidth, height=.25\textwidth, ]{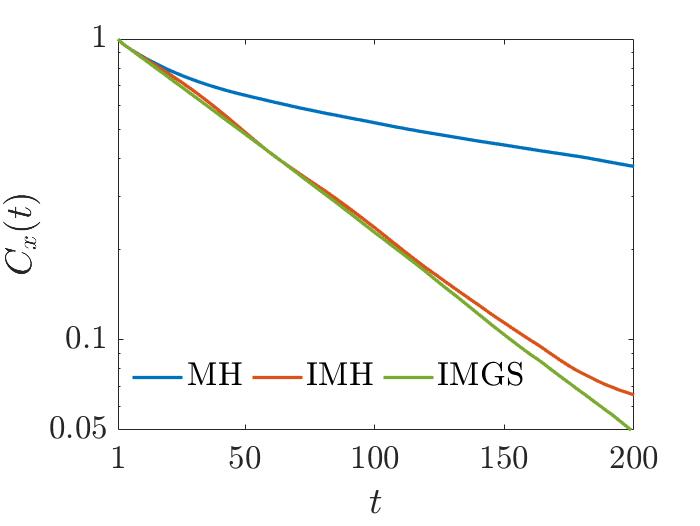}

\includegraphics[width=.32\textwidth, height=.25\textwidth, ]{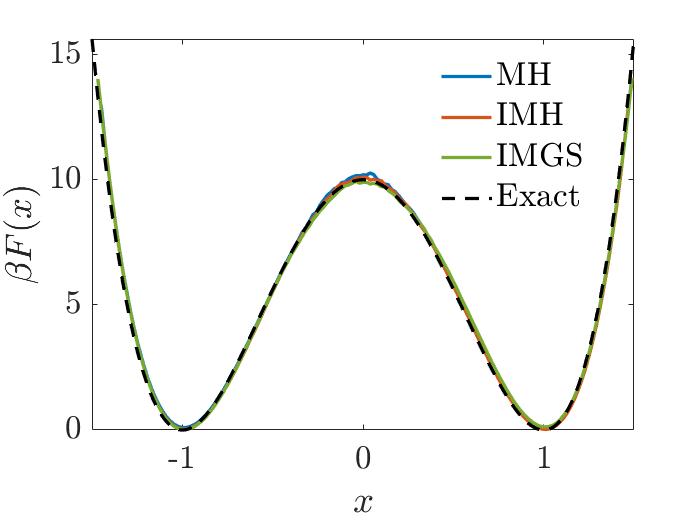}\hfill
\includegraphics[width=.32\textwidth, height=.25\textwidth, ]{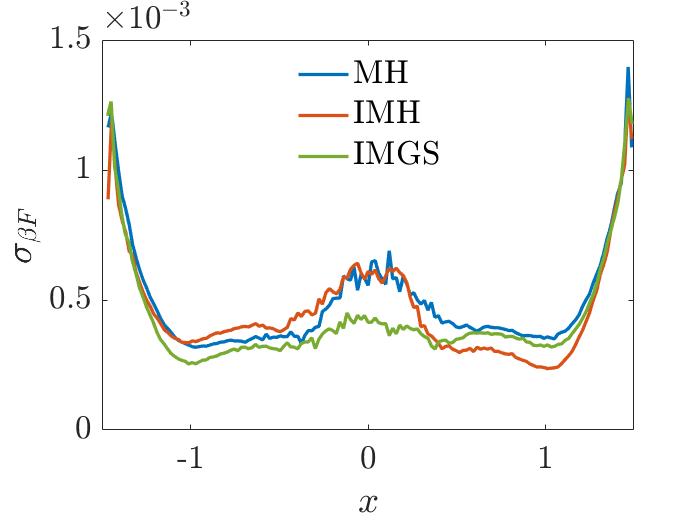}\hfill
\includegraphics[width=.32\textwidth, height=.25\textwidth, ]{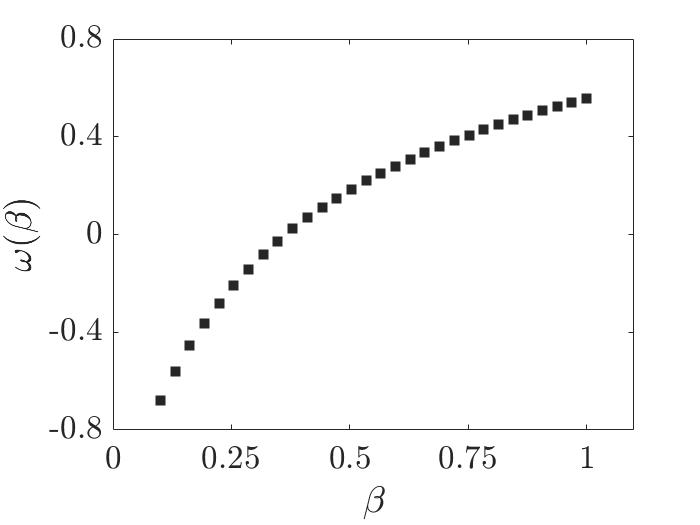}\hfill

\caption{Simulations of a simple system described by the 1D model potential given in \eqref{model potential hamiltonian} for temperature domain size $K = 32$. \textbf{Top row}: History of inverse temperature $\beta$ and position coordinate $x$ as a function of iteration $t$. \textbf{Second row}: Probability of inverse temperatures obtained from frequency distributions. \textbf{Third row:} Autocorrelation functions $C(t)$ for inverse temperature $\beta$, energy $\mathcal{E}(x)$, and position coordinate $x$. \textbf{Bottom row}: Free energy profile $\beta F(x) = -\text{ln}(\pi(x \vert \beta))$ at the coldest temperature $\beta = 1$ (left), standard error on the free energy profile trajectories obtained from 100 independent simulations (center) and the weights $\omega(\beta)$ computed numerically using \eqref{numeric weights} (right). The simulation parameters are : $\Gamma = 10^2$ iterations, $\mathcal{T} = 10^6$ iterations, $\beta_1 = 1$ and $\beta_K = 0.1$ all equally spaced, and energy barrier height $C = 10$ in \eqref{model potential hamiltonian}. The simulations were initialized with $\beta^{(0)} = 1$, $x^{(0)} = -1$ and a random assignment of $\varepsilon \in \lbrace -1,+1 \rbrace$. The deviation parameter $\delta$ is set to 1 for both IMH and IMGS. }
\label{Model_potential_30_K}
\end{figure*}

\begin{figure*}[t!]
\centering

\includegraphics[width=.32\textwidth, height=.25\textwidth, ]{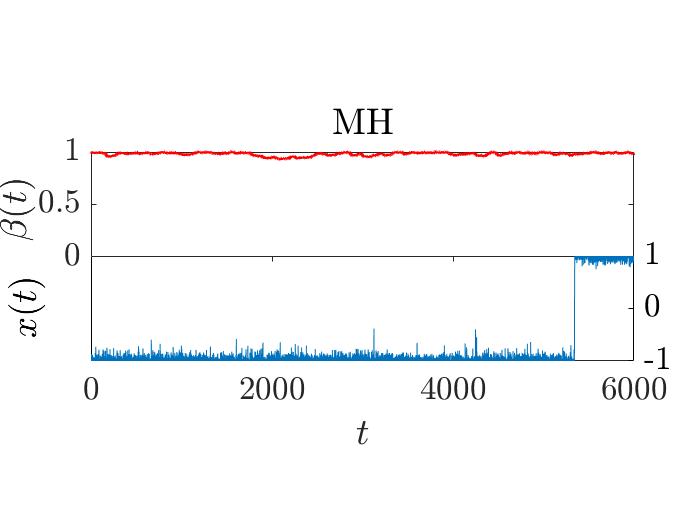}\hfill
\includegraphics[width=.32\textwidth, height=.25\textwidth, ]{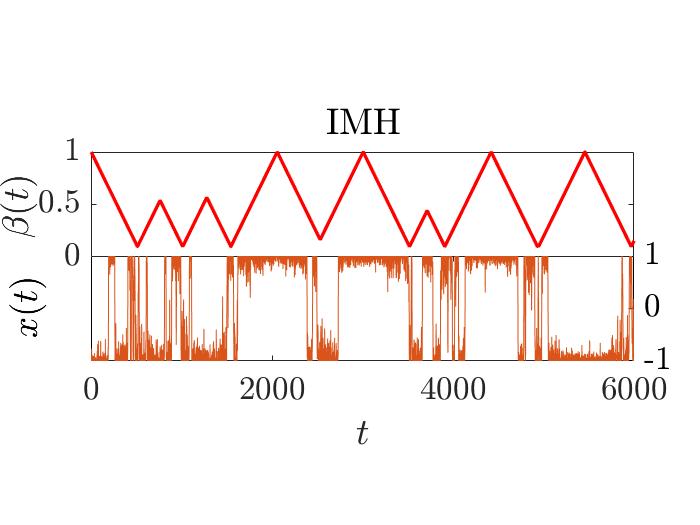}\hfill
\includegraphics[width=.32\textwidth, height=.25\textwidth, ]{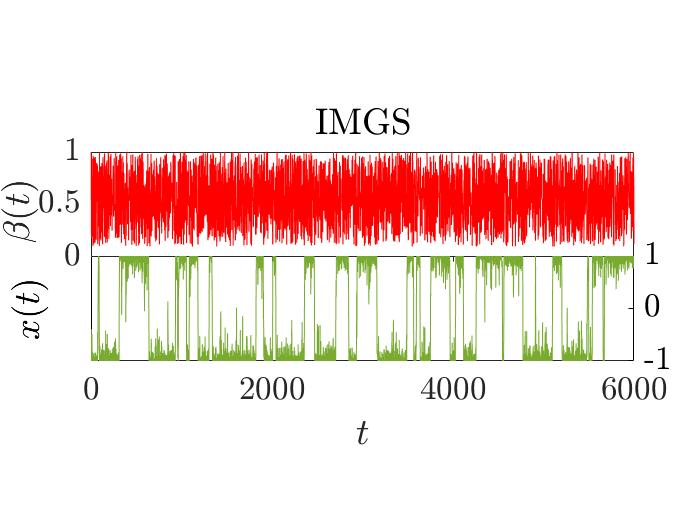}

\includegraphics[width=.32\textwidth, height=.25\textwidth, ]{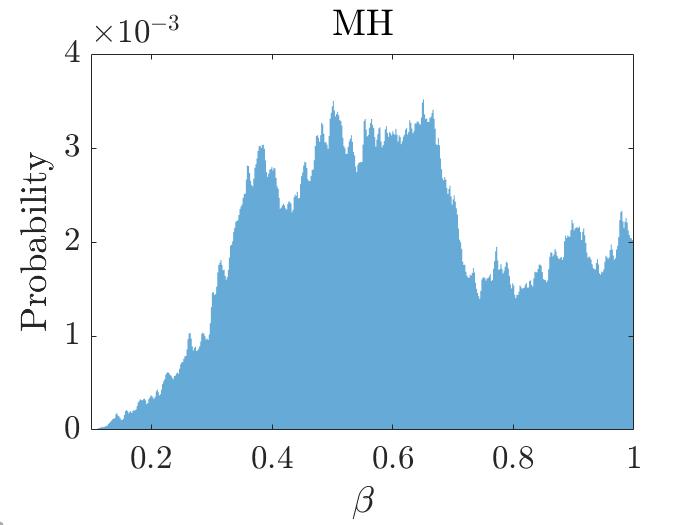}\hfill
\includegraphics[width=.32\textwidth, height=.25\textwidth, ]{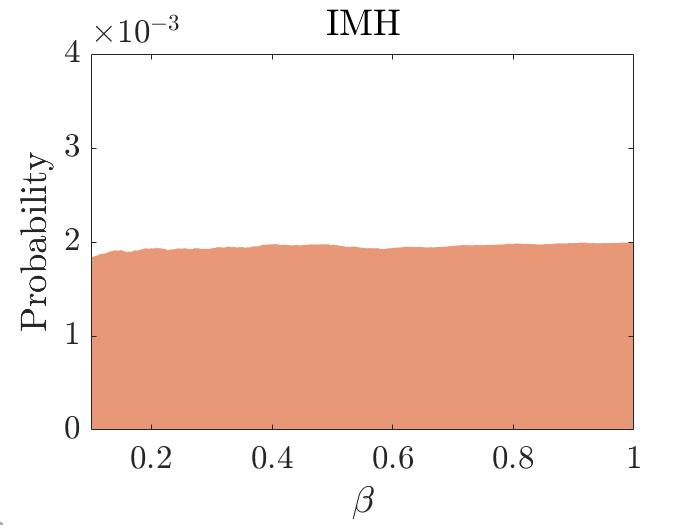}\hfill
\includegraphics[width=.32\textwidth, height=.25\textwidth, ]{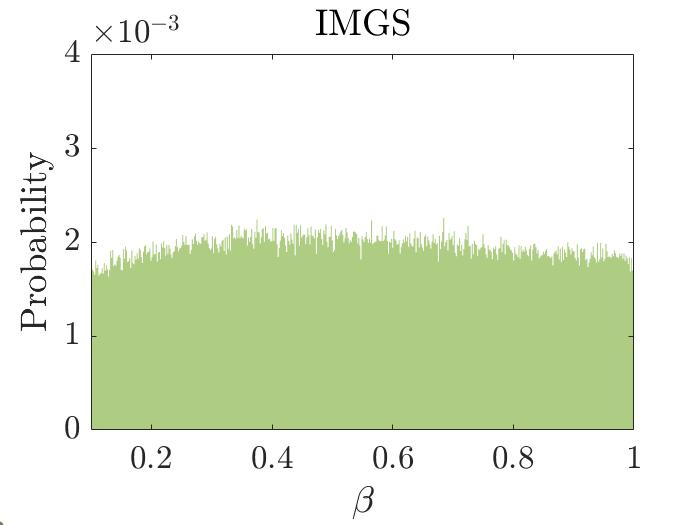}

\includegraphics[width=.32\textwidth, height=.25\textwidth, ]{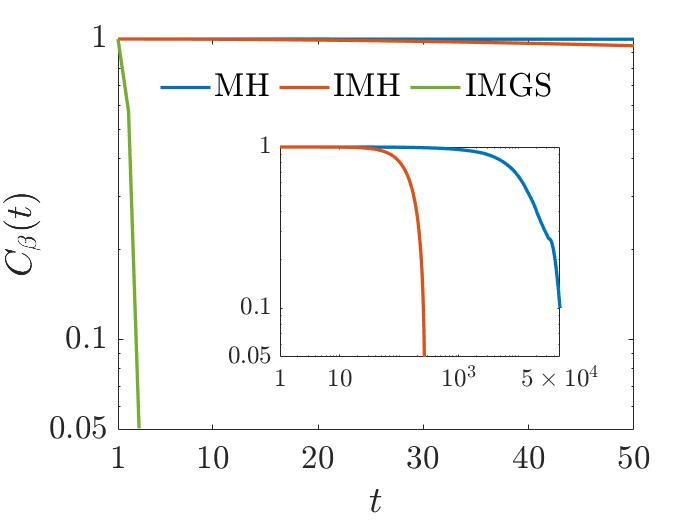}\hfill
\includegraphics[width=.32\textwidth, height=.25\textwidth, ]{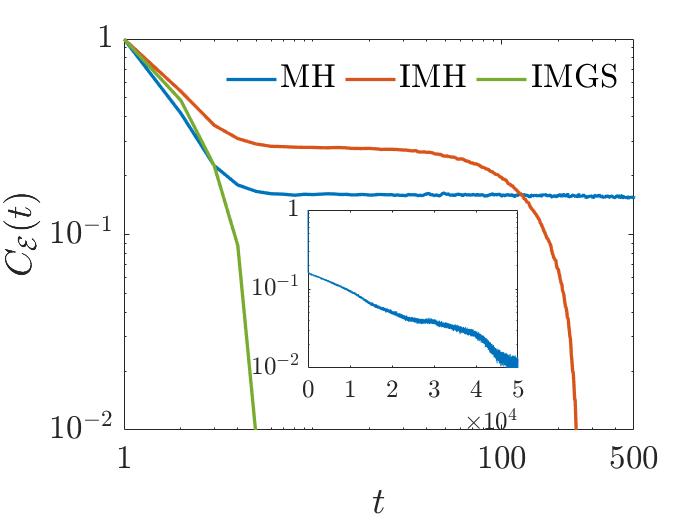}\hfill
\includegraphics[width=.32\textwidth, height=.25\textwidth, ]{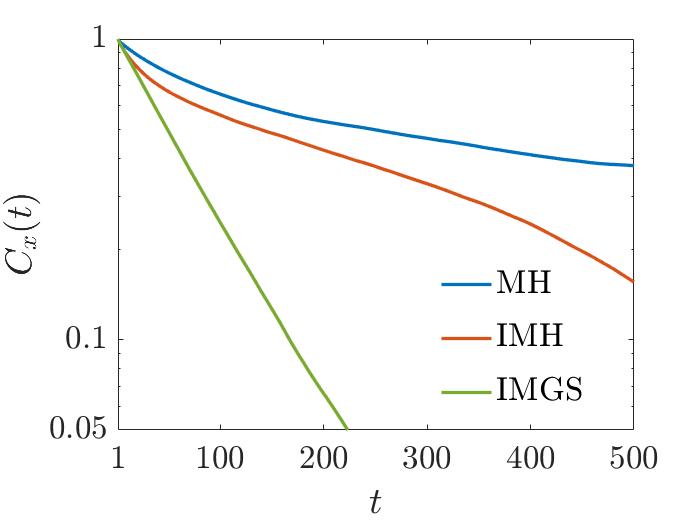}

\includegraphics[width=.32\textwidth, height=.25\textwidth]{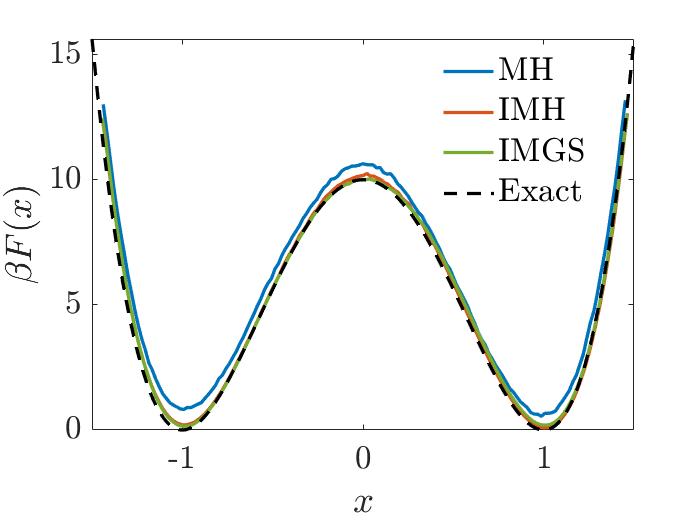}\hfill
\includegraphics[width=.32\textwidth, height=.25\textwidth]{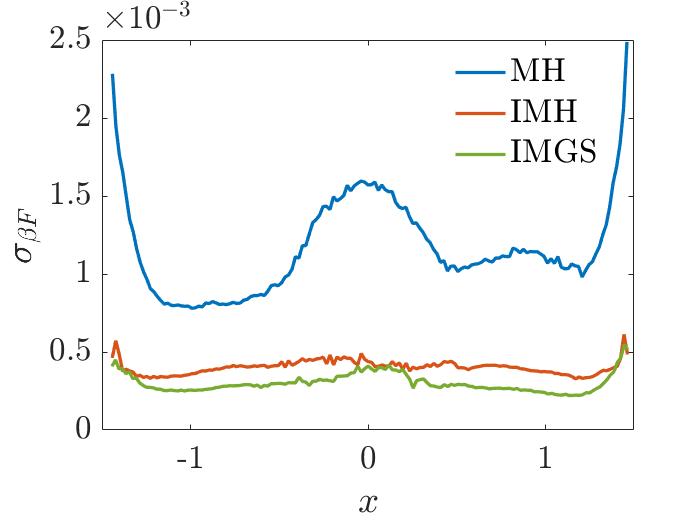}
\includegraphics[width=.32\textwidth, height=.25\textwidth, ]{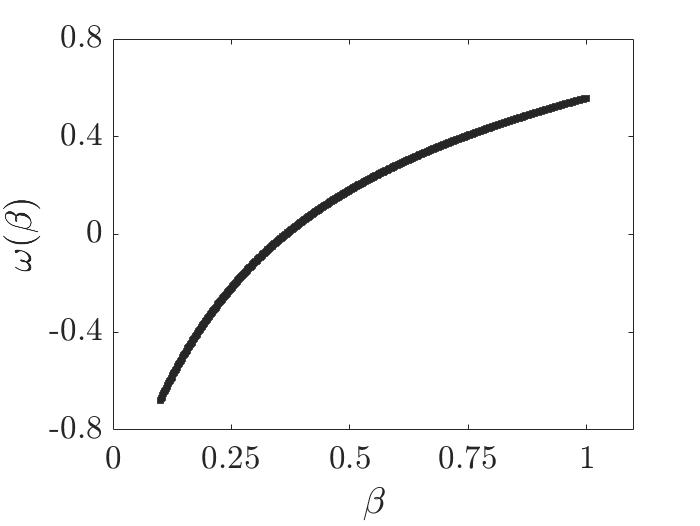}

\caption{Simulations of a simple system described by the 1D model potential given in \eqref{model potential hamiltonian} for temperature domain size $K = 512$. \textbf{Top row}: History of inverse temperature $\beta$ and position coordinate $x$ as function of iteration $t$. \textbf{Second row}: Probability of inverse temperatures obtained from frequency distributions. \textbf{Third row:} Autocorrelation functions $C(t)$ for inverse temperature $\beta$, energy $\mathcal{E}(x)$, and position coordinate $x$. \textbf{Bottom row}: Free energy profile $\beta F(x) = -\text{ln}(\pi(x \vert \beta))$ at the coldest temperature $\beta = 1$ (left), standard error on the free energy profile trajectories obtained from 100 independent simulations (center) and the weights $\omega(\beta)$ computed numerically using \eqref{numeric weights} (right). The simulation parameters are : $\Gamma = 10^2$ iterations, $\mathcal{T} = 10^6$ iterations, $\beta_1 = 1$ and $\beta_K = 0.1$ all equally spaced, and energy barrier height $C = 10$ in \eqref{model potential hamiltonian}. The simulations were initialized with $\beta^{(0)} = 1$, $x^{(0)} = -1$ and a random assignment of $\varepsilon \in \lbrace -1,+1 \rbrace$. The deviation parameter $\delta$ is set to 1 for both IMH and IMGS.}
\label{Model_potential_512_K}
\end{figure*}

\subsection*{1D Model potential}
We consider a simple system described by the 1D double well potential 
\begin{equation}\label{model potential hamiltonian}
U(x) = C(x+1)^2(x-1)^2, \,\,\,\,\,\,  x \in \Omega
\end{equation}
with energy minima at coordinates $x = +1$ and $x = -1$ and an energy barrier of magnitude $C \geqslant 0$ at $x = 0$. To sample from $\tilde{\pi}\left(\bm{\sigma}\vert \beta, \varepsilon \right)$, we perform a Metropolis-Hastings MC simulation at fixed $\beta$ using a Gaussian proposal in the $x$-coordinate,
\begin{equation}
x'_{i+1} = x_i + \varsigma \xi,
\end{equation}
to propose $x' \in \Omega$ for the $(i + 1)^{\text{th}}$ Monte-Carlo time step. $\xi \sim \mathcal{N}(0,1)$ and the standard deviation $\varsigma$ is fixed at 0.05. To sample from $\tilde{\pi}\left( \beta \vert \bm{\sigma}, \varepsilon \right)$ we use a predetermined set of $K$ inverse temperatures that are equally spaced in the range $\beta_1 = 1$ and $\beta_K = 0.1$. The weights are then numerically computed using 
\begin{equation}\label{numeric weights}
w_k = -\text{ln}Z(\beta_k) = -\text{ln}\int_{-\infty}^{+\infty}e^{-\beta_k U(x)}
\end{equation}
Simulated tempering is then performed by alternately sampling from the conditional distributions $\tilde{\pi}\left(\bm{\sigma}\vert \beta, \varepsilon \right)$ and $\tilde{\pi}\left( \beta \vert \bm{\sigma}, \varepsilon \right)$, such that a single iteration $t$ involves $\Gamma$ Monte-Carlo steps to sample from $\tilde{\pi}\left(\bm{\sigma}\vert \beta, \varepsilon \right)$ and a single MC step to sample from $\tilde{\pi}\left( \beta \vert \bm{\sigma}, \varepsilon \right)$.

We have performed two sets of simulated tempering simulations with $K = 30$ and $K = 512$ temperature domain sizes. The results are shown in Figure \ref{Model_potential_30_K} and Figure \ref{Model_potential_512_K} respectively. For comparison we show the performance of IMGS against the standard Metropolis-Hastings algorithm as given in Algorithm.\ref{MH_ST} and the irreversible Metropolis-Hastings (IMH) as proposed by Sakai and Hukushima \cite{Sakai_Hukushima_simulated_tempering}. For both MH and the IMH algorithms we had adopted nearest neighbour exchange scheme for $\beta$. The tunable deviation parameter $\delta$ is set to 1 for both IMH and IMGS. 

 The top row of Figure \ref{Model_potential_30_K} shows the evolution of $\beta$ and the position coordinate $x$ as a function of the iteration $t$.  Expectedly the history of $\beta$ for MH indicates that the system performs a characteristically random walk in the temperature space. The IMH algorithm \cite{Sakai_Hukushima_simulated_tempering} on the other hand was proposed to suppress diffusive behaviour in the temperature space, we therefore observe the typically deterministic exploration of $\beta$ coordinates with visually better mixing rate than MH. The IMGS seems to provide a more ballistic exploration of temperature space as shown on the top right. Unlike MH and IMH which perform optimally with nearest neighbour $\beta$ proposals, the IMGS can perform more distant jumps in the $\beta$ coordinate, thus inducing a more global exploration of temperature space.
 
From the autocorrelation functions $C_{\beta}(t)$ for $\beta$, shown in the third row of Figure \ref{Model_potential_30_K}, we observe that $C_{\beta}(t)$ for IMGS decays more rapidly compared to MH and IMH. In particular we find that $[\tau_{int,\beta}]_{\text{MH}}/[\tau_{int,\beta}]_{\text{IMGS}} \sim 87.6$ and $[\tau_{int,\beta}]_{\text{IMH}}/[\tau_{int,\beta}]_{\text{IMGS}} \sim 5.2$. Generally, an improvement in the relaxation dynamics of $\beta$ is accompanied by an improvement in the relaxation dynamics of system observables. From the history of position coordinate $x$ we clearly observe that an improvement in the mixing rate of $\beta$ induces a more frequent crossing of energy barrier at $x = 0$. The energy and position coordinate autocorrelation functions, $C_{\mathcal{E}}(t)$ and $C_{x}(t)$ respectively, therefore decays most rapidly for the IMGS. Note that $C_x(t)$ for IMGS decays substantially faster than that for MH,  but compared to that for IMH there is little gain. This is in contrast to the autocorrelation function $C_{\mathcal{E}}(t)$ for energy, which decays most rapidly for the IMGS with a distinctly clear gain over that of IMH.  We speculate that perhaps for a given domain size $K$ there could possibly be an optimum mixing rate for $\beta$, such that any further improvement in mixing rate for $\beta$ may not necessarily lead to substantial improvement in the mixing rate for $x$.

In Figure \ref{Model_potential_512_K} we consider simulations for a larger temperature domain, $K = 512$. The improvement in performance of IMGS compared to MH and IMH is now more visibility clear. The history of $\beta$ for MH, as shown in the top row, confirms the expected degradation in exploration of temperature space, which is typical of random walks on domains of increasing size. For the IMH algorithm the deterministic exploration of temperature space in a specific direction (top row, middle) is now visibly clearer than for $K = 32$ case. Notice that this behaviour leads to a very uniform exploration of $\beta$ coordinates as shown in the probability  distribution of inverse temperature, (second row, middle). In contrast the IMGS mixing rate for $\beta$ remains visibly unchanged with increasing temperature domain. In fact notice that the autocorrelation functions $C_{\beta}(t)$, $C_{\mathcal{E}}(t)$ and $C_{x}(t)$ for IMGS are visibly unchanged from the $K = 32$ case. This observation is in agreement with a similar study  by Chodera et al.\cite{Chodera} who show numerically that the integrated autocorrelation time for temperature and position index remain independent of temperature domain size $K$ for algorithms based on Gibbs sampling. In particular we report $[\tau_{int,\beta}]_{\text{MH}}/[\tau_{int,\beta}]_{\text{IMGS}} \sim 1.25 \times 10^4 $ and  $[\tau_{int,\beta}]_{\text{IMH}}/[\tau_{int,\beta}]_{\text{IMGS}} \sim 79.2 $. Consequently the speed up in $\tau_{int,x}$ is $[\tau_{int,x}]_{\text{MH}}/[\tau_{int,x}]_{\text{IMGS}} \sim 14.7 $ and $[\tau_{int,x}]_{\text{IMH}}/[\tau_{int,x}]_{\text{IMGS}} \sim 3.4$. In fact for the MH algorithm the relaxation dynamics of $\beta$ is diffusive and therefore $\tau_{int,\beta}$ scales on the order of $\mathcal{O}(K^2)$. For the IMH algorithm, it has been demonstrated \cite{Sakai_Hukushima_simulated_tempering} that the relaxation dynamics of $\beta$ scales on the order of $\mathcal{O}(K)$, i.e. a square root reduction in the dynamical scaling of relaxation time with respect to temperature domain size. In the next section we have shown using the Ising model as a test bed, that for the IMGS the relaxation dynamics of $\beta$ and system observables is independent of $K$ and scales on the order of $\mathcal{O}(1)$.

Furthermore, we demonstrate in the bottom row of Figures \ref{Model_potential_30_K} and \ref{Model_potential_512_K} that the IMGS leaves the target distribution invariant. We have shown convergence  to the correct free energy profile $\beta F(x) = -\text{ln}(\pi(x\vert \beta))$ for the temperature of interest, the coldest temperature corresponding to $\beta_1 = 1$. The performance of the irreversible Gibbs sampler (IGS) (not shown for this model) is very similar to IMGS. The gain in integrated autocorrelation times for IMGS is only slightly better than those of IGS. Furthermore for this simple model we have not shown comparison of IMGS to that of its reversible counterpart the Metropolised-Gibbs sampler (MGS). The gain in integrated autocorrelation time of $\beta$, $\mathcal{E}$ and $x$ for IMGS is only slightly better than that of MGS.

 We have demonstrated the validity of our method using this simple 1D potential model and have provided first hand comparison with the widely used MH method and the recently proposed IMH algorithm of Sakai and Hukushima \cite{Sakai_Hukushima_simulated_tempering}. The potential gains of IMGS over its reversible counterpart is however not fully captured by this simple model. In the next section we observe that for a more complex and practical system we observe a clear gain of IMGS over its reversible counter-part. 
 
\begin{figure*}[t!]
\centering

\includegraphics[width=.32\textwidth, height=.25\textwidth, ]{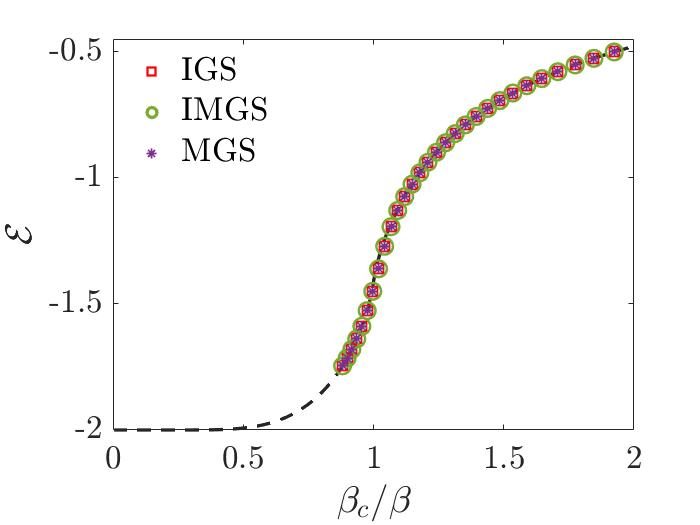}\hfill
\includegraphics[width=.32\textwidth, height=.25\textwidth, ]{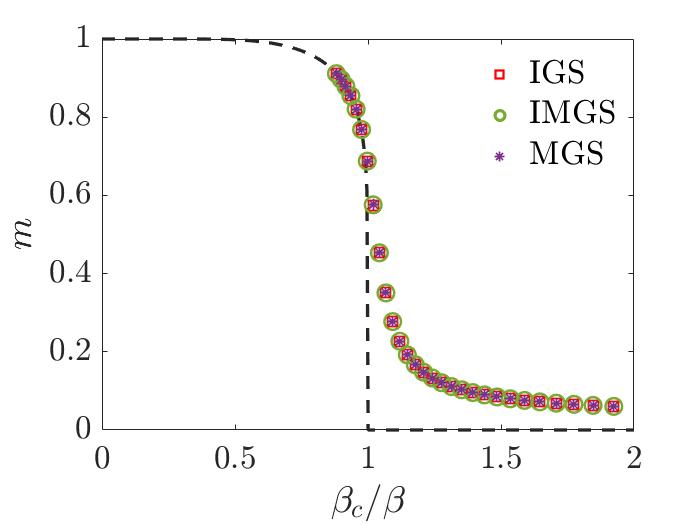}\hfill
\includegraphics[width=.32\textwidth, height=.25\textwidth, ]{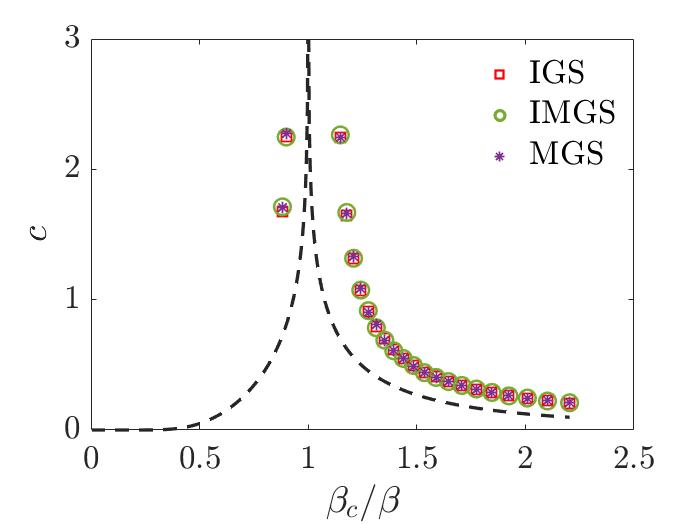}

\includegraphics[width=.32\textwidth, height=.25\textwidth, ]{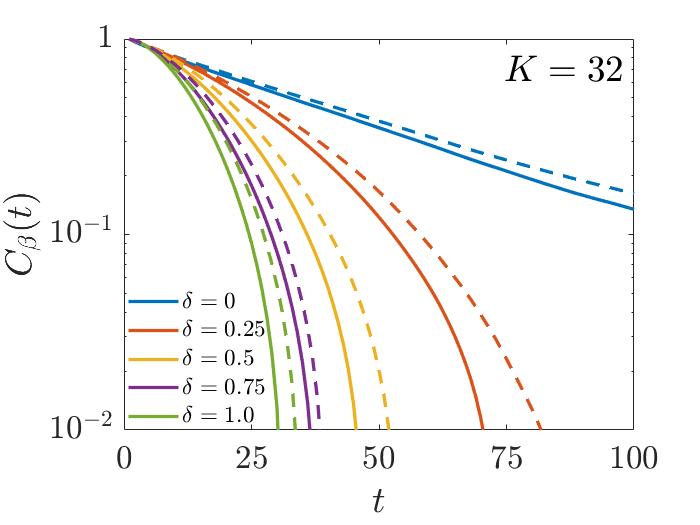}\hfill
\includegraphics[width=.32\textwidth, height=.25\textwidth, ]{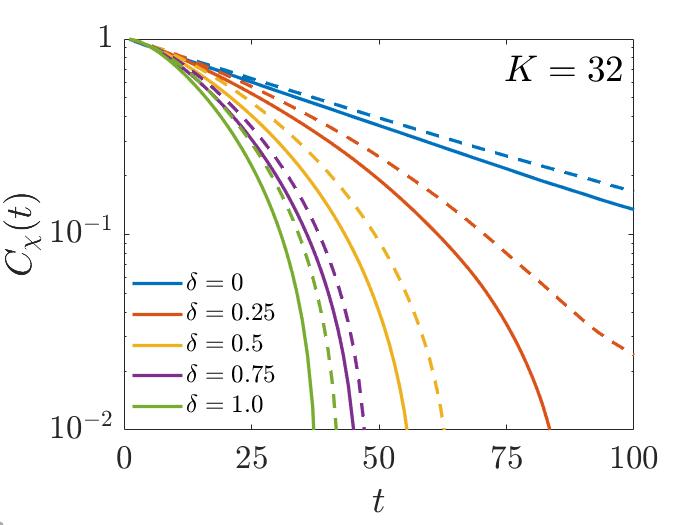}\hfill
\includegraphics[width=.32\textwidth, height=.25\textwidth, ]{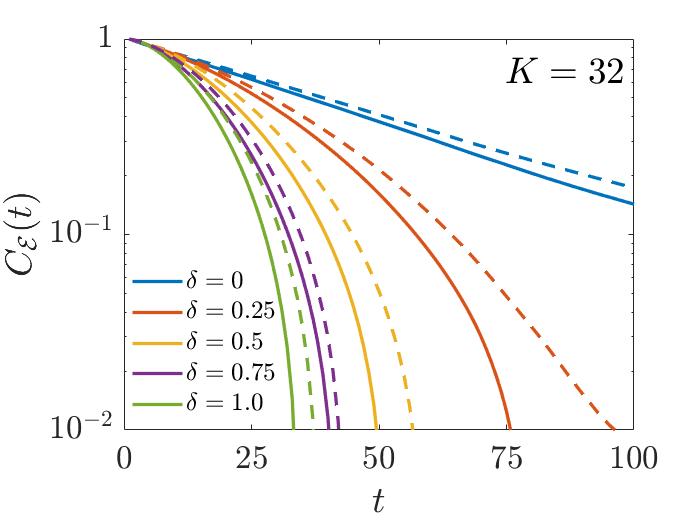}

\includegraphics[width=.32\textwidth, height=.25\textwidth, ]{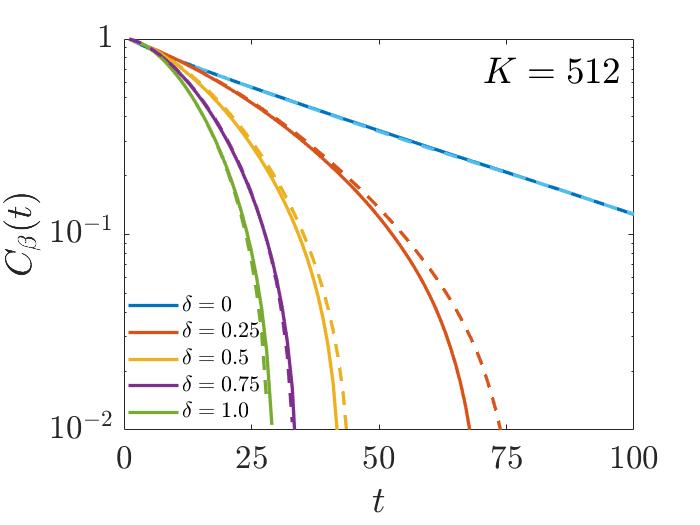}\hfill
\includegraphics[width=.32\textwidth, height=.25\textwidth, ]{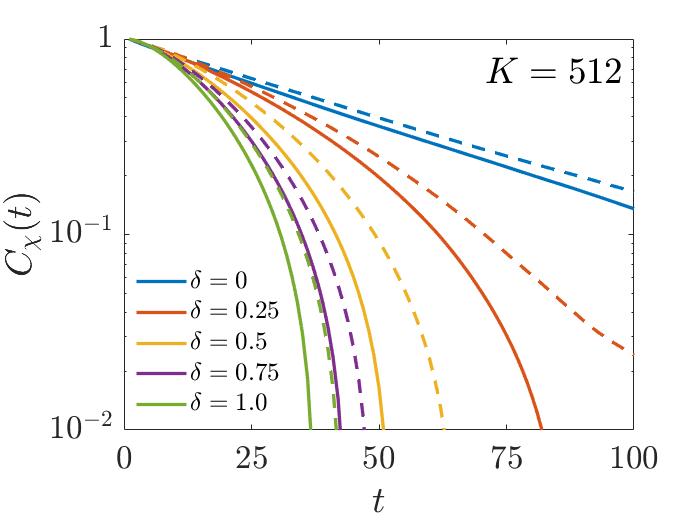}\hfill
\includegraphics[width=.32\textwidth, height=.25\textwidth, ]{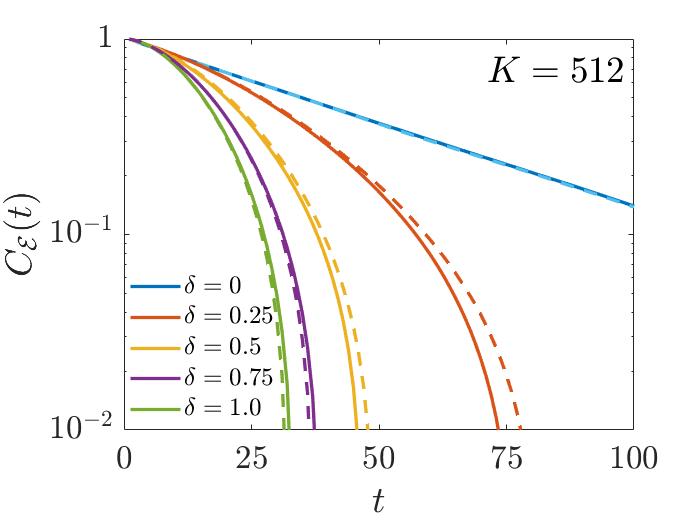}

\includegraphics[width=.32\textwidth, height=.25\textwidth, ]{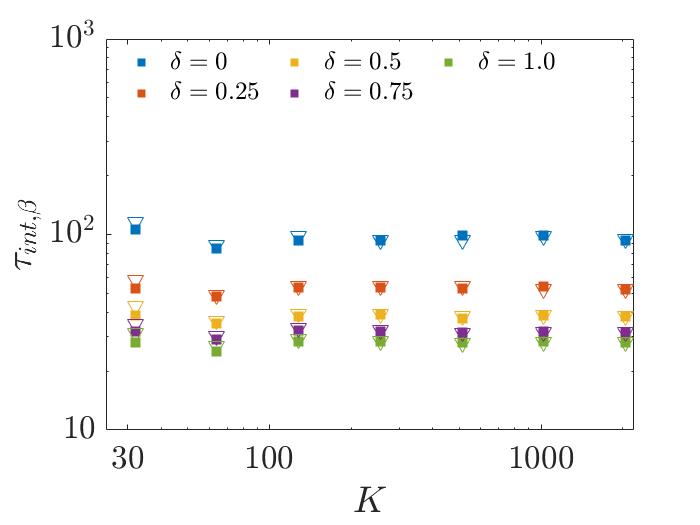}\hfill
\includegraphics[width=.32\textwidth, height=.25\textwidth, ]{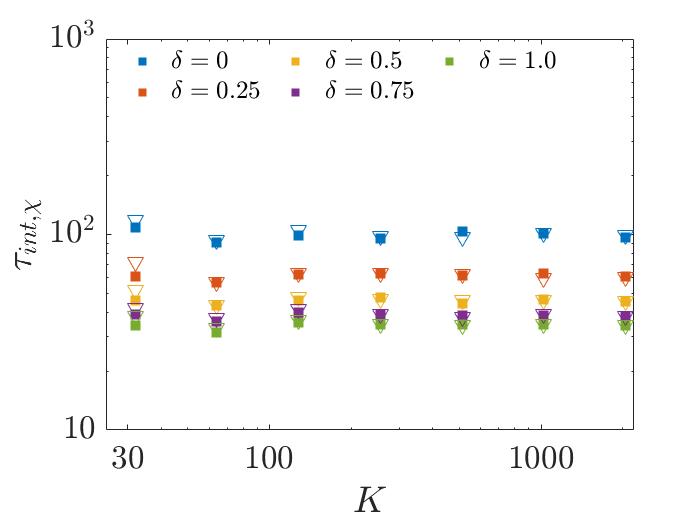}\hfill
\includegraphics[width=.32\textwidth, height=.25\textwidth, ]{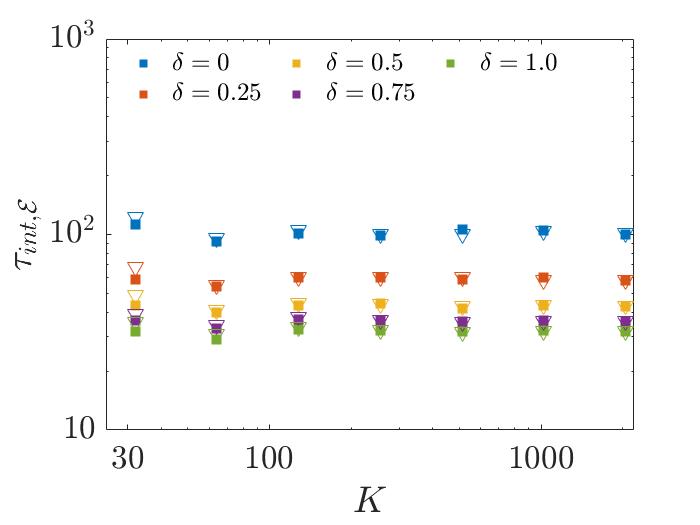}

\caption{Simulation results for the $25 \times 25$ Ising model. The simulation parameters are: $\Gamma = 10^3$ sweeps, $\mathcal{T} = 10^6$ iterations. $\beta_1 = 0.5$ and $\beta_K = 0.2$ all equally spaced. \textbf{Top row:} Energy density $\mathcal{E}$, magnetisation density $m$ and specific heat capacity $c$ obtained from simulated tempering simulations. The values obtained with our methods IGS (red squares) and IMGS (green circles) is in perfect agreement with the well established MGS (purple stars). The dashed lines are Onsager's \cite{Huang} exact solutions for a 2D lattice of infinite dimensions. \textbf{Second and third row:} Autocorrelation functions of $\beta$, $\mathcal{E}$ and $\chi$ for $K = 32$ (second row) and $K = 512$ (third row) for various deviation parameters $\delta$. IMGS (solid lines) and IGS (dashed lines). \textbf{Bottom row:} Integrated autocorrelation times $\tau_{int}$ with respect to temperature domain size $K$. IMGS (squares) and IGS (inverted triangles).}
\label{25by25_system}
\end{figure*}

\begin{figure*}[t!]
\centering

\includegraphics[width=.32\textwidth, height=.25\textwidth, ]{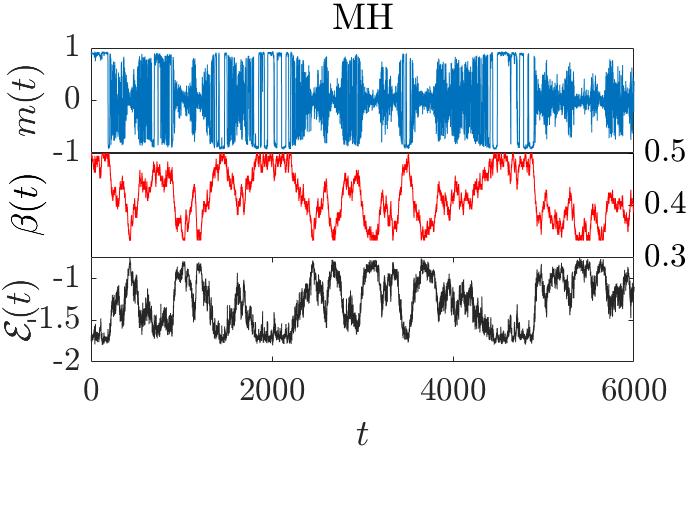}\hfill
\includegraphics[width=.32\textwidth, height=.25\textwidth, ]{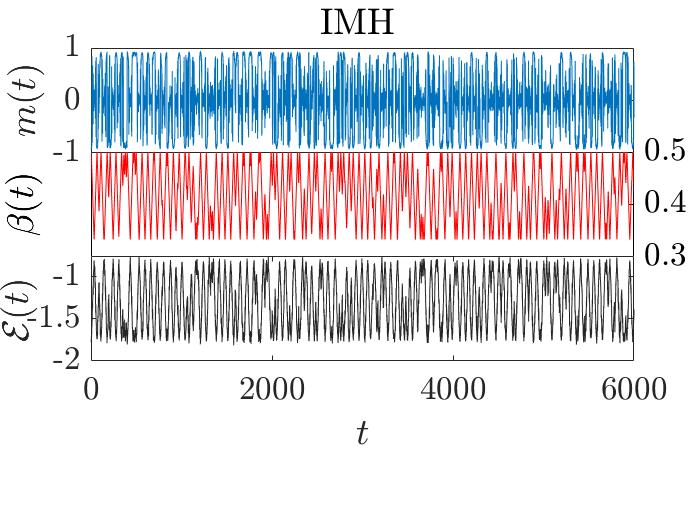}\hfill
\includegraphics[width=.32\textwidth, height=.25\textwidth, ]{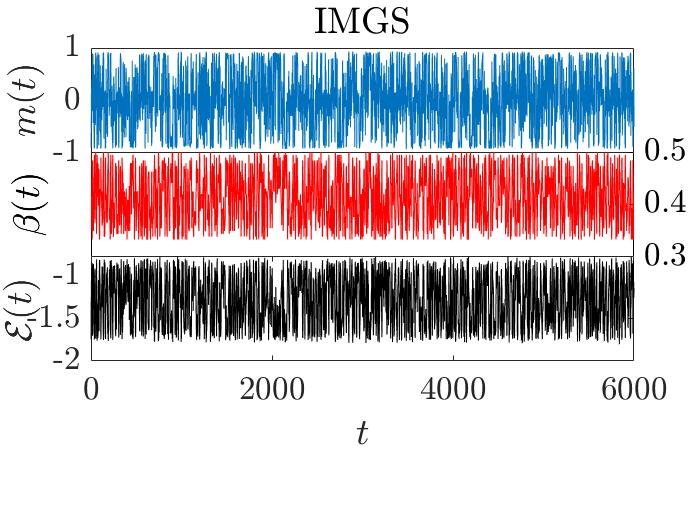}

\includegraphics[width=.32\textwidth, height=.25\textwidth, ]{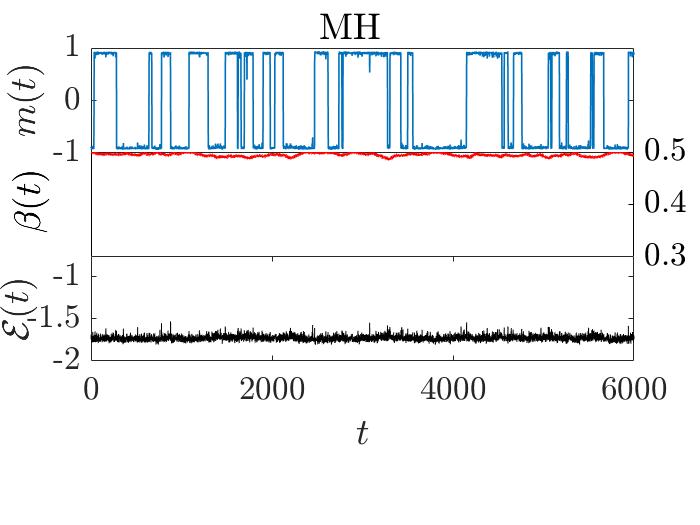}\hfill
\includegraphics[width=.32\textwidth, height=.25\textwidth, ]{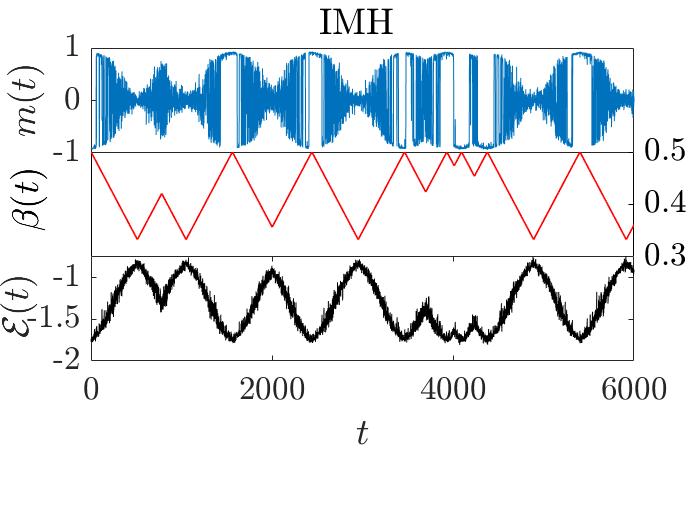}\hfill
\includegraphics[width=.32\textwidth, height=.25\textwidth, ]{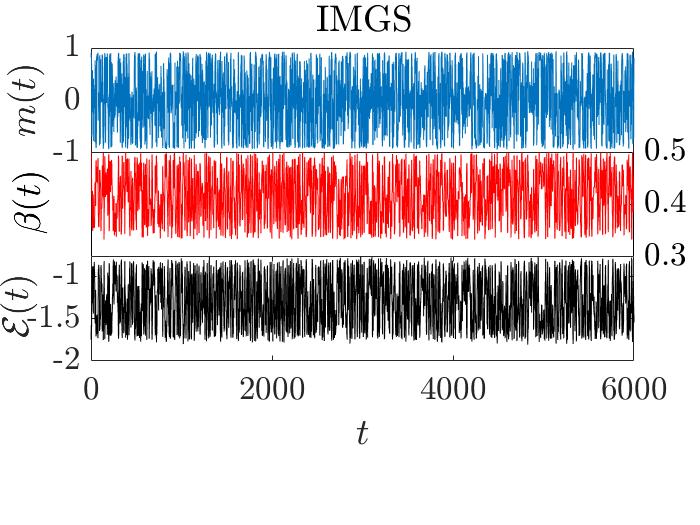}

\includegraphics[width=.32\textwidth, height=.25\textwidth, ]{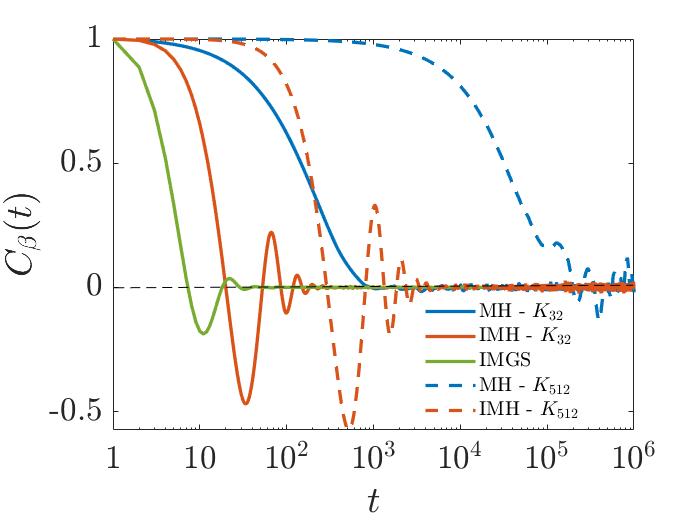}\hfill
\includegraphics[width=.32\textwidth, height=.25\textwidth, ]{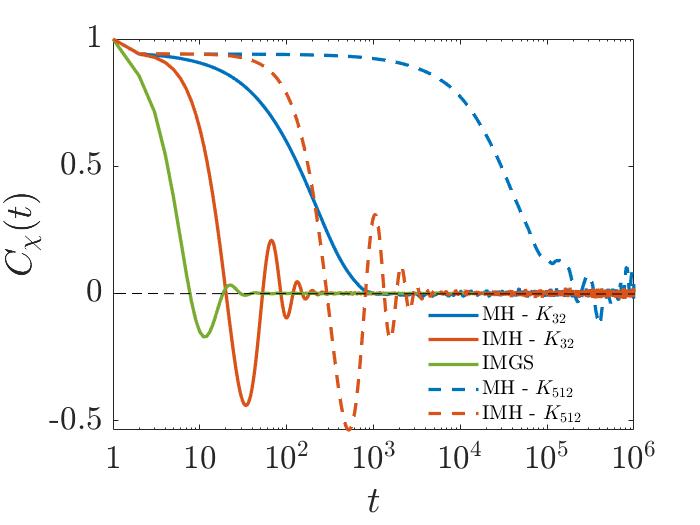}\hfill
\includegraphics[width=.32\textwidth, height=.25\textwidth, ]{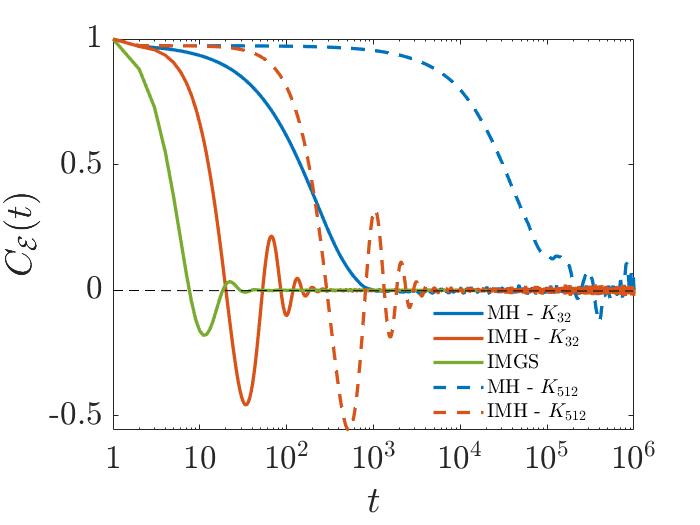}

\includegraphics[width=.32\textwidth, height=.25\textwidth, ]{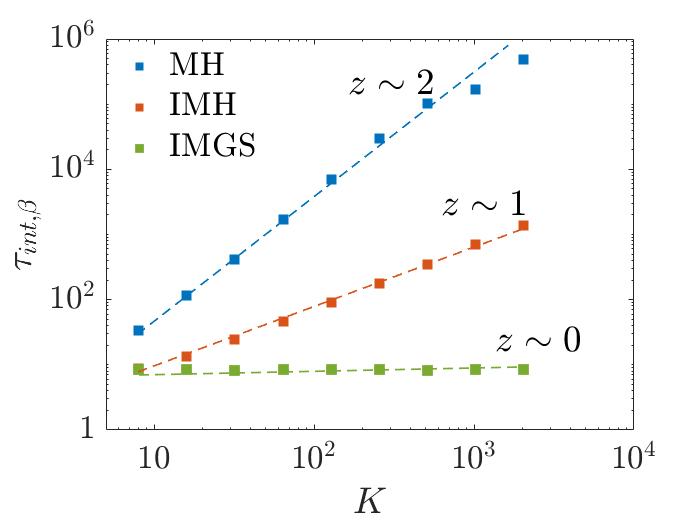}\hfill
\includegraphics[width=.32\textwidth, height=.25\textwidth, ]{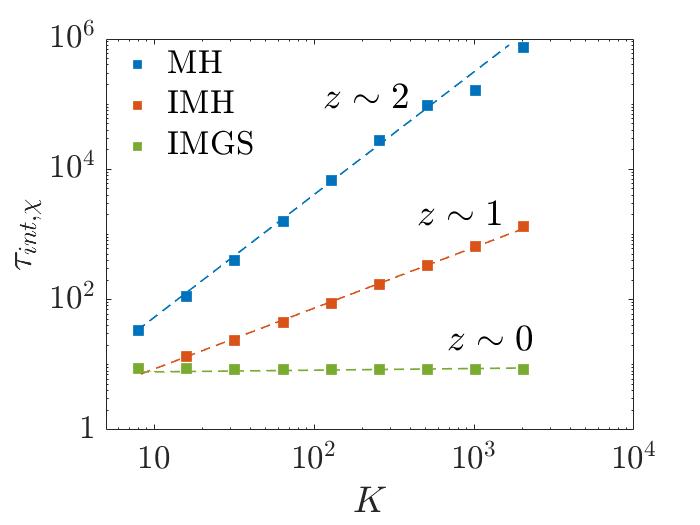}\hfill
\includegraphics[width=.32\textwidth, height=.25\textwidth, ]{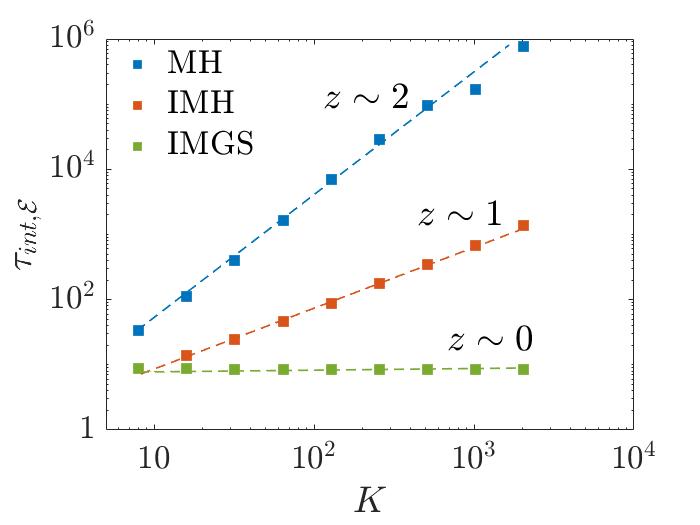}

\caption{Simulations results for the $12 \times 12$ Ising model. The simulation parameters are: $\Gamma = 10^2$ sweeps, $\mathcal{T} = 10^7$ iterations. $\beta_1 = 0.5$ and $\beta_K = 0.33$ all equally spaced. The deviation parameter $\delta = 1 $ for both IMH and IMGS. \textbf{First and second row:} Histories of inverse temperature $\beta$, magnetisation and energy density $m$ and $\mathcal{E}$ respectively shown for the first 6000 iterations $t$ for $K = 32$ (top row) and $K = 512$ (second row) temperatures. \textbf{Third row:} Autocorrelation functions $C_{\beta}(t)$, $C_{\chi}(t)$ and $C_{\mathcal{E}}(t)$ for the temperature domain size $K = 32$ (solid lines) and $K = 512$ (dashed lines). The autocorrelation functions of IMGS for the two domain sizes heavily overlap. \textbf{Bottom row:} Dynamical scaling of the integrated autocorrelation times for $\beta$, $\chi$ and $\mathcal{E}$ with respect to temperature domain size $K$. The dynamical scaling exponents are retrieved using the asymptotic relationship $\tau_{int} \sim K^{z}$. The dashed lines are the least squares fit to the data.}
\label{12by12_Ising}
\end{figure*}
 
\subsection*{Ising model}
In this section we test our methods on a $L \times L$ 2D Ising model \cite{MCMC_physics2} with periodic boundary conditions. The Hamiltonian $H(\bm{\sigma})$ of the Ising model is given by 
\begin{equation}
H(\bm{\sigma}) = -\sum\limits_{\langle i,j \rangle}J_{ij}\, \sigma_i\sigma_j
\end{equation}
where the notation $\langle i,j \rangle$ indicates that spins $\sigma_i$, $\sigma_j \in \lbrace +1,-1 \rbrace$ are nearest neighbours. We set the interaction strength to constant so that  $J_{ij} = J = 1$. A given configuration or state of the model is defined by the state vector $\bm{\sigma} = \left(\sigma_1, ..., \sigma_N\right) \in \Omega$ with $N = L^2$ spins. The discrete state space $\Omega = \lbrace 1, ..., S \rbrace$ therefore consists of $S = 2^N$ configurations. We consider a 2D lattice of row index $m$ and column index $n$, the energy density $\mathcal{E}$ of the system is then defined by
\begin{equation}
\mathcal{E} =  -J/N\sum\limits_{m,n}\left( \sigma_{m,n}\sigma_{m+1,n} + \sigma_{m,n}\sigma_{m,n+1}\right),
\end{equation} 
where periodic boundary conditions are imposed so that $\sigma_{m,L+1} = \sigma_{m,1}$ and $\sigma_{L+1,n} = \sigma_{1,n}$. The magnetisation density of the system is defined as $m = 1/N\sum_i \sigma_i$ and the magnetic susceptibility $\chi$ is given by 
\begin{equation}
\chi = \frac{1}{N}\parallel\sum_{i}\sigma_i\parallel^2,
\end{equation}

We perform simulated tempering by alternately sampling from the distributions $\tilde{\pi}(\bm{\sigma}\vert \beta, \varepsilon)$ and $\tilde{\pi}(\beta \vert \bm{\sigma}, \varepsilon)$. We define one sweep of the 2D spin lattice as $N$ MC trials to update individual spins. To update individual spins, we make use of the Metropolis Monte Carlo algorithm \cite{MCMC_physics2} with sequential updating scheme \cite{Ren}, whereby individual spins are updated in a fixed sequential order. $\Gamma$ sweeps are performed at fixed $\beta$ to sample from $\tilde{\pi}(\bm{\sigma}\vert \beta, \varepsilon)$, taking measurements at each sweep, before attempting a single Monte Carlo step to sample from $\tilde{\pi}(\beta \vert \bm{\sigma}, \varepsilon)$. The weights $\omega(\beta)$ are determined using the method proposed by Park and Pande \cite{weights_4}.
  
We have checked the correctness of our irreversible simulated tempering algorithms IGS and IMGS and have obtained near perfect agreement for energy density $\mathcal{E}$, magnetisation density $m$ and specific heat capacity $c$ with the Metropolized-Gibbs sampler \cite{Gibbs_ST, MGS}.  see Figure \ref{25by25_system} (top row). Notice that the critical inverse temperature $\beta_c \simeq 0.4407$ of the model is within our predetermined set of temperatures for the simulation: $\beta_1 = 0.5$ to $\beta_K = 0.2$ all equally spaced. We are reminded that the deviation parameter $\delta$ in the skewness function given in \eqref{skewness function} determines the extend to which DBC is violated, $\delta = 0$ recovers DBC. In this light, we are reminded that with $\delta  = 0$ IGS as given in Algorithm \ref{IGS ST algo} decomposes to its reversible counter-part ,the standard Gibbs sampler with transition probability given in \eqref{Gibbs ST}. Likewise the IMGS with transition in \eqref{IMGS SDBC transition} breaks down to its reversible counter-part with DBC: Metropolized-Gibbs sampler with reversible transition probability in \eqref{MGS ST}.   In Figure \eqref{25by25_system} (second and third row) we show the autocorrelation functions for $\beta$, $\mathcal{E}$ and $\chi$ in two temperature domain sizes, $K = 32$ and $K = 512$ for both of our methods IGS (dahed lines) and IMGS (solid lines). From the autocorrelation functions we observe that in both methods deviation from the DBC ($\delta = 0$) accelerates the relaxation dynamics of all three variables. Furthermore the IMGS seems to provide a visibility modest improvement over IGS for all deviation parameters. Since the implementation of IMGS comes with no additional computational cost it is therefore recommended to choose IMGS over IGS.

 To quantify the relaxation dynamics of $ \beta$, $\mathcal{E}$ and $\chi$  we have computed the integrated autocorrelation times $\tau_{int,\beta}, \tau_{int, \chi}, \tau_{int, \mathcal{E}}$. We show these in Figure 5 for various temperature domain sizes $K$ for both IGS (inverted triangles) and IMGS (squares). The numerical gain in relaxation dynamics over their reversible counterparts ($\delta = 0$) is clear for both IGS and IMGS. In particular we observe that $[\tau_{int}]_{\delta = 0}/[\tau_{int}]_{\delta = 1} \sim 3.3$ for all three variables $\beta$, $\mathcal{E}$ and $\chi$ for both IGS and IMGS. Reportedly for all values of $\delta$ the integrated autocorrelation times remain fairly independent of temperature domain size. This observation is consistent with a similar study with the Gibbs sampler \cite{Gibbs_ST}. The seemingly independence of $\tau_{int}$ with respect to $K$ is in  contrast to both MH and IMH whose sampling efficiency degrades with temperature domains of increasing size $K$, as shown in  Figure \ref{12by12_Ising} (bottom row).
 
  Since the IMGS seems to be the best of our two methods, we will henceforth provide performance analysis only with IMGS. In the bottom row of Figure \ref{12by12_Ising}  we compare the integrated autocorrelation times of IMGS with those obtained from simulated tempering with the standard MH of Algorithm \ref{MH_ST} and the IMH algorithm of Sakai and Hukushima \cite{Sakai_Hukushima_simulated_tempering}. The scaling of $\tau_{int,\beta}$ with respect to the temperature domain size $K$ reveals the expectedly diffusive relaxation dynamics of $\beta$ for the MH algorithm, whereby $\tau_{int,\beta}$ scales on the order of $\mathcal{O}(K^2)$. 
  It was numerically demonstrated \cite{Sakai_Hukushima_simulated_tempering} that the IMH algorithm,  which breaks detailed balance, provides a square root reduction in the mixing time for $\beta$, we have reproduced this result to confirm that the relaxation dynamics of $\tau_{int,\beta}$ scales on the order of $\mathcal{O}(K)$ for IMH. The scaling of $\tau_{int,\beta}$ with respect to $K$ may be asymptotically modelled with $\tau_{int} \sim K^z$, in which case we observe that the dynamical scaling exponent $z$ of the IMGS is effectively $\sim 0$ compared to $z \sim 2$ of MH and $z \sim 1$ of IMH. In other words for the IMGS $\tau_{int,\beta}$ scales on the order of $\mathcal{O}(1)$. A similar dynamical scaling behaviour is observed for $\chi$ and $\mathcal{E}$ as shown in Figure \ref{12by12_Ising} (bottom row, second and third columns respectively). Compared to the conventionally used MH algorithm the IMGS provides a decisive gain in the relaxation dynamics of all three variables for all values of $K$, note that the gain accelerates with increasing $K$. Likewise compared to the IMH algorithm we clearly observe a decisive gain in relaxation dynamics with increasing $K$. Even for practically small $K$ values the integrated autocorrelation times are shortened, but by a modest factor, compared to IMH. As for example, the autocorrelation functions for $K = 32$ in Figure \ref{12by12_Ising} (third row)  show that $C(t)$ decays faster compared to both MH and IMH for all three variables $\beta$, $\mathcal{E}$ and $\chi$. 
  
In Figure \ref{12by12_Ising}, the histories of $\beta$, $m$, and $\mathcal{E}$ for $K = 32$ (top row) and $K = 512$ (second row) show a similar pattern to those in Figures \ref{Model_potential_30_K} and \ref{Model_potential_512_K}. While the sampling efficiency of both MH and IMH degrades with increasing values of $K$, that of IMGS seems independent of $K$.  The IMGS provides a numerical gain in relaxation dynamics of $\beta$, $\chi$ and $\mathcal{E}$ compared to its reversible counterpart, the Metropolized-Gibbs sampler. Furthermore, unlike MH and IMH it seems insensitive to increasing temperature domain size. The implementation of IMGS in simulated tempering simulations  may therefore be of interest, particularly so in simulations that may require large temperature domain size $K$. 

\section*{Performance analysis with MD simulations}
In this section we test our methods using Molecular dynamics simulations to sample from the distribution $\pi(\bm{\sigma} \vert \beta, \varepsilon)$. As well as testing the flexibility of our methods we intend to examine applicability to simulations of bimolecular interest. We do this by making use of Alanine Pentapeptide (ALA5) as a test system. Under the assumption of the previous section that IMGS appears to be slightly better than IGS,  we will therefore provide comparison analysis of IMGS with some conventionally used simulated tempering algorithms. In this section we set $\beta = 1/k_BT$ where the Boltzmann constant is given by $k_B  \simeq 1.38 \times10^{-23}$ JKelvin$^{-1}$.

\begin{figure*}[t!]
\centering

\includegraphics[width=.32\textwidth, height=.25\textwidth, ]{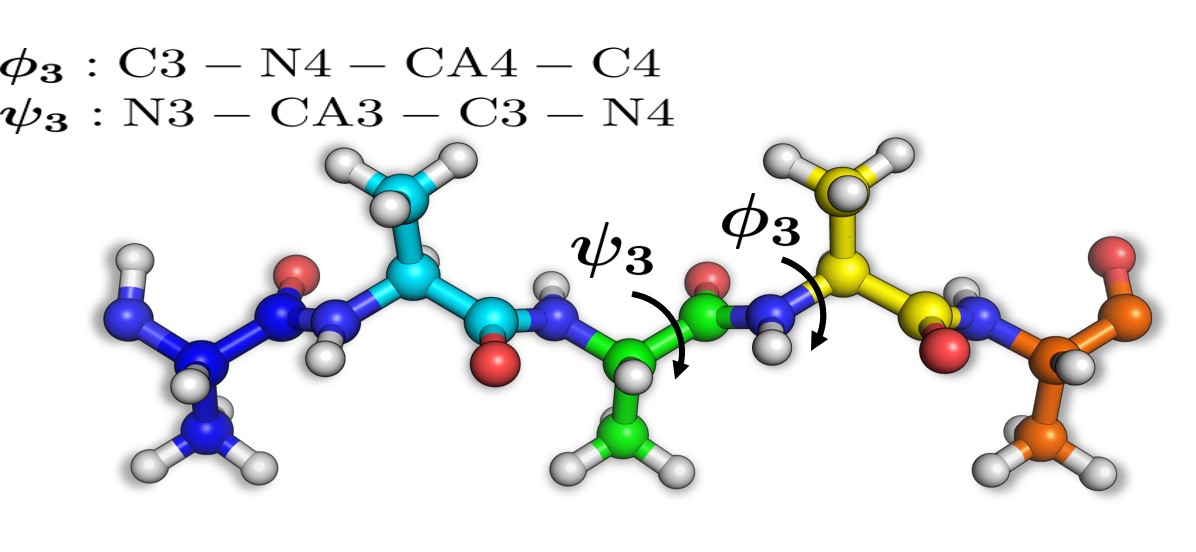}\hfill
\includegraphics[width=.32\textwidth, height=.25\textwidth, ]{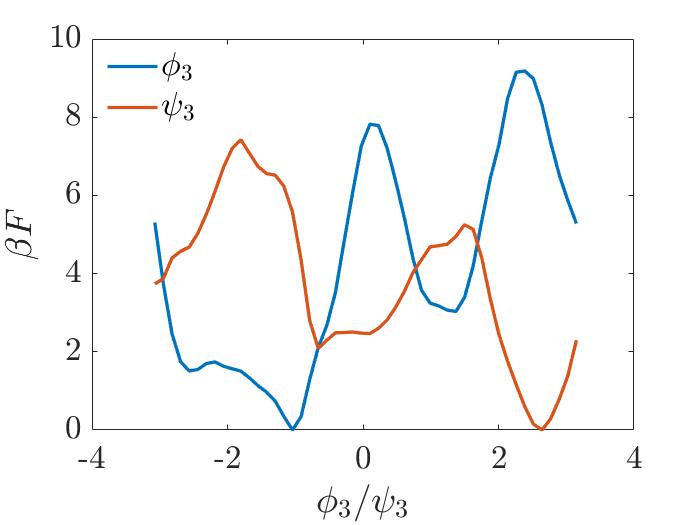}\hfill
\includegraphics[width=.32\textwidth, height=.25\textwidth, ]{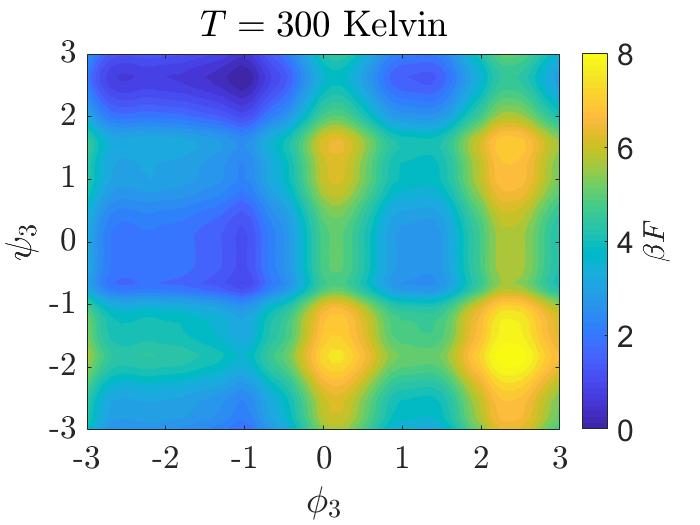}

\caption{\textbf{Left:} Alanine pentapeptide with the slowest relaxing dihedral angles $\phi_3: \text{C}3-\text{N}4-\text{CA}4-\text{C}4$ and $\psi_3: \text{N}3-\text{CA}3-\text{C}3-\text{N}4$. \textbf{Center}: Free energy profiles $\beta F = -\text{ln}(\pi(. \vert \beta))$  in $\phi_3$ and $\psi_3$ obtained from a very long free MD simulation at 300 Kelvin. \textbf{Right:} 2D free energy landscape in $\phi_3$ and $\psi_3$ for ALA5. Units of free energy are in kcal/mol.}

\label{MD_sim_0}
\end{figure*}

\subsection*{Setup}
 We constructed a simple linear model of Alanine pentapeptide (ALA5) whereby the peptide was capped with NTER at the N terminus and CTER at the C terminus. We have performed MD simulations of our ALA5 model (see Figure \ref{MD_sim_0}) with the CHARMM36 Force Field \cite{Forcefield} in explicit water using CHARMM-GUI \cite{CHARMM-GUI} to set up the system. The system was solvated in a rectangular truncated box size of 10 Å edge distance where we have used 3582 TIP3 water molecules and had added 3 K$^+$ and 3 Cl$^-$ counterions to account for a 0.15 M KCl concentration. The simulations were run using NAMD \cite{NAMD} with a time step of 2 fs using a Langevin thermostat with a damping coefficient of 1/ps. We used the Particle Mesh Ewald method \cite{mesh-ewald} in the periodic boundary conditions with a standard cut-off values given in the CHARMM-GUI\cite{CHARMM-GUI} protocols. We had used the standard protocol for the equilibration step of CHARMM-GUI before performing any production run. 

Simulated tempering was performed to alternately sample from the distributions $\tilde{\pi}(\bm{\sigma}\vert \beta, \varepsilon)$ and $\tilde{\pi}(\beta \vert \bm{\sigma}, \varepsilon)$. The distribution $\tilde{\pi}(\bm{\sigma}\vert \beta, \varepsilon)$ was sampled at a fixed temperature for $\Gamma = 0.6$ ps before attempting a single MC trial to sample from $\tilde{\pi}(\beta \vert \bm{\sigma}, \varepsilon)$, i.e. a single trial to update the temperature. Simulated tempering trajectories for $K = 32$ and $K = 512$ temperatures equally spaced between 300 to 500 Kelvin were therefore ran for $\mathcal{T} = 2\times 10^5$ temperature swap iterations. This consisted of $\Gamma = 0.6$ ps of MD sampling per temperature swap iteration, therefore totalling 120 ns of MD sampling per Markov chain. For both $K = 32$ and $K = 512$ temperatures we had performed 6 independent experiments with the same starting structure. Dihedral angles used as descriptors of the system were recorded at every step (2 fs resolution) of the MD simulations. For comparison we have constructed a baseline free energy profile in the slowest relaxing dihedral angle $\phi_3: \text{C}3-\text{N}4-\text{CA}4-\text{C}4$, whereby a single 2 $\mu$s long simulation at the coldest temperature was run with free MD simulation (no simulated tempering) to sample enough for every possible configuration of ALA5 (see Figure \ref{MD_sim_0}). The total set of dihedral angles of the system are: $\phi_1: \text{C}1-\text{N}2-\text{CA}2-\text{C}2$, $\phi_2: \text{C}2-\text{N}3-\text{CA}3-\text{C}3$, $\phi_3: \text{C}3-\text{N}4-\text{CA}4-\text{C}4$, $\phi_4: \text{C}4-\text{N}5-\text{CA}5-\text{C}5$ and $\psi_1: \text{N}1-\text{CA}1-\text{C}1-\text{N}2$, $\psi_2: \text{N}2-\text{CA}2-\text{C}2-\text{N}3$, $\psi_3: \text{N}3-\text{CA}3-\text{C}3-\text{N}4$, $\psi_4: \text{N}4-\text{CA}4-\text{C}4-\text{N}5$.

\begin{figure*}[t!]
\centering

\includegraphics[width=.32\textwidth, height=.25\textwidth, ]{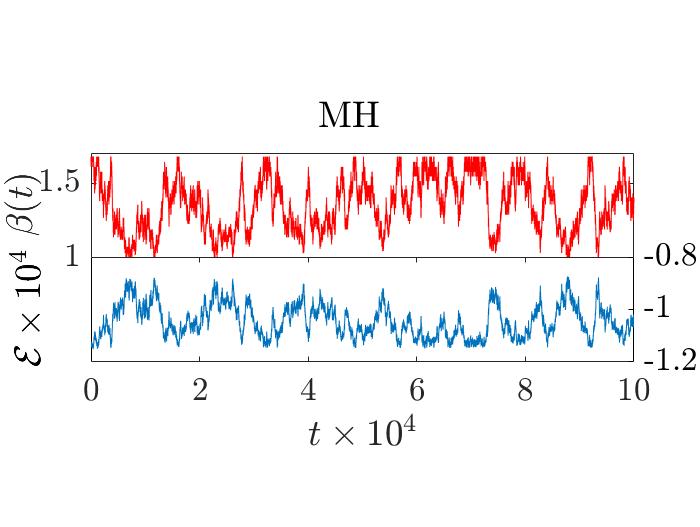}\hfill
\includegraphics[width=.32\textwidth, height=.25\textwidth, ]{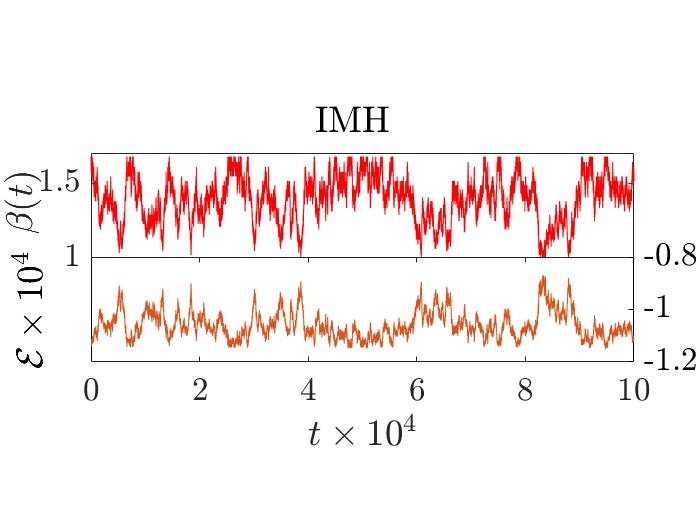}\hfill
\includegraphics[width=.32\textwidth, height=.25\textwidth, ]{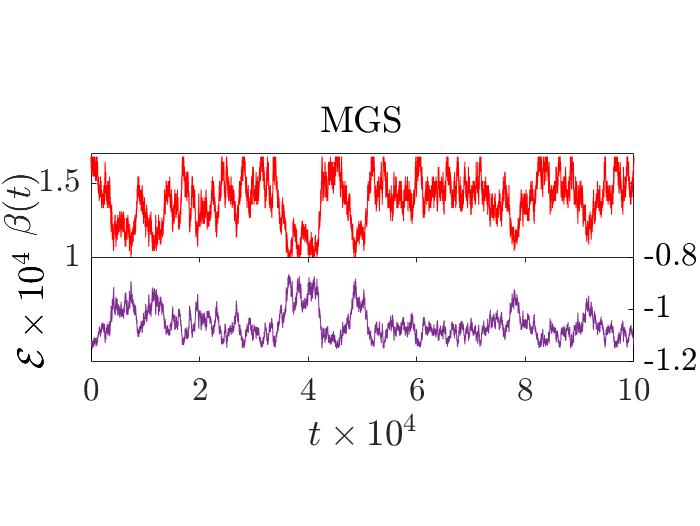}

\includegraphics[width=.32\textwidth, height=.25\textwidth, ]{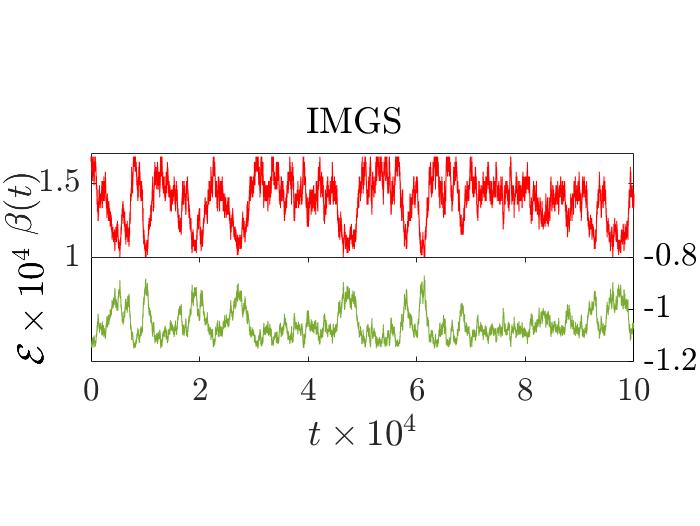}\hfill
\includegraphics[width=.32\textwidth, height=.25\textwidth, ]{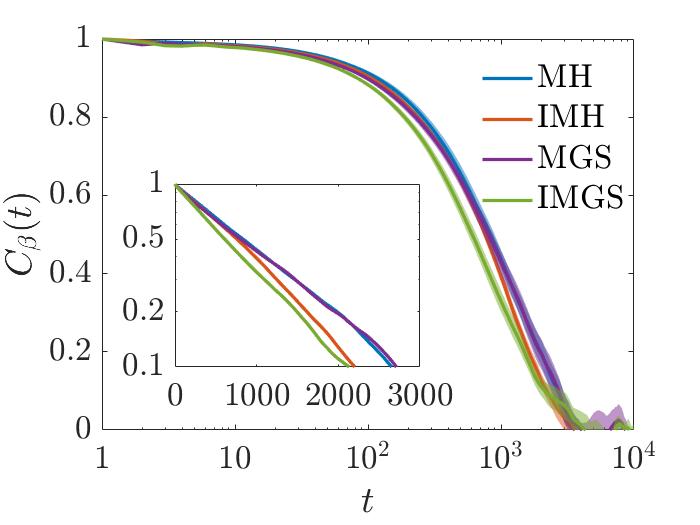}\hfill
\includegraphics[width=.32\textwidth, height=.25\textwidth, ]{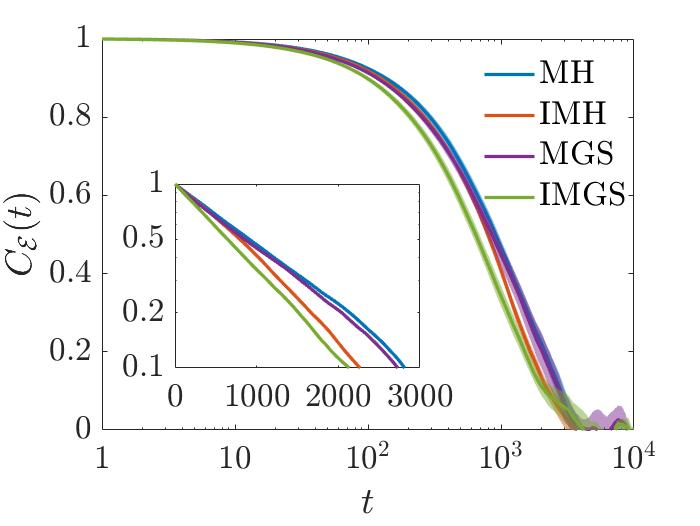}

\includegraphics[width=.32\textwidth, height=.25\textwidth, ]{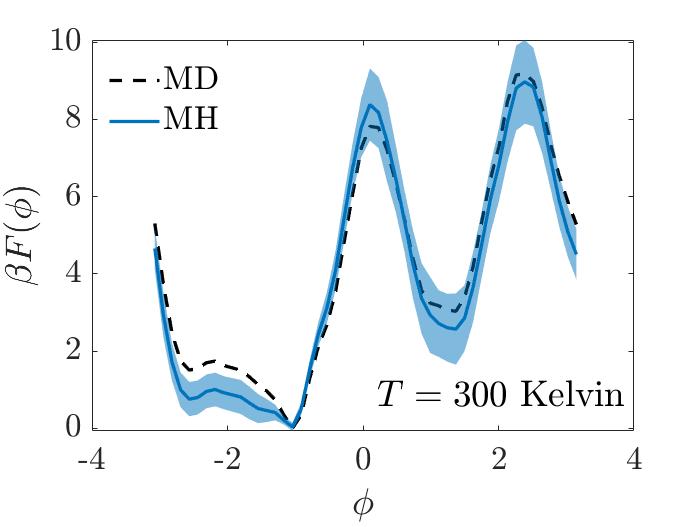}\hfill
\includegraphics[width=.32\textwidth, height=.25\textwidth, ]{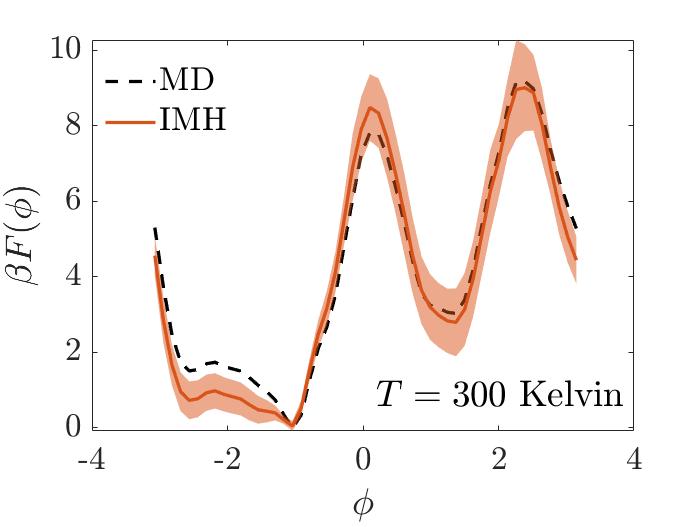}\hfill
\includegraphics[width=.32\textwidth, height=.25\textwidth, ]{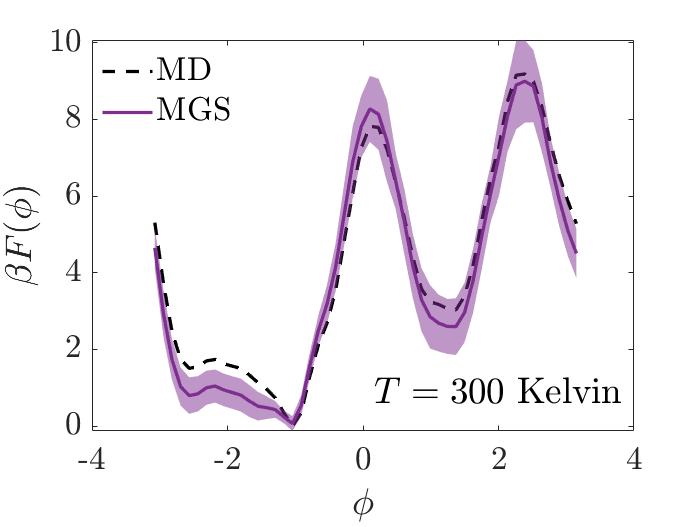}

\includegraphics[width=.32\textwidth, height=.25\textwidth, ]{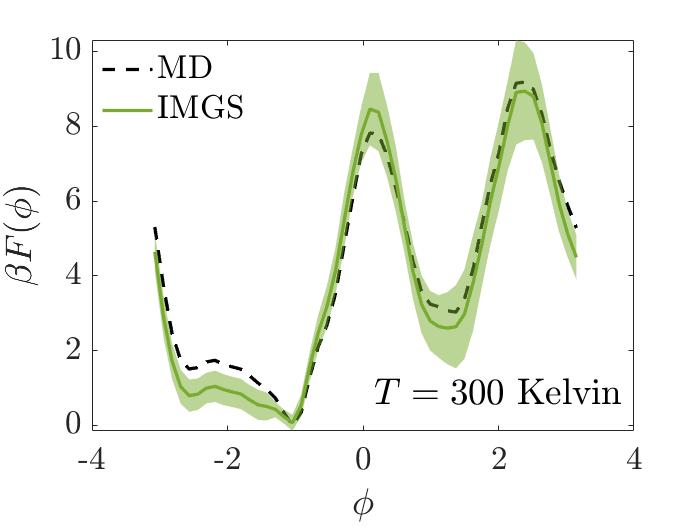}\hfill
\includegraphics[width=.32\textwidth, height=.25\textwidth, ]{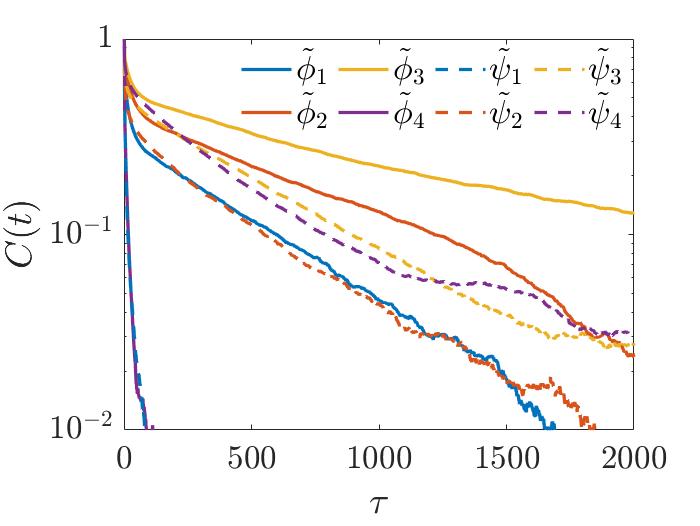}\hfill
\includegraphics[width=.32\textwidth, height=.25\textwidth, ]{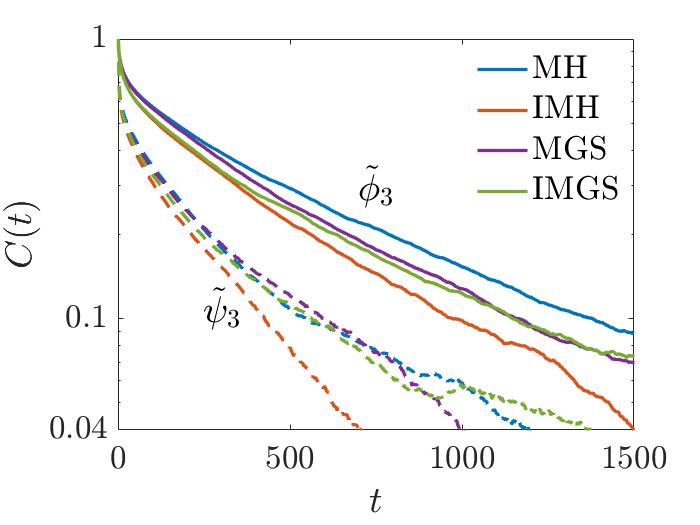}\hfill

\caption{Simulation results for ALA5 with $K = 32$ temperatures. The deviation parameter $\delta = 1 $ for both IMH and IMGS. \textbf{First row and second row, left:} Histories of inverse temperature $\beta$ and total energy $\mathcal{E}$ shown for the first $10^5$ iterations $t$. \textbf{Second row, center and right:} Average trajectory for the autocorrelation functions $C_{\beta}(t)$ and $C_{\mathcal{E}}(t)$, the shaded colours indicate standard error on the average trajectory. The insets  show the same plots with y-axis in the logarithmic scale. \textbf{Third row and fourth row, left:} Average free energy profile trajectories in $\phi_3$ at the lowest temperature, the shaded colours indicate the standard error on the average trajectory. The dashed line serves as a baseline comparison, obtained from a much longer MD simulation with no simulated tempering ($\sim$ 16.7 times longer simulation time). Units of free energy are in kcal/mol. \textbf{Bottom row, center:} Autocorrelation functions of dihedral angles (see main text) obtained from a long MD simulation with no simulated tempering. \textbf{Bottom row, right:}. Autocorrelation functions of the slowest relaxing dihedral angles $\tilde{\phi}_3$ (solid lines) and $\tilde{\psi}_3$ (dashed lines) obtained from simulated tempering.}

\label{MD_sim_1}
\end{figure*}

\begin{figure*}[t!]
\centering

\includegraphics[width=.32\textwidth, height=.25\textwidth, ]{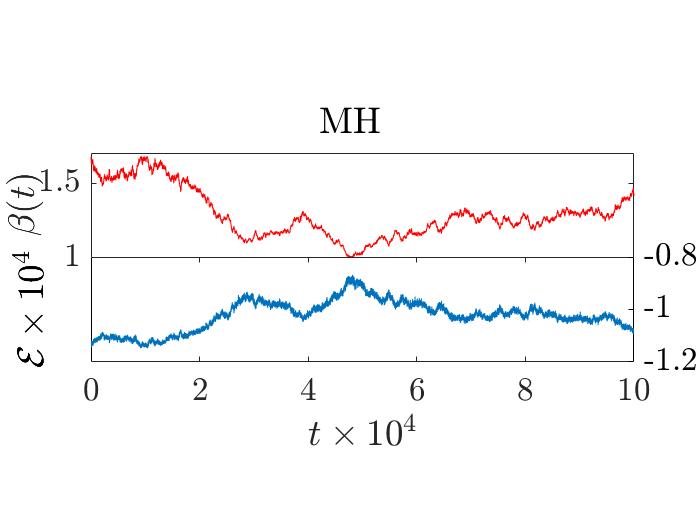}\hfill
\includegraphics[width=.32\textwidth, height=.25\textwidth, ]{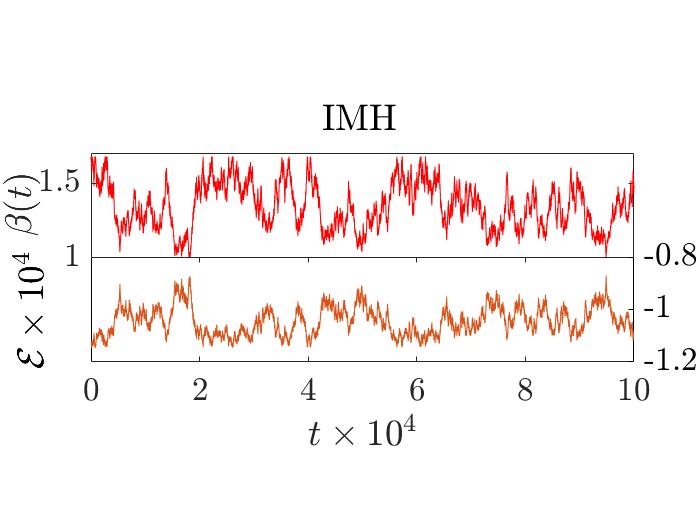}\hfill
\includegraphics[width=.32\textwidth, height=.25\textwidth, ]{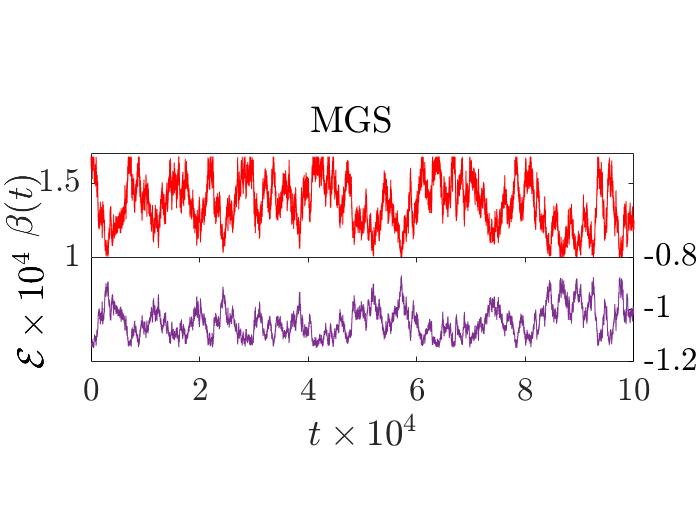}

\includegraphics[width=.32\textwidth, height=.25\textwidth, ]{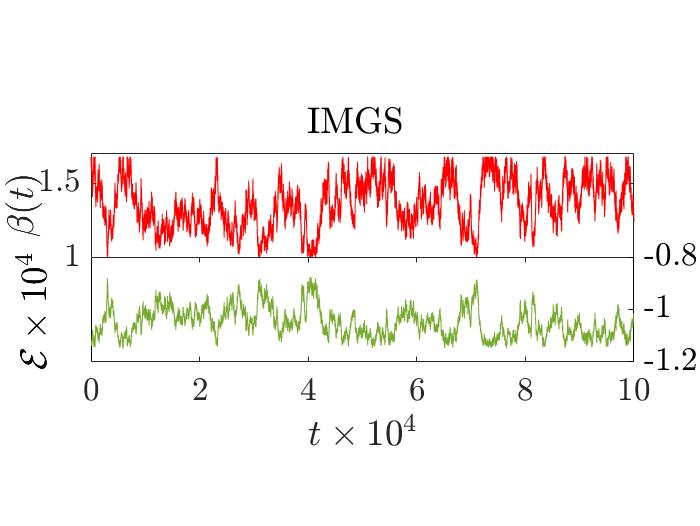}\hfill
\includegraphics[width=.32\textwidth, height=.25\textwidth, ]{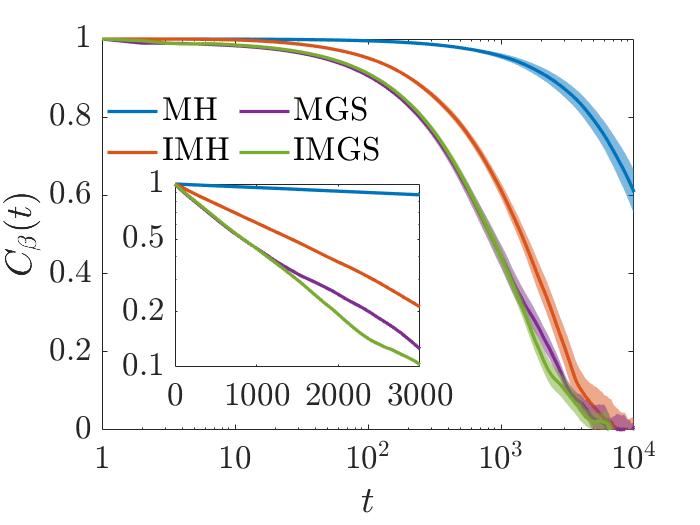}\hfill
\includegraphics[width=.32\textwidth, height=.25\textwidth, ]{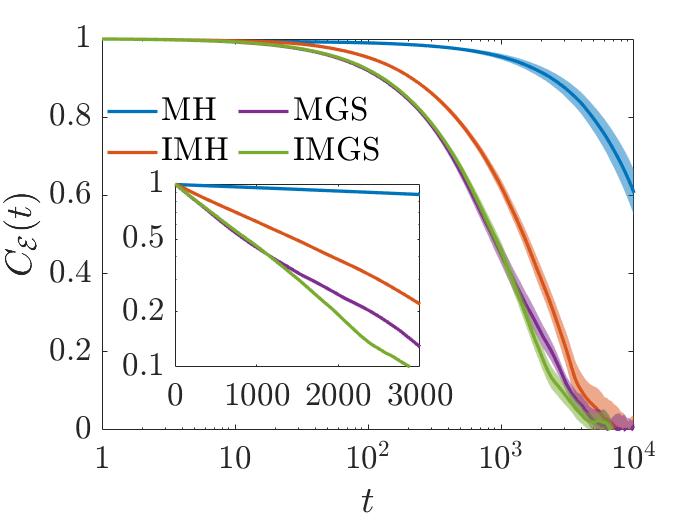}

\includegraphics[width=.32\textwidth, height=.25\textwidth, ]{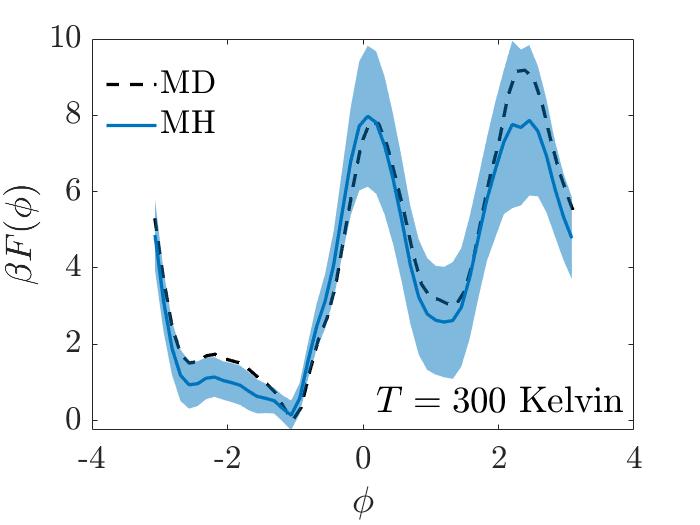}\hfill
\includegraphics[width=.32\textwidth, height=.25\textwidth, ]{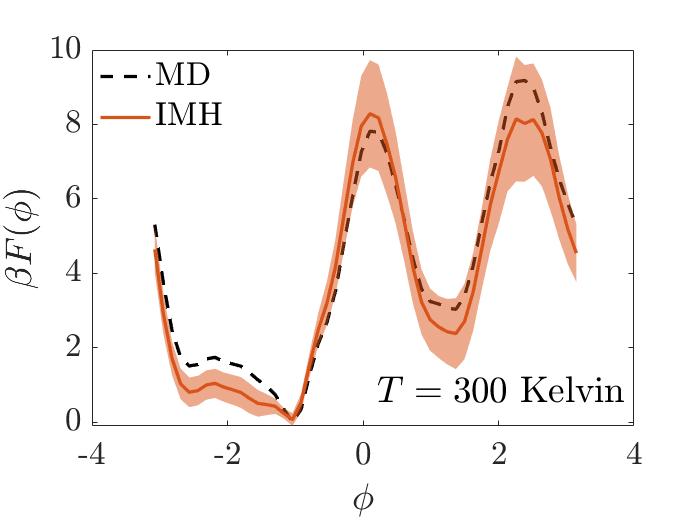}\hfill
\includegraphics[width=.32\textwidth, height=.25\textwidth, ]{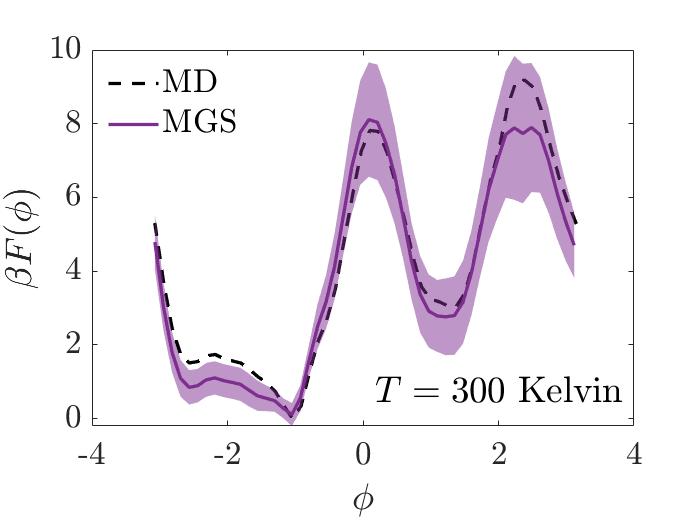}

\includegraphics[width=.32\textwidth, height=.25\textwidth, ]{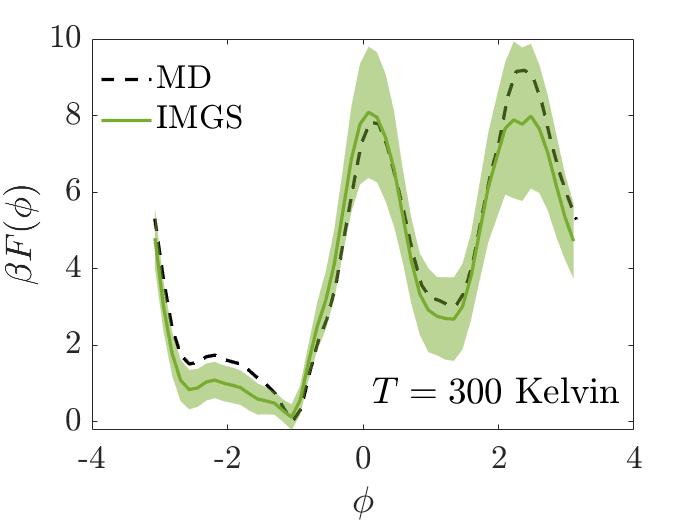}\hfill
\includegraphics[width=.32\textwidth, height=.25\textwidth, ]{dihedral_autocorrelation_sinusoidal.jpg}\hfill
\includegraphics[width=.32\textwidth, height=.25\textwidth, ]{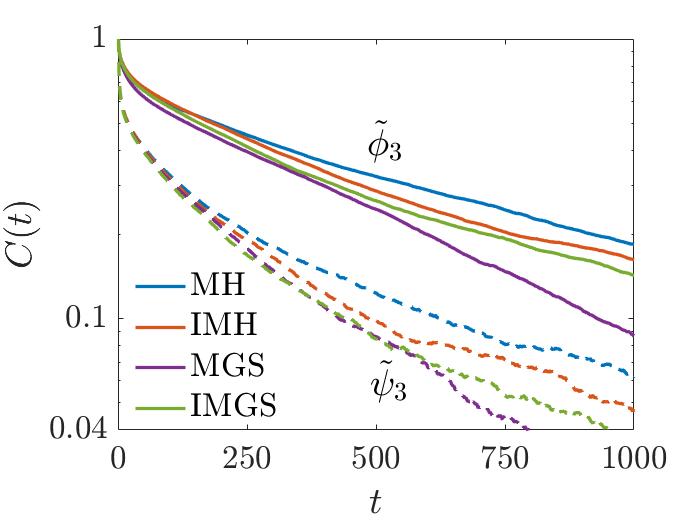}\hfill

\caption{Simulation results for ALA5 with $K = 512$ temperatures. The deviation parameter $\delta = 1 $ for both IMH and IMGS. \textbf{First row and second row, left:} Histories of inverse temperature $\beta$ and total energy $\mathcal{E}$ shown for the first $10^5$ iterations $t$. \textbf{Second row, center and right:} Average trajectory for the autocorrelation functions $C_{\beta}(t)$ and $C_{\mathcal{E}}(t)$, the shaded colours indicate standard error on the average trajectory. The insets  show the same plots with y-axis in the logarithmic scale. \textbf{Third row and fourth row, left:} Average Free energy profile trajectories in $\phi_3$ at the lowest temperature, the shaded colours indicate the standard error on the average trajectory. The dashed line serves as a baseline comparison, obtained from a much longer MD simulation with no simulated tempering ($\sim$ 16.7 times longer simulation time).  Units of free energy are in kcal/mol. \textbf{Bottom row, center:} Autocorrelation functions of dihedral angles (see main text) obtained from a long MD simulation with no simulated tempering. \textbf{Bottom row, right:} Autocorrelation functions of the slowest relaxing dihedral angles $\tilde{\phi}_3$ (solid lines) and $\tilde{\psi}_3$ (dashed lines) obtained from simulated tempering.}

\label{MD_sim_2}
\end{figure*} 

\begin{table}
\setlength{\tabcolsep}{5pt}
\begin{tabular}{c c c c c  }
\toprule
\toprule
&  & &  &  \\
           & \multicolumn{4}{c}{\textbf{Integrated autocorrelation times ($\times 10^3$)}}\\
           &  & &  &  \\
    & $\tau_{int,\beta}$ & $\tau_{int,\mathcal{E}}$ & $\tau_{int, \tilde{\phi}_3}$ & $\tau_{int, \tilde{\psi_3}}$  \\
\midrule
               & \multicolumn{4}{c}{\textbf{K = 32}}\\
                \midrule
    MH     & $2.4 \pm 0.2$    & $2.5 \pm 0.2$    & $1.3 \pm 0.4$  & $0.6 \pm 0.2$  \\
   IMH     & $2.0 \pm 0.1$    & $2.06 \pm 0.09$    & $0.7 \pm 0.1$ & $0.29 \pm 0.03$ \\
    MGS     & $2.3 \pm 0.2$    & $2.4 \pm 0.3$    & $0.8 \pm 0.1$  & $0.38 \pm 0.03$  \\
  IMGS     & $2.0 \pm 0.2$    & $2.1 \pm 0.2$   & $0.9 \pm 0.3$  & $0.42 \pm 0.03$ \\
        \midrule
                 & \multicolumn{4}{c}{\textbf{K = 512}}\\
                \midrule
   MH     & $33.0 \pm 4.0$    & $30.0 \pm 4.0$   & $1.9 \pm 0.7$ & $0.52 \pm 0.07$ \\
 IMH     & $3.9 \pm 0.5$    & $3.9 \pm 0.4$    & $1.2 \pm 0.2$  & $0.44 \pm 0.09$  \\
 MGS     & $2.9 \pm 0.2$   & $2.9 \pm 0.2$   & $0.67 \pm 0.09$ & $0.30 \pm 0.03$  \\
IMGS     & $2.6 \pm 0.2$    & $2.6 \pm 0.2$    & $1.0 \pm 0.2$  & $0.35 \pm 0.08$ \\
    \bottomrule
 &  & &  &  \\
           & \multicolumn{4}{c}{\textbf{Relative speedup}}\\
    &  & &  &  \\
    \midrule
               & \multicolumn{4}{c}{\textbf{K = 32}}\\
                \midrule
    MH    &  $1.0$ & $1.0$  & $1.0$ & $1.0$ \\
   IMH      & $1.2 \pm 0.2$ & $1.2 \pm 0.2$  & $1.9 \pm 0.8$ & $2.1 \pm 0.9$ \\
    MGS     & $1.0 \pm 0.2$ & $1.0 \pm 0.2$  & $1.6 \pm 0.7$ & $1.6 \pm 0.7$ \\
  IMGS     & $1.2 \pm 0.2$ & $1.2 \pm 0.2$  & $1.4 \pm 0.9$ & $1.4 \pm 0.6$ \\
        \midrule
                 & \multicolumn{4}{c}{\textbf{K = 512}}\\
                \midrule
   MH     & $1.0$ & $1.0$ & $1.0$ & $1.0$\\
 IMH     & $8.5 \pm 2.1$ & $7.7 \pm 1.8$  & $1.6 \pm 0.8$ & $1.2 \pm 0.4$  \\
 MGS     & $11.4 \pm 2.2$ & $10.3 \pm 2.1$  & $2.8 \pm 1.4$ & $1.7 \pm 0.4$ \\
IMGS    & $12.7 \pm 2.5$ & $11.5 \pm 2.4$  & $1.9 \pm 1.1$ & $1.5 \pm 0.5$ \\ 
    \bottomrule
    \bottomrule
\end{tabular}

\caption{The integrated autocorrelation times (in units of $\beta$ update iterations $t$) for inverse temperature $\beta$, total energy $\mathcal{E}$ and slowest relaxing dihedral angles $\tilde{\phi}_3$ and $\tilde{\psi_3}$. The relative speed up for a given variable $f$ is defined with respect to the corresponding value $\tau_{int,f}$ of MH.}

\label{table_1}
\end{table}

\subsection*{Results}

In Figure \ref{MD_sim_1} and Figure \ref{MD_sim_2} we show the simulation results for ALA5 for respective $K = 32$ and $K = 512$ temperature domain sizes. For $K = 32$ in Figure  \ref{MD_sim_1}, the histories of $\beta$ for all algorithms exhibit a random walk exploration of temperature space. Note that for IMH and IMGS this observation is in contrast to that for the Ising model, where the IMH seemed to have a more deterministic exploration of temperature space and that of IMGS was ballistic. For both temperature domain sizes, we have plotted an average trajectory (obtained from six independent experiments) for the autocorrelation functions of $\beta$ and total energy of the system $\mathcal{E}$. For $K = 32$ (Figure \ref{MD_sim_1}, second row) the autocorrelation functions of $\beta$ and $\mathcal{E}$ suggest a modest gain for both IMH and IMGS over MH and MGS. The corresponding integrated autocorrelation times $\tau_{int,\beta}$ and $\tau_{int,\mathcal{E}}$ recorded in Table \ref{table_1} indicate a modest gain for IMGS over MH. However no statistically conclusive gain is observed over IMH and MGS. On the other hand, for a larger temperature domain size of $K = 512$, notice from the histories of $\beta$ (Figure \ref{MD_sim_2}, top row) that the mixing rate of $\beta$ and $\mathcal{E}$ for MH is drastically poor compared to the other three algorithms. Examining the corresponding integrated autocorrelation times in Table \ref{table_1} we notice that IMGS returns smaller integrated autocorrelation times for $\beta$ and $\mathcal{E}$ than both MH and IMH, however compared to MGS, yet again, no conclusive gain is observed. it therefore seems that concerning the mixing times of $\beta$  and $\mathcal{E}$, the superiority of IMGS over IMH becomes more distinct in large temperature domain sizes. This is particularly clear when we compare the autocorrelation function for $K = 32$ (Figure \ref{MD_sim_1} second row, center and right) to those for $K = 512$ (Figure \ref{MD_sim_2} second row, center and right).

The dihedral angles for ALA5 are considered the slowest relaxing variables of the system. We had determined the slowest relaxing dihedral angles $(\phi,\psi)$ from a very long free MD simulation (no simulate tempering), see Figure \ref{MD_sim_1} (bottom row, middle). To demonstrate convergence to the correct target distribution we have constructed the free energy profiles in the slowest relaxing dihedral angle $\phi_3$ at the lowest temperature of 300 Kelvin, see Figures \ref{MD_sim_1} and \ref{MD_sim_2} third row. The trajectories shown are the average of six independent experiments, each of which was plotted by re-weighting profiles at all temperatures with respect to the coldest temperature. As a baseline for comparison we have used the free energy profile that is constructed using a very long free MD simulation. The baseline trajectory (shown in dashed line) consist of $\sim 16.7$ times longer MD simulation time than the trajectories constructed using simulated tempering. In Figure \ref{MD_sim_1} (bottom row, right) we show the autocorrelation functions corresponding to the slowest dihedral angles, $\phi_3: \text{C}3-\text{N}4-\text{CA}4-\text{C}4$ and $\psi_3: \text{N}3-\text{CA}3-\text{C}3-\text{N}4$, obtained from simulated tempering simulations. Due to the circular nature of the dihedral coordinates we have chosen to define the sinusoidal functions $\tilde{\phi}_3 = 1/2(\text{cos} \, \phi_3 + \text{sin} \, \phi_3)$ and $\tilde{\psi}_3 = 1/2(\text{cos} \, \psi_3 + \text{sin} \, \psi_3)$ and have plotted in Figure  \ref{MD_sim_1} (bottom row, right) the average autocorrelation functions $C_{\tilde{\phi}_3}(t)$ and $C_{\tilde{\psi}_3}(t)$ obtained from six independent experiments. To quantify the relaxation dynamics of the dihedral angles $\tilde{\phi}_3$ and $\tilde{\psi}_3$ we have computed the corresponding integrated autocorrelation times in Table \ref{table_1}. Keeping in mind the standard error on the mean values we observe a modest gain with IMGS over MH in both temperature domain sizes. However no statistically conclusive gain is observed over IMH and MGS.

\section*{Discussion}
In this paper we have generalized our recently introduced irreversible Gibbs sampler (IGS) and its variant the irreversible Metropolized-Gibbs sampler (IMGS) \cite{EIMCS} to the simulated tempering method. In particular IGS and IMGS, which break DBC but satisfy skewed detailed balance (SDBC), are adapted for the update scheme of inverse temperature $\beta$ for a fixed configuration $\bm{\sigma}$. We have tested the correctness of our methods on a simple system described by a 1D model potential, whose exact weight factors $\omega(\beta)$ can be numerically computed. With this simple system we have demonstrated that our methods provide a significant improvement in the relaxation dynamics of inverse temperature and some system observables over the conventionally used simulated tempering with the Metropolis-Hastings scheme. When compared to the irreversible Metropolis-Hastings (IMH) method of Sakai and Hukushima \cite{Sakai_Hukushima_simulated_tempering}, we observe that the improvement in the mixing time of $\beta$ and system observables accelerates with increasing temperature domain size $K$.

Furthermore, we have tested our methods on the Ising model and have shown that both IGS and IMGS provide a decisive gain in the relaxation dynamics of $\beta$, magnetic susceptibility $\chi$ and energy density $\mathcal{E}$, by as much as 3.3 times when compared to their reversible counter-parts with DBC, respectively the Gibbs sampler and the Metropolized-Gibbs sampler (MGS). We further demonstrate that in both samplers the furthest deviation from the DBC produces the shortest mixing time for $\beta$, $\chi$ and $\mathcal{E}$. For this system too we have provided comparison with MH and IMH. The integrated autocorrelation times $\tau_{int,\beta}, \tau_{int,\chi}, \tau_{int,\mathcal{E}}$ for our methods scale (with respect to $K$) on the order of $\mathcal{O}(1)$. We compare this to MH and IMH which respectively scale on the order of $\mathcal{O}(K^2)$ and $\mathcal{O}(K)$. Our methods outperform the conventionally used MH for all domain sizes $K$. The gain in relaxation times over IMH is modest for small $K$ values, but accelerates with increasing $K$. In summary  when compared to their respective reversible counter-parts, our methods seem to provide a near fixed numerical gain in relaxation times at all temperature domain sizes $K$. However the dynamical scaling of the integrated autocorrelation times suggest that the gain in sampling efficiency over both MH and IMH increases with increasing $K$.

\begin{figure*}[t!]
\centering

\includegraphics[width=.45\textwidth, height=.35\textwidth, ]{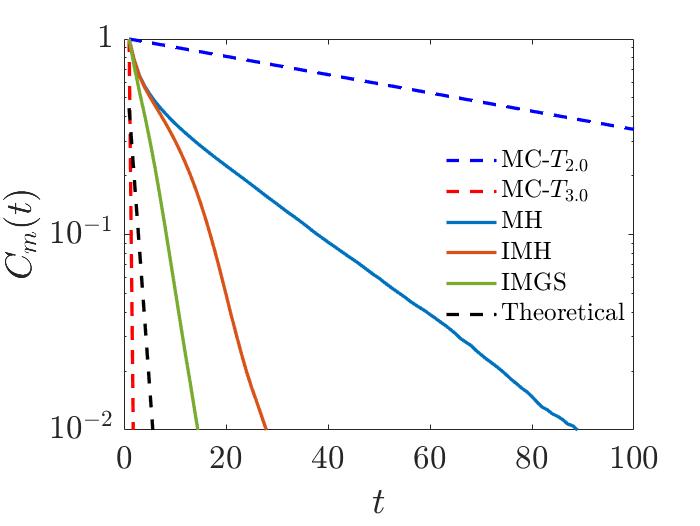}\hfill
\includegraphics[width=.45\textwidth, height=.35\textwidth, ]{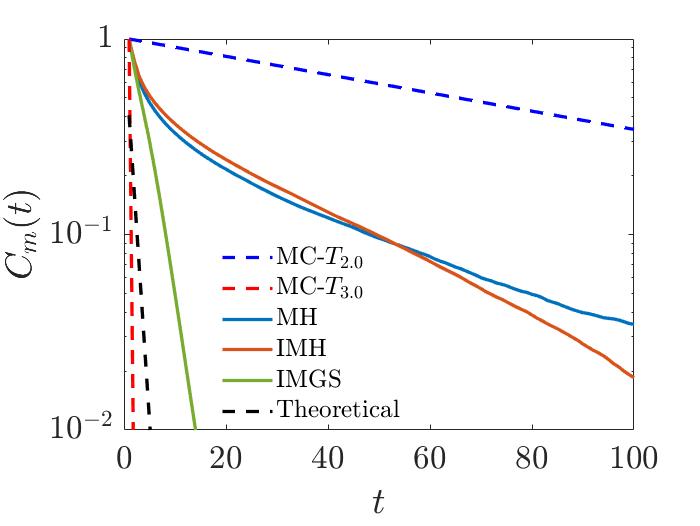}

\caption{Autocorrelation functions of the magnetisation density in the $12 \times 12$ Ising model for $K = 32$ (left) and $K = 512$ (right) inverse temperatures between $\beta_1 = 0.5$ and $\beta_K = 0.33$ all equally spaced. The blue and red dashed lines are respectively the autocorrelation functions obtained from a very long MC simulations at 300 and 500 Kelvin with no simulated tempering. The solid lines are those obtained from simulated tempering simulations. The black dashed line is the theoretically predicted optimum value derived by Rosta and Hummer\cite{Rosta_and_Hummer_1}. The simulation parameters are: $\Gamma = 10^2$ sweeps, $\mathcal{T} = 10^7$ iterations. The deviation parameter $\delta = 1 $ for both IMH and IMGS.}

\label{discussion fig}
\end{figure*} 

We have also performed simulated tempering MD simulations on Alanine 5 (ALA5) with temperature domain sizes, $K = 32$ and $K = 512$, equally spaced between 300 and 500 Kelvin. Guided by the performance on previous two systems we have chosen to test the best of our two methods, namely the IMGS, to compare performance with existing methods. For $K = 32$, the relaxation times of $\beta$ and total energy $\mathcal{E}$ indicate modest improvements for IMGS over MH, but no conclusive gain over IMH and MGS. However at a larger temperature domain size, $K = 512$, IMGS provides a distinct improvement in the relaxation times of $\beta$ and $\mathcal{E}$ over both MH and IMH, but no conclusive gain is observed over its reversible counter-part MGS. This is typical of both MH and IMH which perform optimally with nearest neighbour temperature swaps, therefore with increasing temperature domain size, one expects a less efficient sampling of the temperature space. As for example the integrated autocorrelation time of $\beta$ for MH algorithm scales on the order of $\mathcal{O}(K^2)$, as is expected of a random walk on domains of increasing size, while that of IMH scales on the order of $\mathcal{O}(K)$, as was shown by Sakai and Hukushima \cite{Sakai_Hukushima_simulated_tempering} for the Ising model and reproduced in this paper. The IMGS and its reversible counter-part the MGS however, are not restricted to nearest neighbour temperature swaps and perform a more global exploration of temperature space that provide a better mixing rate.

 The relaxation dynamics of the slowest relaxing dihedral angles $\phi_3: \text{C}3-\text{N}4-\text{CA}4-\text{C}4$ and $\psi_3: \text{N}3-\text{CA}3-\text{C}3-\text{N}4$ indicate that for all three algorithms, IMH, IMGS and MGS, $\phi_3$ and $\psi_3$ relax faster than with the conventionally used MH algorithm. However, the integrated autocorrelation times of the dihedral angles indicate that IMGS performs fairly similarly to IMH and MGS, with no statistically conclusive gain over either method in both temperature domains ($K = 32$ and $K = 512$). The slowest relaxing dihedral angles therefore do not distinguish the performance of our method from IMH and MGS with statistical significance. 
  
A reason for this could be that in conventional simulated tempering, the relaxation rate of a given variable, say the magnetisation of the system in the Ising model, cannot be slower than that at the coldest temperature, and equally it cannot be faster than that at the hottest temperature. The slowest relaxation time of the system in simulated tempering therefore lies somewhere between those at the coldest and hottest temperatures. This follows from Rosta and Hummer's \cite{Rosta_and_Hummer_1} work who had derived an expression for the maximum efficiency gain in simulated tempering simulations with ideally fast mixing rates. We consider the Ising model due to its relative simplicity and ease of generating large amounts of data.  Following Rosta and Hummer's work \cite{Rosta_and_Hummer_1}, we show in Figure \ref{discussion fig} the theoretical prediction for the optimum autocorrelation function of the magnetisation density in our simulated tempering simulations. Also shown are the autocorrelation functions at the coldest and hottest temperatures of the set obtained from very long free MC simulations (no simulated tempering), and those obtained from simulated tempering with  MH, IMH and IMGS. Note that the theoretical prediction lies between those of the coldest and hottest temperatures and the autocorrelation from IMGS lies closest to the theoretically predicted optimum function. We wish to point out with this example that likewise for simulated tempering with ALA5 there exists an optimum efficiency gain concerning the relaxation time of the slowest dihedral angles. It is then feasible that the mixing time of the dihedral angles obtained from all algorithms are relatively close to the optimum predicted value, therefore the other three algorithms (IMH, MGS and IMGS), which have demonstrated substantial gains over MH in the Ising model, now seem to produce only modest gains over MH. Alternatively, the modest speedup could also be because for some systems, it is possible that varying the temperature may not be the optimal collective variable to observe a speedup, and a Hamiltonian-based simulated tempering \cite{HRPM_00, HRPM_01, HRPM_1, HRPM_2, HRPM_3, HRPM_example1} is more suitable with a better chosen collective variable.

An extensive literature exists on techniques for enhancing the sampling efficiency of simulated and parallel tempering simulations. Among several studies, proposals have been made on determining the optimal temperature spacing,\cite{Rosenthal, Kofke, Vousden} frequency of temperature exchange attempts \cite{sindhikara,Abraham} and optimum range and number of temperatures.\cite{Rosta_and_Hummer_1, Rosta_and_Hummer_2} In this paper we have introduced two irreversible methods for temperature swaps that are essentially based on Gibbs sampling techniques. For fair comparison to the widely used MH scheme and other existing methods we have therefore kept other important aspects of the simulation such as temperature spacing and range, number of temperatures and frequency of temperature swaps constant across the different algorithms. However, it is of interest for a further study to also explore some of these aspects  for our irreversible methods. Some mentioned aspects which have been thoroughly studied for the widely used MH scheme, may provide further guidance and isolate optimum parameters for our methods, so as to further boost sampling efficiency.

Some remarks are due in regards to comparison of simulated tempering with parallel tempering. Parallel tempering has often been a more appealing choice because unlike in simulated tempering one does not require the determination of weights $\omega(\beta)$, which are determined either by short trial simulations \cite{weights_4} or continually adjusted throughout the main simulation \cite{weights_5} to ensure uniform sampling of the temperature space. However some studies have provided efficiency comparison between the two methods \cite{ST-PT_compare_1, ST-PT_compare_2, ST-PT_compare_3} outlining some arguments for the use of one method over the other in some cases. Both methods therefore remain in common use. In our current work in progress \cite{Faizi} we implement our algorithms introduced here for both temperature and Hamiltonian-based parallel tempering.

Conventional simulated tempering methods with the MH scheme perform optimally with nearest neighbour temperature swaps so that the temperature change may be accepted with a reasonable probability. Given a fixed temperature range the number of temperatures $K$ needed scales as $\mathcal{O}(\mathcal{N}^{1/2})$ \cite{HRPM_01, REM_spin_glass} for a system with degrees of freedom $\mathcal{N}$. For large complicated systems, such as biomolecules, a large number of temperatures $K$ is therefore required to ensure optimum acceptance probability for temperature swaps. It is particularly for the simulation of large systems that conventional simulated tempering with the MH scheme proves to be inefficient at exploring temperature space, as is expected for a random walk in domains of increasing size. Although Sakai and Hukushima \cite{Sakai_Hukushima_simulated_tempering} had shown with the Ising model that the IMH algorithm provides a square root reduction in the mixing time of inverse temperature as compared to MH. Our study here on the Ising model demonstrates that the mixing times with Gibbs sampling techniques including our irreversible methods scale on the order of $\mathcal{O}(1)$. In addition our methods provide a further numerical gain ($\sim 3.3$ times in the case of a $25 \times 25$ Ising model) in relaxation times over the standard reversible Gibbs sampling techniques. Breaking DBC can therefore pay off. In summary our methods cannot only provide an efficiency gain over the conventionally used MH scheme in all practical temperature domain sizes, but particularly for simulations of large systems that may require large number of temperatures our methods can be  more efficient alternatives to both MH and IMH, which suffer from dynamical scaling with respect to $K$.

 Further practical applications of our methods could extend to larger and/or more complex systems. In applications to ALA5 and biomolecular systems in general, it is worth investigating further if more distinct gains in sampling efficiency can be obtained with our methods using an alternative dynamical variable other than temperature in simulated tempering. For example a dynamical variable in Hamiltonian-based simulated tempering that may be more effective than temperature in flattening the free energy landscape in $\phi/\psi$.

\begin{acknowledgments}
F.F. is supported by the EPSRC Centre for Doctoral Training in Cross-Disciplinary Approaches to Non-Equilibrium Systems (EPSRC reference: EP/L015854/1). E.R. acknowledges support from EPSRC (EP/R013012/1) and the ERC (Project No. 757850 BioNet). The authors acknowledge use of the research computing facility at King’s College London, Rosalind (https://rosalind.kcl.ac.uk).
\end{acknowledgments}

\appendix

\section{The conditional $\tilde{\pi}(\beta,\varepsilon \vert \bm{\sigma})$ satisfies the balance condition}
Here we show that the conditional $\tilde{\pi}(\beta,\varepsilon \vert \bm{\sigma})$ satisfies the balance condition. We do this by writing the balance condition explicitly:
\begin{align}\label{BC for IGS algo}
\tilde{\pi}(\beta, \varepsilon \vert \bm{\sigma}) &= \sum\limits_{\varepsilon'}\sum\limits_{\beta'}\tilde{\pi}(\beta',\varepsilon' \vert \bm{\sigma})\mathcal{G}\left(\beta, \varepsilon, \bm{\sigma} \vert \beta', \varepsilon', \bm{\sigma}\right) \nonumber \\
&= \sum\limits_{\varepsilon'}\sum\limits_{\beta' \neq \beta}\tilde{\pi}(\beta',\varepsilon' \vert \bm{\sigma})\mathcal{G}\left(\beta, \varepsilon, \bm{\sigma} \vert \beta', \varepsilon', \bm{\sigma}\right)\\ \nonumber
& + \sum\limits_{\varepsilon'}\sum\limits_{\beta' = \beta}\tilde{\pi}(\beta',\varepsilon' \vert \bm{\sigma})\mathcal{G}\left(\beta, \varepsilon, \bm{\sigma} \vert \beta', \varepsilon', \bm{\sigma}\right) 
\end{align} 
The first term on the right hand side decomposes to 
\begin{equation}\label{first term proof}
\sum\limits_{\beta' \neq \beta}\tilde{\pi}\left(\beta',\varepsilon \vert \bm{\sigma}\right)\mathcal{G}\left(\beta, \varepsilon, \bm{\sigma} \vert \beta', \varepsilon, \bm{\sigma}\right)
\end{equation}
since $\mathcal{G}\left(\beta, \varepsilon, \bm{\sigma} \vert \beta', -\varepsilon, \bm{\sigma}\right) = 0, \,\,\,\, \forall \,\, \beta' \neq \beta$. The second term on the right hand side of \eqref{BC for IGS algo} breaks down to 
\begin{align}\label{second term proof}
&\tilde{\pi}\left(\beta, \varepsilon \vert \bm{\sigma}\right)\mathcal{G}\left(\beta, \varepsilon, \bm{\sigma} \vert \beta, \varepsilon, \bm{\sigma}\right) + \tilde{\pi}\left(\beta, -\varepsilon \vert \bm{\sigma}\right)\mathcal{G}\left(\beta, \varepsilon, \bm{\sigma} \vert \beta, -\varepsilon, \bm{\sigma}\right) \nonumber \\
 &= \tilde{\pi}(\beta, \varepsilon \vert \bm{\sigma})\gamma^{(\varepsilon)}\left[1 - \frac{1}{\gamma^{(\varepsilon)}}\Lambda\left(\beta,-\varepsilon,\bm{\sigma} \vert \beta, \varepsilon, \bm{\sigma}\right)\right] \nonumber \\
 & \,\,\,\,\,\,\,\,\,\,\,\,\,\,\,\,\,\,\,\,\, + \tilde{\pi}\left(\beta, -\varepsilon \vert \bm{\sigma}\right)\gamma^{(-\varepsilon)}\left[\frac{1}{\gamma^{(-\varepsilon)}}\Lambda\left(\beta,\varepsilon,\bm{\sigma} \vert \beta, -\varepsilon, \bm{\sigma}\right)\right] \nonumber \\
 &= \tilde{\pi}\left(\beta, \varepsilon \vert \bm{\sigma}\right)\gamma^{(\varepsilon)} - \tilde{\pi}\left(\beta, \varepsilon \vert \bm{\sigma}\right)\Lambda\left(\beta, -\varepsilon, \bm{\sigma} \vert \beta, \varepsilon, \bm{\sigma}\right) \nonumber \\
 & \,\,\,\,\,\,\,\,\,\,\,\,\,\,\,\,\,\,\,\,\, + \tilde{\pi}\left(\beta, -\varepsilon \vert \bm{\sigma}\right) \Lambda \left(\beta, \varepsilon, \bm{\sigma} \vert \beta, -\varepsilon, \bm{\sigma}\right)
\end{align}
where $\gamma^{(\varepsilon)} = 1 - \sum\limits_{\beta' \neq \beta}\mathcal{G}\left(\beta',\varepsilon,\bm{\sigma} \vert \beta, \varepsilon, \bm{\sigma}\right) $. We therefore combine \eqref{first term proof} and \eqref{second term proof} to write
\begin{align}
\tilde{\pi}\left(\beta, \varepsilon \vert \bm{\sigma}\right) &= \sum\limits_{\beta'}\pi\left(\beta', \varepsilon \vert \bm{\sigma}\right)\mathcal{G}\left(\beta, \varepsilon, \bm{\sigma} \vert \beta', \varepsilon, \bm{\sigma}\right)\\ \nonumber
& \,\,\,\,\,\,\,\,\,\,\,\,\,\,\,\,\,\,\,\,\, - \tilde{\pi}\left(\beta, \varepsilon, \vert \bm{\sigma}\right)\Lambda\left(\beta, -\varepsilon, \bm{\sigma} \vert \beta, \varepsilon, \bm{\sigma}\right) \nonumber\\
 &  \,\,\,\,\,\,\,\,\,\,\,\,\,\,\,\,\,\,\,\,\,\,\,\,\,\,\,\,\,\,\,\,\,\,\,\,\,\,\,\,\,\,\, + \tilde{\pi}\left(\beta, -\varepsilon \vert \bm{\sigma}\right) \Lambda \left(\beta, \varepsilon, \bm{\sigma} \vert \beta, -\varepsilon, \bm{\sigma}\right) \nonumber \\
 &= \sum\limits_{\beta'}\tilde{\pi}\left(\beta, \varepsilon\vert \bm{\sigma}\right)\mathcal{G}\left(\beta',\varepsilon,\bm{\sigma} \vert \beta, \varepsilon, \bm{\sigma}\right) \nonumber \\
 &= \tilde{\pi}\left(\beta, \varepsilon \vert \bm{\sigma}\right) \nonumber \\
 & = \frac{1}{2}\pi(\beta \vert \bm{\sigma})
\end{align}
where we have obtained the second equality by invoking the balance condition in \eqref{BC ST}.


\begin{thebibliography}{0}%
\makeatletter
\providecommand \@ifxundefined [1]{%
 \@ifx{#1\undefined}
}%
\providecommand \@ifnum [1]{%
 \ifnum #1\expandafter \@firstoftwo
 \else \expandafter \@secondoftwo
 \fi
}%
\providecommand \@ifx [1]{%
 \ifx #1\expandafter \@firstoftwo
 \else \expandafter \@secondoftwo
 \fi
}%
\providecommand \natexlab [1]{#1}%
\providecommand \enquote  [1]{``#1''}%
\providecommand \bibnamefont  [1]{#1}%
\providecommand \bibfnamefont [1]{#1}%
\providecommand \citenamefont [1]{#1}%
\providecommand \href@noop [0]{\@secondoftwo}%
\providecommand \href [0]{\begingroup \@sanitize@url \@href}%
\providecommand \@href[1]{\@@startlink{#1}\@@href}%
\providecommand \@@href[1]{\endgroup#1\@@endlink}%
\providecommand \@sanitize@url [0]{\catcode `\\12\catcode `\$12\catcode
  `\&12\catcode `\#12\catcode `\^12\catcode `\_12\catcode `\%12\relax}%
\providecommand \@@startlink[1]{}%
\providecommand \@@endlink[0]{}%
\providecommand \url  [0]{\begingroup\@sanitize@url \@url }%
\providecommand \@url [1]{\endgroup\@href {#1}{\urlprefix }}%
\providecommand \urlprefix  [0]{URL }%
\providecommand \Eprint [0]{\href }%
\providecommand \doibase [0]{http://dx.doi.org/}%
\providecommand \selectlanguage [0]{\@gobble}%
\providecommand \bibinfo  [0]{\@secondoftwo}%
\providecommand \bibfield  [0]{\@secondoftwo}%
\providecommand \translation [1]{[#1]}%
\providecommand \BibitemOpen [0]{}%
\providecommand \bibitemStop [0]{}%
\providecommand \bibitemNoStop [0]{.\EOS\space}%
\providecommand \EOS [0]{\spacefactor3000\relax}%
\providecommand \BibitemShut  [1]{\csname bibitem#1\endcsname}%
\let\auto@bib@innerbib\@empty
\end{thebibliography}%


\begin{thebibliography}{9}

\bibitem{Metropolis}  N. Metropolis, A.W. Rosenbluth, M.N. Rosenbluth, A.H. Teller, and E. Teller,``Equation of State Calculations by Fast Computing Machines", J. Chem. Phys. \textbf{21}, 1087 (1953).

\bibitem{Hastings}  W.K. Hastings,``Monte Carlo sampling methods using Markov chains and their applications", Biometrika \textbf{57}, 97-109 (1970).

\bibitem{MCMC_physics1} D.P. Landau and K. Binder, \textit{A Guide to Monte Carlo Simulations in Statistical Physics}, 2nd ed. (Cambridge University Press, Cambridge, 2005). 

\bibitem{MCMC_physics2}  M.E.J. Newman and G.T. Barkema, \textit{Monte Carlo Methods in Statistical Physics} (Oxford University Press, New York, 2001).

\bibitem{MCMC_biochemistry}  U.H.E. Hansmann and Y. Okamoto, ``New Monte Carlo algorithms for protein folding", Curr. Opin. Struct. Biol. \textbf{9}, 177-183 (1999).

\bibitem{MCMC_biochemistry2} A. Kolinski and J. Skolnick, ``Monte carlo simulations of protein folding. II. Application to protein A, ROP, and crambin." Proteins: Struct. Funct. Genet. \textbf{18}, 353-366 (1994).

\bibitem{MCMC_finance} R. W. Shonkwiler, \textit{Finance with Monte Carlo} (Springer: New York, 2013). 

\bibitem{Gibbs_sampler} S. Geman and D. Geman, ``Stochastic Relaxation, Gibbs Distributions, and the Bayesian Restoration of Images." IEEE Trans. Pattern Anal. Mach. Intell. \textbf{6}, 721-741 (1984).

\bibitem{Generalized_ensemble} A. Mitsutake, Y. Sugita, and Y. Okamoto, ``Generalized-ensemble algorithms for molecular simulations of biopolymers."  Biopolymers \textbf{60}, 96-123 (2001).

\bibitem{MUCA_1} B.A. Berg and T. Neuhaus, ``Multicanonical algorithms for first order phase transitions", Phys. Lett. B. \textbf{267}, 249-253 (1991).

\bibitem{MUCA_2} B.A. Berg and T. Neuhaus,  ``Multicanonical ensemble: A new approach to simulate first-order phase transitions", Phys. Rev. Lett. \textbf{68}, 9-12 (1992).

\bibitem{STM} E. Marinari and G. Parisi, `` Simulated Tempering: A New Monte Carlo Scheme." EPL \textbf{19}, 451-458 (1992).

\bibitem{REM_1} R.H. Swendsen and J.S. Wang, ``Replica Monte Carlo Simulation of Spin-Glasses", Phys. Rev. Lett. \textbf{57}, 2607 (1986).

\bibitem{REM_2} U.H.E Hansmann, `` Parallel tempering algorithm for conformational studies of biological molecules", Chem. Phys. Lett. \textbf{281}, 140-150 (1997).

\bibitem{REM_3} Y. Sugita and Y. Okamoto, ``Replica-exchange molecular dynamics method for protein folding", Chem. Phys. Lett. \textbf{314}, 141-151 (1999).

\bibitem{BC_sufficiency1} L. Tierney, `` Markov Chains for Exploring Posterior Distributions", Ann. Statist. \textbf{22}, 1701-1728 (1994).

\bibitem{BC_sufficiency2} S. P. Meyn, R. L. Tweedie, \textit{Markov Chains and Stochastic Stability} (Springer-Verlag: London, 1993). 

\bibitem{BC_sufficiency3} V.I. Manousiouthakis and M.W. Deem, ``Strict detailed balance is unnecessary in Monte Carlo simulation," J. Chem. Phys. \textbf{110}, 2753 (1999).

\bibitem{Diaconis}  P. Diaconis, S. Holmes, and R.M. Neal, ``Analysis of a nonreversible Markov chain sampler", Ann. Appl. Probab. \textbf{10}, 726-752 (2000).

\bibitem{Chen} F. Chen, L. Lov\'asz, I. Pak, "Lifting Markov chains to speed up mixing". Proceedings of the 31st Annual ACM Symposium on Theory of Computing, Atlanta, GA, USA, May 1-4, 1999, Association for Computing Machinery: NY, USA, 1999, 275-281. 

\bibitem{Barkema}  R.D. Schram and G.T. Barkema, ``Monte Carlo methods beyond detailed balance", Physica A \textbf{418}, 88-93 (2015).

\bibitem{Suwa-Todo} H. Suwa and S. Todo, ``Markov Chain Monte Carlo Method without Detailed Balance", Phys. Rev. Lett. \textbf{105}, 120603 (2010).

\bibitem{Turitsyn} K.S. Turitsyn, M. Chertkov, and M. Vucelja, ``Irreversible Monte Carlo algorithms for efficient sampling", Physica D \textbf{240}, 410-414 (2011).

\bibitem{Weigel} H.C. Fernandes and M. Weigel, ``Non-reversible Monte Carlo simulations of spin models",  Comput. Phys. Commun. \textbf{182}, 1856-1859 (2011).

\bibitem{Sakai_Hukushima_1D} Y. Sakai and K. Hukushima, `` Dynamics of One-Dimensional Ising Model without Detailed Balance Condition", J. Phys. Soc. Jpn. \textbf{82}, 064003 (2013).

\bibitem{Sakai_Hukushima_2D} K. Hukushima and Y. Sakai, ``An irreversible Markov-chain Monte Carlo method with skew detailed balance conditions", J. Phys. Conf. Ser. \textbf{473}, 012012 (2013).

\bibitem{Sakai_Hukushima_eigenvalue} Y. Sakai and K. Hukushima, ``Eigenvalue analysis of an irreversible random walk with skew detailed balance conditions", Phys. Rev. E \textbf{93}, 043318 (2016).

\bibitem{Sakai_Hukushima_simulated_tempering} Y. Sakai and K. Hukushima, ``Irreversible Simulated Tempering", J. Phys. Soc. Jpn. \textbf{85}, 104002 (2016).

\bibitem{ECMC_continuous_spins}  M. Michel, J. Mayer, and W. Krauth, ``Event-chain Monte Carlo for classical continuous spin models", EPL \textbf{112}, 20003 (2015).

\bibitem{ECMC_heisenberg} Y. Nishikawa, M. Michel, W. Krauth, and K. Hukushima, ``Event-chain algorithm for the Heisenberg model: Evidence for $z \simeq 1$ dynamic scaling", Phys. Rev. E \textbf{92}, 063306 (2015).

\bibitem{hot_topic_1} A. Ichiki and M. Ohzeki, ``Violation of detailed balance accelerates relaxation", Phys. Rev. E \textbf{88}, 020101 (2013).

\bibitem{hot_topic_2} M. Kaiser, R.L. Jack, and J. Zimmer, `` Acceleration of Convergence to Equilibrium in Markov Chains by Breaking Detailed Balance", J. Stat.Phys. \textbf{168}, 259-287 (2017).

\bibitem{MGS} J.S. Liu, ``Peskun's theorem and a modified discrete-state Gibbs sampler", Biometrika \textbf{83}, 681-682 (1996).

\bibitem{EIMCS} F. Faizi, G. Deligiannidis, and E. Rosta, ``Efficient irreversible Monte Carlo samplers", J. Chem. Theory Comput. \textbf{16}, 2124-2138 (2020).

\bibitem{Rosta_1} E. Rosta and G. Hummer, `` Error and efficiency of simulated tempering simulations", J. Chem. Phys. \textbf{132}, 034102 (2010).

\bibitem{Gibbs_ST} J.D. Chodera and M.R. Shirts, ``Replica exchange and expanded ensemble simulations as Gibbs sampling: Simple improvements for enhanced mixing", J. Chem. Phys. \textbf{135}, 194110 (2011).

\bibitem{ST_Suwa-Todo} Y. Mori and H. Okumura, ``Simulated tempering based on global balance or detailed balance conditions: Suwa–Todo, heat bath, and Metropolis algorithms", J. Comput. Chem. \textbf{36}, 2344-2349 (2015).

\bibitem{weights_1} E. Marinari, G. Parisi, J. Ruiz-Lorenzo, ``Numerical Simulations of Spin Glass Systems", (1997), arXiv:cond-mat/9701016 [cond-mat.dis-nn]. arXive e-prints. https://arxiv.org/abs/cond-mat/9701016 (accessed: Apr 13, 2020).

\bibitem{weights_2} A. Irb\"{a}ck and F. Potthast, ``Studies of an off‐lattice model for protein folding: Sequence dependence and improved sampling at finite temperature", J. Chem. Phys. \textbf{103}, 10298 (1995).

\bibitem{weights_3} U.H.E. Hansmann and Y. Okamoto, ``Numerical comparisons of three recently proposed algorithms in the protein folding problem", J. Comput. Chem. \textbf{18}, 920-933 (1997).

\bibitem{weights_4} S. Park and V.S. Pande, ``Choosing weights for simulated tempering", Phys. Rev. E. \textbf{76}, 016703 (2007).

\bibitem{weights_5}  P.H. Nguyen, Y. Okamoto, and P. Derreumaux, ``Communication: Simulated tempering with fast on-the-fly weight determination", J. Chem. Phys. \textbf{138}, 061102 (2013).

\bibitem{weights_6} A. Mitsutake and Y. Okamoto, ``Replica-exchange simulated tempering method for simulations of frustrated systems", Chem. Phys. Lett. \textbf{332}, 131-138 (2000).

\bibitem{Barker} A.A. Barker, ``Monte Carlo calculations of the radial distribution functions for a proton-electron plasma", Aust. J. Phys. \textbf{18}, 119-134 (1965).

\bibitem{Metropolized-Gibbs} L. Pollet, S.M.A. Rombouts, K.V. Houcke, and K. Heyde, Phys. Rev. E. \textbf{70}, 056705 (2004).

\bibitem{Hamiltonian_REM}  H. Fukunishi, O. Watanabe, and S. Takada, ``On the Hamiltonian replica exchange method for efficient sampling of biomolecular systems: Application to protein structure prediction", J. Chem. Phys. 116, 9058 (2002).

\bibitem{Chodera} J.D. Chodera, W.C. Swope, J.W. Pitera, C. Seok, and K.A. Dill, ``Use of the Weighted Histogram Analysis Method for the Analysis of Simulated and Parallel Tempering Simulations", J. Chem. Theory Comput. \textbf{3}, 26-41 (2006).

\bibitem{WHAM}  S. Kumar, J.M. Rosenberg, D. Bouzida, R.H. Swendsen, and P.A. Kollman, ``The weighted histogram analysis method for free‐energy calculations on biomolecules. I. The method", J. Comput. Chem. \textbf{13}, 1011-1021 (1992).

\bibitem{Ren} R. Ren and G. Orkoulas, ``Acceleration of Markov chain Monte Carlo simulations through sequential updating", J. Chem. Phys. \textbf{124}, 064109 (2006).

\bibitem{Huang} K. Huang, Statistical Mechanics, 2nd ed. (New York, Wiley, 1987).

\bibitem{HRPM_00} Y. Sugita, A. Kitao, and Y. Okamoto, `` Multidimensional replica-exchange method for free-energy calculations", J. Chem. Phys. \textbf{113}, 6042 (2000).

\bibitem{HRPM_01}  H. Fukunishi, O. Watanabe, and S. Takada, ``On the Hamiltonian replica exchange method for efficient sampling of biomolecular systems: Application to protein structure prediction", J. Chem. Phys. \textbf{116}, 9058 (2002).

\bibitem{HRPM_1} S.G. Itoh and H. Okumura, `` Hamiltonian replica-permutation method and its applications to an alanine dipeptide and amyloid‐$\beta$(29–42) peptides", J. Comput. Chem. \textbf{34}, 2493-2497 (2013).

\bibitem{HRPM_2} M. Meli and G. Colombo, ``A Hamiltonian Replica Exchange Molecular Dynamics (MD) Method for the Study of Folding, Based on the Analysis of the Stabilization Determinants of Proteins", Int. J. Mol. Sci \textbf{14}, 12157-12169 (2013).

\bibitem{HRPM_3} J. Hritz and C. Oostenbrink, ``Hamiltonian replica exchange molecular dynamics using soft-core interactions",  J. Chem. Phys. \textbf{128}, 144121 (2008).

\bibitem{HRPM_example1} S. Jang, S. Shin, and Y. Pak, ``Replica-Exchange Method Using the Generalized Effective Potential", Phys. Rev. Lett. \textbf{91}, 058305 (2003).

\bibitem{REM_spin_glass} K. Hukushima and K. Nemoto, ``Exchange Monte Carlo Method and Application to Spin Glass Simulations," J. Phys. Soc. Jpn. \textbf{65}, 1604-1608 (1996).

\bibitem{Faizi} Faizi,  F.; Deligiannidis,  G.;  Rosta, E. in preparation.

\bibitem{Rosta_and_Hummer_1} E. Rosta and G. Hummer, ``Error and efficiency of simulated tempering simulations", J. Chem. Phys. \textbf{132}, 034102 (2010).

\bibitem{Rosta_and_Hummer_2} E. Rosta and G. Hummer, ``Error and efficiency of replica exchange molecular dynamics simulation", J. Chem. Phys. \textbf{131}, 165102 (2009).

\bibitem{Rosenthal}  Y.F. Atchad\'{e}, G.O. Roberts, and J.S. Rosenthal, ``Towards optimal scaling of metropolis-coupled Markov chain Monte Carlo",  Stat Comput \textbf{21}, 555-568 (2011).

\bibitem{Kofke} A. Kone and D.A. Kofke, ``Selection of temperature intervals for parallel-tempering simulations", J. Chem. Phys. \textbf{122}, 206101 (2005).

\bibitem{Vousden} W.D. Vousden, W.M. Farr, and I. Mandel, ``Dynamic temperature selection for parallel tempering in Markov chain Monte Carlo simulations", Mon. Not. R. Astron. Soc \textbf{455}, 1919-1937 (2016).

\bibitem{sindhikara} D. Sindhikara, Y. Meng, and A.E. Roitberg, ``Exchange frequency in replica exchange molecular dynamics", J. Chem. Phys. \textbf{128}, 024103 (2008).

\bibitem{Abraham} M.J. Abraham and J.E. Gready, ``Ensuring Mixing Efficiency of Replica-Exchange Molecular Dynamics Simulations", J. Chem. Theory Comput. \textbf{4}, 1119-1128 (2008).

\bibitem{ST-PT_compare_1} S. Park, `` Comparison of the serial and parallel algorithms of generalized ensemble simulations: An analytical approach",  Phys. Rev. E \textbf{77}, 016709 (2008).

\bibitem{ST-PT_compare_2} C. Zhang and J. Ma, ``Comparison of sampling efficiency between simulated tempering and replica exchange", J. Chem. Phys. \textbf{129}, 134112 (2008).

\bibitem{ST-PT_compare_3} C.E. Fiore and M.G.E.D. Luz, ``Comparing parallel- and simulated-tempering-enhanced sampling algorithms at phase-transition regimes", Phys. Rev. E \textbf{82}, 031104 (2010).

\bibitem{CHARMM-GUI} S. Jo, T. Kim, V.G. Iyer, and W. Im, ``CHARMM-GUI: A Web-Based Graphical User Interface for CHARMM", J. Comput. Chem \textbf{29}, 1859-1865 (2008).

\bibitem{NAMD} J.C. Phillips, R. Braun, W. Wang, J. Gumbart, E. Tajkhorshid, E. Villa, C. Chipot, R.D. Skeel, L. Kal\'{e}, and K. Schulten, `` Scalable molecular dynamics with NAMD", J. Comput. Chem \textbf{26}, 1781-1802 (2005).

\bibitem{Forcefield} J. Huang, S. Rauscher, G. Nawrocki, T. Ran, M. Feig, B.L.D. Groot, H. Grubm\"{u}ller, and A.D. Mackerell, ``CHARMM36m: An Improved Force Field for Folded and Intrinsically Disordered Proteins",  Nat. Methods \textbf{14}, 71-73 (2017).

\bibitem{mesh-ewald} T. Darden, D. York, and L. Pedersen, ``Particle mesh Ewald: An Nlog(N) method for Ewald sums in large systems", J. Chem. Phys. \textbf{98}, 10089 (1993).










\end{thebibliography}
\end{document}